\documentclass[11pt,a4paper]{article}

\usepackage[utf8]{inputenc}
\usepackage[T1]{fontenc}
\usepackage[english]{babel}
\usepackage{amsmath,amssymb,amsthm}
\usepackage{newpxtext,newpxmath}
\usepackage{parskip}
\usepackage{mathtools}
\usepackage[margin=2.5cm]{geometry}
\usepackage{hyperref}
\usepackage{cleveref}
\usepackage{enumitem}
\usepackage{tikz}
\usepackage{tikz-cd}
\usepackage{booktabs}
\usepackage{amssymb}
\usepackage{graphicx}
\usepackage[title,titletoc]{appendix}
\usepackage{array}
\usepackage{pgfplots}
\pgfplotsset{compat=1.18}
\usetikzlibrary{positioning, fit}

\usetikzlibrary{babel}
\usetikzlibrary{decorations.pathreplacing}
\usetikzlibrary{arrows.meta, calc, angles, quotes, backgrounds,shapes.geometric}
\usepackage{fancyhdr}
\fancypagestyle{plain}{
	\fancyhf{}
	\fancyfoot[R]{\raisebox{2mm}{\thepage}}
	
}

\hypersetup{
	colorlinks = true,
	linkcolor  = blue!70!black,
	citecolor  = green!50!black,
	urlcolor   = blue!60!black
}

\usepackage{xcolor}

\newtheoremstyle{customthm}
{8pt}{8pt}
{\itshape}
{0pt}
{\bfseries}
{.}
{.5em}
{}

\theoremstyle{customthm}
\newtheorem{definition}{Definition}[section]

\newtheorem{theorem}[definition]{Theorem}
\newtheorem{proposition}[definition]{Proposition}
\newtheorem{lemma}[definition]{Lemma}
\newtheorem{corollary}[definition]{Corollary}
\newtheoremstyle{customrem}
{8pt}{8pt}
{\normalfont}
{0pt}
{\bfseries}
{.}
{.5em}
{}

\theoremstyle{customrem}
\newtheorem{remark}[definition]{Remark}

\setlist[enumerate]{label=(\arabic*)}

\newcommand{\R}{\mathbb{R}}

\newcommand{\N}{\mathbb{N}}
\newcommand{\SW}{\mathrm{SW}}
\newcommand{\RW}{\mathrm{RW}}
\newcommand{\pers}[1]{\operatorname{pers}(#1)}

\DeclareMathOperator{\Lip}{Lip}

\newcommand{\norm}[1]{\left\|#1\right\|}

\newcommand{\dd}{\mathrm{d}}

\makeatletter
\renewcommand{\@seccntformat}[1]{\csname the#1\endcsname\quad}
\makeatother

\title{Projected and Reflected Sliced Wasserstein Kernels for Persistence Diagrams}
\author{\textbf{Julián Uribe Castañeda}\thanks{Email: \texttt{juribec4@estudiante.uc.cl}}\ \\
	\small Faculty of Mathematics, Pontificia Universidad Católica de Chile, Santiago, Chile
	\and
	\textbf{Mircea Petrache}\thanks{Email: \texttt{mpetrache@uc.cl}}\\
	\small Faculty of Mathematics, Pontificia Universidad Católica de Chile, Santiago, Chile}

\date{\today}

\begin{document}
	\maketitle
	
	\begin{abstract}
		\noindent Persistence diagrams do not naturally live in a Hilbert space.
		Carrière, Cuturi, and Oudot embedded them in kernel methods through the
		\emph{Sliced Wasserstein kernel}, built from the Sliced Wasserstein distance
		$\SW_1$, which is conditionally negative definite (CND). We study the projected
		($\SW_p$) and reflected ($\RW_p$) Sliced Wasserstein distances for every order
		$p\ge1$, and compare them with the $p$-Wasserstein distance $d_p$ and with a
		reflected variant $d_p^r$, in which projection onto the diagonal is replaced by
		reflection across it. We show that $2^{1/p}d_p\le d_p^r\le 2\,d_p$ with sharp
		constants, so that $d_1^r=2\,d_1$ and $d_p^r$ inherits the stability of $d_p$,
		although it is a metric only for $p=1$. The distances $\SW_1$ and $\RW_1$ are
		CND metrics, and $\RW_1$ gives rise to a new positive definite kernel, stable
		with respect to $d_1$. For $p>1$, neither $\SW_p$ nor $\RW_p$ satisfies the
		triangle inequality, and rigorous interval-arithmetic certificates prove that
		$\SW_2$ and $\RW_2$ are not CND; numerical experiments suggest that the CND
		property fails for every $p$ above some $p^*\in(1,1.2]$. On diagrams with at most
		$N$ points, $\SW_p$ and $\RW_p$ are metrically equivalent to $d_p$ and $d_p^r$,
		with sharp upper constants. On an orbit recognition benchmark, the kernels built
		from $\SW_1$ and $\RW_1$ reach about $85\%$ accuracy, outperform the persistence
		scale space kernel, perform on par with the persistence weighted Gaussian kernel,
		and are about one percentage point below the kernel built from $\SW_2$, which is
		not positive definite.
	\end{abstract}
	
	
	\newpage
	\setcounter{tocdepth}{2}
	\tableofcontents
	\newpage

	\section{Introduction}
	\label{sec:introduccion}
	
	To use persistence diagrams in kernel methods, one needs a positive
	definite kernel. Several works construct one through an explicit feature
	map into a Hilbert space
	\cite{Reininghaus2015,Kwitt2015,Kusano2016,Kusano2018,Bubenik2015,AdamsPersistenceImages2017}.
	However, such maps cannot preserve the bottleneck distance, not even
	coarsely, since the space of persistence diagrams endowed with this
	distance does not admit a coarse embedding into a Hilbert space
	\cite{BubenikWagner2020}.
	Another approach is to exponentiate a distance $f$: by
	Schoenberg's theorem \cite{Schoenberg1938a,Schoenberg1938b}, the kernel
	$e^{-tf}$ is positive definite for all $t>0$ if and only if $f$ is
	conditionally negative definite (CND). Carrière, Cuturi, and Oudot
	\cite{Carriere2017} followed this approach with the Sliced Wasserstein
	distance $\SW_1$ between diagrams, which projects the diagrams onto lines
	and, to equalize their masses, adds to each diagram the projections onto
	the diagonal of the points of the other diagram. Moreover, they proved
	that $\SW_1$ is CND and defined the kernel $\exp(-\SW_1/(2\sigma^2))$.
	
	Kolouri et al.\ \cite{Kolouri2016} proved that, for probability measures
	on $\R^d$, the square of the Sliced Wasserstein distance of
	order $2$ is CND. However, that distance differs from the one relevant for persitence diagrams, i.e. $\SW_2$:
	the former compares two measures of equal mass directly, whereas in $\SW_p$ the
	measure associated with a diagram $A$ also depends on the other diagram
	$B$. Hence, their argument does not apply to diagrams, and for $p>1$ it
	was not known whether $\SW_p$ is a metric or whether it is CND.
	
	We consider four families of distances between diagrams: the classical
	distance $d_p$, which is a stable metric
	\cite{CohenSteiner2007,CohenSteiner2010,Mileyko2011}; a reflected
	variant $d_p^r$, in which each diagram is completed with the reflection
	of the other diagram across the diagonal (instead of with points on the
	diagonal); the distance $\SW_p$, which is the order-$p$ version of the
	distance of Carrière, Cuturi, and Oudot; and its reflected variant
	$\RW_p$. We determine for which $p$ the last three are metrics, study the
	CND property of $\SW_p$ and $\RW_p$, and compare them with $d_p$ to
	obtain stability results.
	
	The reflected variants are natural for two reasons. First, if each diagram is
	completed with the reflections of the points of the other diagram, instead of
	their projections onto the diagonal, the comparison becomes one between two sets
	of the same cardinality, \(A\cup r_\Delta(B)\) and \(B\cup r_\Delta(A)\), in
	which the diagonal plays no role. Then \(d_p^r\) is an optimal matching distance
	between these two sets, and \(\RW_p\) is the Sliced Wasserstein distance between
	their counting measures (Lemma~\ref{lem:estructura_medidas_p}). Second,
	reflection across the diagonal is already used in persistence kernels, for
	instance in the persistence scale space kernel \cite{Reininghaus2015}, where it
	makes the points on the diagonal contribute nothing. For \(p=1\) nothing is lost
	by reflecting, since \(d_1^r=2\,d_1\), and \(\RW_1\) has the same guarantees as
	\(\SW_1\), so that it provides a second positive definite kernel, different from
	\(k_{\SW_1}\), with the same theoretical properties.
	
	\subsection*{Contributions}
	\label{subsec:contribuciones}
	
	\begin{enumerate}
		\item \underline{Comparison and stability of the reflected distances.}
		For every $p\in[1,\infty]$, we have
		$2^{1/p}d_p\le d_p^r\le 2\,d_p$, with sharp constants. In particular,
		we have $d_1^r=2\,d_1$ and $d_\infty^r(A,B)\le 2\|f-g\|_\infty$
		whenever $A,B$ are the diagrams of two tame functions $f,g$. Moreover,
		the function $d_p^r$ is not a metric for $p>1$.
		
		\item \underline{Metric classification.} The distances $\SW_1$ and
		$\RW_1$ are metrics, whereas for $p>1$ neither $\SW_p$ nor $\RW_p$ is a
		metric. The paper \cite{Carriere2017} treats only $\SW_1$.
		
		\item \underline{A new positive definite kernel.} The distance $\RW_1$
		is CND, so the kernel $k_{\RW_1}$ is positive definite. Moreover, the distance $\RW_1$ is
		stable with respect to $d_1$, since $\RW_1\le (4\sqrt2/\pi)\,d_1$,
		which is the same bound as for $\SW_1$.
		
		\item \underline{$\SW_2$ and $\RW_2$ are not CND.} We prove this by
		means of a rigorous interval-arithmetic certificate that uses the same
		four single-point diagrams for both distances.
		
		\item \underline{Metric equivalence.} On the set $\mathcal{D}_{\le N}$ of diagrams with at
		most $N$ points, the
		distances $\SW_p$ and $\RW_p$ are metrically equivalent to $d_p$ and
		$d_p^r$, with sharp upper constants. In particular, for $p=1$, our
		constants $2/(\pi\Lambda_N)$ and $4\sqrt2/\pi$, where $\Lambda_N=1+2N(2N-1)$, improve the
		constants
		$1/(2\Lambda_N)$ and $2\sqrt2$ of \cite{Carriere2017}.
		
		\item \underline{Experiments.} The failure of the CND property is
		observed numerically for $p>p^*$, with $p^*\in(1,1.2]$. On the orbit
		recognition benchmark, the kernels $k_{\SW_1}$ and $k_{\RW_1}$ achieve
		about $85\%$ accuracy, outperform $k_{\mathrm{PSS}}$, perform on par
		with $k_{\mathrm{PWG}}$, and remain about one percentage point below
		$k_{\SW_2}$, which is not positive definite.
	\end{enumerate}
	
	It remains open whether $\SW_p$ and $\RW_p$ are CND for $p\in(1,2)$ and
	for $p>2$, although the experiments in Subsection~\ref{subsec:exp_csdn}
	suggest that the CND property fails for every $p$ above a threshold
	$p^*\in(1,1.2]$.
	
	\paragraph{Organization of the paper.}
	Section~\ref{sec:preliminares} recalls the distances $d_p$ between
	diagrams, introduces the reflected distance $d_p^r$, and collects the facts
	on one-dimensional optimal transport that we use.
	Section~\ref{sec:kernel} defines $\SW_p$ and $\RW_p$, compares $d_p^r$ with
	$d_p$, and proves that $\SW_1$ and $\RW_1$ are metrics.
	Section~\ref{sec:sw_rw_no_metricas} shows that these properties are lost
	for $p>1$: the triangle inequality fails for $\SW_p$ and $\RW_p$, and
	$\SW_2$ and $\RW_2$ are not CND. Section~\ref{sec:equiv_metrica} turns to a
	property that holds for every $p$, namely the metric equivalence with $d_p$
	and $d_p^r$ on diagrams of bounded cardinality, and deduces the stability
	of $d_\infty^r$. Section~\ref{sec:CND_sw1_rw1} proves that $\SW_1$ and
	$\RW_1$ are CND and discusses what the resulting kernels guarantee in
	learning, and Section~\ref{sec:experimentos} reports the numerical
	experiments. The proof of Theorem~\ref{thm:dp_equivalencias} and some
	auxiliary proofs are deferred to the appendices.
	
	\section{Preliminaries}
	\label{sec:preliminares}
	
	In this section, we fix notation and recall the two ingredients that the
	rest of the paper combines: distances between persistence diagrams and
	the Wasserstein distance on the line. All of this material is classical,
	except for the reflected distance $d_p^{r}$ of
	Definition~\ref{def:dp_r}, which we introduce here. Diagrams and the
	distances $d_p$ are defined as in
	\cite{CohenSteiner2007,CohenSteiner2010,Mileyko2011}, and
	Subsection~\ref{subsec:wasserstein} follows
	\cite{Villani2009,Santambrogio2015}.

	\subsection{Distances between persistence diagrams}
	\label{subsec:distancias}

	Comparing two persistence diagrams amounts to matching their points, but
	the two diagrams need not have the same number of points. The standard
	remedy is to allow a point to be matched with the diagonal, which
	represents features of zero persistence. We make this precise and then
	introduce a variant in which projection onto the diagonal is replaced by
	reflection across it.
	
	\begin{definition}[\textbf{Finite diagrams}]
		\label{def:diagrama_finito}
		Let $\Omega\coloneqq\{(x,y)\in\R^2:x<y\}$. A \emph{finite persistence
			diagram} is a finite set $A$ of triples $(b,d,j)$, with $(b,d)\in\Omega$ and
		$j\in\N$. We denote by $\mathcal{D}_{\mathrm{fin}}$ the set of all such
		diagrams, up to the equivalence of Remark~\ref{rem:diagramas_equivalentes},
		and for $N\in\N$ we set
		$\mathcal{D}_{\le N}\coloneqq\{A\in\mathcal{D}_{\mathrm{fin}}:|A|\le N\}$.
	\end{definition}
	
	\begin{remark}[\textbf{Equivalent diagrams}]
		\label{rem:diagramas_equivalentes}
		The label $j$ serves only to distinguish copies of the same point. For a
		diagram $A$ and $(b,d)\in\Omega$, let
		\[
		m_A(b,d)\coloneqq\#\{j\in\N:(b,d,j)\in A\}
		\]
		be the number of times $(b,d)$ appears in $A$. Two diagrams $A$ and $B$
		are \emph{equivalent}, written $A\sim B$, if $m_A(b,d)=m_B(b,d)$ for every
		$(b,d)\in\Omega$, that is, if the first two coordinates of their triples
		give the same points, counted with multiplicity. For example,
		\[
		\{(1,2,3),(3,4,5),(3,4,6)\}\sim\{(1,2,1),(3,4,7),(3,4,8)\},
		\]
		since both yield the points $(1,2),(3,4),(3,4)$. This is an equivalence
		relation, and equivalent diagrams have the same cardinality. Moreover, all
		the distances in this paper depend only on the first two coordinates of
		the triples, so they take the same value on equivalent diagrams. We
		therefore identify equivalent diagrams, and in statements about metrics
		we write $A=B$ to mean $A\sim B$.
	\end{remark}

	To allow points to be left unmatched, we enlarge a diagram with the
	diagonal, taking countably many copies of each diagonal point.
	
	\begin{definition}[\textbf{Augmented diagram}]
		\label{def:diagrama_aumentado}
		Let $A$ be a finite persistence diagram. The \emph{augmented diagram}
		associated with $A$ is the set $\overline{A} \coloneqq A \cup (\Delta \times \N)$, where $\Delta = \{(x,x): x\in\R\}$ is the diagonal and $
		\Delta \times \N = \{\, (x,x,n) \mid x\in\R,\; n\in\N \,\}$.
		
	\end{definition}

	The third coordinate never enters the norm, and the cheapest diagonal
	partner of a point is its orthogonal projection onto the diagonal.
	
	\begin{remark}[\textbf{Computation of $\norm{x - y}_\infty$ for $x\in\overline{A}$, $y\in\overline{B}$}]
		\label{rem:norma_diagonal}
		For $x=(u,v,w)\in\overline{A}$ and $y=(u',v',w')\in\overline{B}$, only
		the first two components matter:
		$\norm{x-y}_\infty=\max\{|u-u'|,|v-v'|\}$.
		In particular, if $x=(u,v,w)\in A$ and $y=(z,z,n)\in\Delta\times\N$, then
		$\norm{x-y}_\infty=\max\{|u-z|,|v-z|\}$, which is minimized at
		$z=\tfrac{u+v}{2}$, that is, when $(z,z)$ is the orthogonal projection of $(u,v)$ onto
		the diagonal.
	\end{remark}

	A matching is then a bijection between two augmented diagrams that moves
	only finitely many points, and we measure its cost in two ways.
	
	\begin{definition}[\textbf{Partial bijection and cost}]
		\label{def:biyeccion_parcial}
		Let $A,B$ be two persistence diagrams and $\overline{A},\overline{B}$
		their augmented diagrams. A \emph{partial bijection} between $A$ and
		$B$ is a bijection $
		\gamma:\overline{A}\to\overline{B}$
		such that the set $
		\{x\in\overline{A} : \|x-\gamma(x)\|_{\infty}>0\}$
		is finite. Thus, each point of $A$ is matched either with a point of
		$B$ or with a diagonal triple in $\Delta\times\N$, and
		similarly for the points of $B$.
		
		\begin{itemize}
			\item The \emph{cost} of $\gamma$ in $\ell^\infty$ is $c(\gamma) \coloneqq \sup\limits_{x\in\overline{A}} \|x-\gamma(x)\|_\infty$.
			
			\item For $p\in[1,\infty)$, the $\ell^p$ cost is $\displaystyle c_p(\gamma) \coloneqq \Bigl( \sum\limits_{x\in\overline{A}} \|x-\gamma(x)\|_\infty^p \Bigr)^{1/p}$. Since $\gamma$ moves only finitely many points, the sum above has
			only finitely many nonzero terms.
		\end{itemize}
	\end{definition}

	Optimizing over all such matchings yields the classical family of
	distances between diagrams.
	
	\begin{definition}[\textbf{Projected distance $d_p$}]
		\label{def:dp}
		The \emph{bottleneck distance} between $A$ and $B$ is $d_\infty(A,B) \coloneqq \inf\limits_{\gamma} c(\gamma)$,
		where the infimum is taken over all partial bijections $
		\gamma:\overline{A}\to\overline{B}$
		between $A$ and $B$. For $p\in[1,\infty)$, the \emph{projected
			$p$-distance between diagrams} is $d_p(A,B) \coloneqq \inf \limits_{\gamma} c_p(\gamma)$.
	\end{definition}

	\begin{remark}[\textbf{The role of the third component in the augmented diagonal}]
		\label{rem:tercera_componente}
		Like the label $j$ of a diagram, the third component $n\in\N$ of
		$\Delta\times\N$ only distinguishes copies of the same point. Since $\N$
		is infinite, each diagonal point $(z,z)$ has infinitely many copies
		$(z,z,1),(z,z,2),\dots$, all at distance $0$ from each other. This is
		what allows $\gamma$ to be a bijection. Indeed, several points of $A$
		with the same projection $(z,z)$ are sent to different copies of
		$(z,z)$, and the remaining copies are sent to copies of the same point,
		at cost $0$.
	\end{remark}
	
	The two costs of Definition~\ref{def:biyeccion_parcial} are related, and
	consequently the bottleneck distance is the limit of $d_p$ as
	$p\to\infty$.
	
	\begin{remark}[\textbf{The limit $p\to\infty$}]
		\label{rem:p_infinito}
		Let $\gamma:\overline{A}\to \overline{B}$ be a partial bijection between finite diagrams $A$ and $B$, and
		let $k$ be the number of $x\in\overline{A}$ with $\|x-\gamma(x)\|_\infty>0$. Since
		$c(\gamma)^p\le\sum\limits_{x\in\overline{A}}\|x-\gamma(x)\|_\infty^p\le k\,c(\gamma)^p$, we have
		\[
		c(\gamma)\le c_p(\gamma)\le k^{1/p}\,c(\gamma),
		\]
		and the cost $c_p(\gamma)$ is nonincreasing in $p$. Hence
		$\lim\limits_{p\to\infty} c_p(\gamma)= c(\gamma)$.
		Moreover, we may modify $\gamma$ on the triples of $\Delta\times\N$ that
		are not sent to points of $B$, so that each of them is sent to a copy of
		the same diagonal point (for each diagonal point, infinitely many copies
		remain on both sides). This does not increase $c(\gamma)$ or
		$c_p(\gamma)$, and afterwards each $x$ with $\|x-\gamma(x)\|_\infty>0$
		is either a point of $A$ or a triple of $\Delta\times\N$ sent to a point
		of $B$; hence $k\le|A|+|B|$. Taking the infimum over $\gamma$, we obtain
		\[
		d_\infty(A,B)\le d_p(A,B)\le(|A|+|B|)^{1/p}\,d_\infty(A,B),
		\]
		and hence $d_p(A,B)$ is nonincreasing in $p$ and
		$d_\infty(A,B)=\lim\limits_{p\to\infty}d_p(A,B)$.
	\end{remark}

	Instead of sending an unmatched point to the diagonal, one may send it
	across the diagonal. This yields a second family of distances, which is
	the one adapted to the reflected Sliced Wasserstein distance of
	Section~\ref{sec:kernel}; we compare it with $d_p$ in
	Subsection~\ref{subsec:dp_dpr}.
	
	\begin{definition}[\textbf{Reflected distance $d_p^{r}$}]
		\label{def:dp_r}
		Let $A,B\in\mathcal{D}_{\mathrm{fin}}$. Consider the augmented sets $P \coloneqq A \cup r_\Delta(B)$ and $Q \coloneqq B \cup r_\Delta(A)$,
		where
		$r_\Delta(b,d)=(d,b)$ denotes the reflection across the diagonal. For
		$1\le p<\infty$, we define $\displaystyle
		d_p^{r}(A,B)
		\coloneqq
		\inf \limits_{\gamma:P\to Q}
		\left(
		\sum\limits_{x\in P} \|x - \gamma(x)\|_\infty^p
		\right)^{\!1/p}$,
		where the infimum is taken over all bijections $\gamma$ between
		$P$ and $Q$. For $p=\infty$, we define $
		d_\infty^{r}(A,B)\coloneqq\inf \limits_{\gamma:P\to Q}\,\max \limits_{x\in P}\|x-\gamma(x)\|_\infty $.
		Since $P$ and $Q$ are finite, the infima are minima. Unlike for $d_p$,
		the diagonal plays no role here, since each point can be matched either
		with a point of the other diagram or with a reflected point.
	\end{definition}
	
	As with $d_p$, the case $p=\infty$ is the limit of the cases $p<\infty$.
	
	\begin{remark}[\textbf{The limit $p\to\infty$ for $d_p^{r}$}]
		\label{rem:dpr_infinito}
		Let $A,B\in\mathcal{D}_{\mathrm{fin}}$, so that $|P|=|Q|=|A|+|B|$. As in
		Remark~\ref{rem:p_infinito}, every bijection $\gamma:P\to Q$ satisfies
		\[
		\max_{x\in P}\|x-\gamma(x)\|_\infty
		\le\Bigl(\sum_{x\in P}\|x-\gamma(x)\|_\infty^p\Bigr)^{1/p}
		\le(|A|+|B|)^{1/p}\max_{x\in P}\|x-\gamma(x)\|_\infty ,
		\]
		and the middle term is nonincreasing in $p$. Taking the minimum over the
		finitely many bijections $\gamma$, we obtain
		\[
		d_\infty^{r}(A,B)\le d_p^{r}(A,B)\le(|A|+|B|)^{1/p}\,d_\infty^{r}(A,B),
		\]
		and hence $d_p^{r}(A,B)$ is nonincreasing in $p$ and
		$d_\infty^{r}(A,B)=\lim\limits_{p\to\infty}d_p^{r}(A,B)$.
	\end{remark}

	Both constructions rely on the following elementary properties of
	$\pi_\Delta$ and $r_\Delta$, which we use repeatedly.
	
	\begin{remark}[\textbf{Geometric properties of the diagonal}]
		\label{rem:prop_diagonal}
		For each point $u=(u_1,u_2)\in\R^2$, we write $\pi_\Delta(u) \coloneqq \left(\frac{u_1+u_2}{2},\frac{u_1+u_2}{2}\right)$ and
		$r_\Delta(u)\coloneqq(u_2,u_1)$. Then the following hold.
		\begin{enumerate}
			\item  $ \|u-\pi_\Delta(u)\|_\infty
			\leq
			\|u-(t,t)\|_\infty$ for all
			$t\in\R$ and all $u=(u_1,u_2)\in\R^2$.

			\item $\|\pi_\Delta(u)-\pi_\Delta(v)\|_\infty
			\le
			\|u-v\|_\infty$ for all $u,v\in \R^2$.
			
			\item  $\|u-r_\Delta(v)\|_\infty\ge\|u-\pi_\Delta(u)\|_\infty+\|r_\Delta(v)-\pi_\Delta(r_\Delta(v))\|_\infty$ for all $u,v\in \Omega$.

			\item The map $r_\Delta$ is an isometry of
			$(\R^2,\|\cdot\|_\infty)$, and
			$\|u-r_\Delta(u)\|_\infty=|u_1-u_2|=2\,\|u-\pi_\Delta(u)\|_\infty$.
		\end{enumerate}
	\end{remark}

	We close this subsection with two reformulations of $d_p$. In the first,
	the diagonal copies are chosen explicitly; in the second, they are
	eliminated in favor of a partial matching between $A$ and $B$, together
	with a penalty for the points it leaves unmatched.
	
	\begin{theorem}[\textbf{Equivalent formulations of $d_p$}]
		\label{thm:dp_equivalencias}
		Let $A,B\in\mathcal{D}_{\mathrm{fin}}$ and $p\in[1,\infty)$, and for
		$x\in\Omega\times\N$ let $\delta(x)\coloneqq\|x-\pi_\Delta(x)\|_\infty$. Then the distance
		$d_p$ of Definition~\ref{def:dp} admits the following two
		formulations.
		\begin{enumerate}
			\item $\displaystyle
			d_p(A,B) = \min\limits_{\substack{A_0,B_0 \subseteq \Delta \times \N \\ |A_0|=|A|,\, |B_0|=|B|}} \;
			\min\limits_{\substack{s : A \cup B_0 \to A_0 \cup B \\ s \text{ bijection}}} \;
			\Bigl( \sum\limits_{x \in A \cup B_0} \|x - s(x)\|_\infty^p \Bigr)^{\!1/p}$.
			
			\item $\displaystyle d_p(A,B)
			=
			\min_{\substack{A_1\subseteq A,\ A_2\coloneqq A\setminus A_1\\
					B_1\subseteq B,\ B_2\coloneqq B\setminus B_1\\
					\gamma:A_1\to B_1\ \text{bijection}}}
			\Bigl(
			\sum \limits_{a\in A_1}\|a-\gamma(a)\|_\infty^p
			+\sum \limits_{a\in A_2}\delta(a)^p
			+\sum \limits_{b\in B_2}\delta(b)^p
			\Bigr)^{1/p}$.
		\end{enumerate}
		Both formulations also hold for $p=\infty$, with each sum of $p$-th
		powers replaced by a maximum and $d_p$ replaced by $d_\infty$. In
		particular, we have $\displaystyle
		d_\infty(A,B)=\min_{A_1,B_1,\gamma}\
		\max\Bigl\{\max_{a\in A_1}\|a-\gamma(a)\|_\infty,\
		\max_{a\in A_2}\delta(a),\ \max_{b\in B_2}\delta(b)\Bigr\}$.
	\end{theorem}
	
	Up to notation, and with an infimum in place of the minimum,
	formulation~(1) is the definition of the $p$-Wasserstein distance for the
	metric pair $(\Omega\cup\Delta,\|\cdot\|_\infty,\Delta)$ given in
	\cite[Def.~4.12]{BubenikElchesen2022}, and formulation~(2) corresponds to
	the reformulation stated without proof in
	\cite[Lemma~4.13]{BubenikElchesen2022}. See also
	\cite{DivolLacombe2021}, where these distances are written as optimal
	partial transport problems. A related reduction, namely to a perfect
	matching between $A\cup\pi_\Delta(B)$ and $B\cup\pi_\Delta(A)$ in which
	pairs of diagonal points have zero cost, is the Reduction Lemma of
	\cite[Ch.~VIII.4]{EdelsbrunnerHarer2010}; see also
	\cite[Lemma~2.2]{Kerber2017}. Since several proofs below use the fact that
	the minima are attained, we prove Theorem~\ref{thm:dp_equivalencias} in
	Appendix~\ref{app:D}, by means of one lemma for each formulation.

	\subsection{Wasserstein distance on $\R$}
	\label{subsec:wasserstein}

	The Wasserstein distances measure the minimal transport cost between two
	positive Borel measures of equal total mass. In the one-dimensional case,
	these distances admit a closed-form expression in terms of the quantile
	functions of the measures involved, which makes them particularly well
	suited to the purposes of this work.

	\begin{definition}[\textbf{Borel measures of fixed mass}]
		\label{def:medidas_masa_fija}
		Given a real number $r>0$, we denote by $\mathcal{M}_r(\R)$ the set
		of all positive Borel measures $\mu$ on $\R$ such that
		$\mu(\R)=r$. For $\mu\in\mathcal{M}_r(\R)$, we write
		$F_\mu(x)\coloneqq\mu((-\infty,x])$ for its distribution function and
		$F_\mu^{-1}(t)\coloneqq\inf\{x\in\R:F_\mu(x)\ge t\}$ for its generalized
		inverse.
	\end{definition}

	\begin{definition}[\textbf{$p$-Wasserstein distance}]
		\label{def:wasserstein_p}
		Let $p\in[1,\infty)$, $r>0$ and $\mu,\nu\in\mathcal{M}_r(\R)$. The
		\emph{$p$-Wasserstein distance} between $\mu$ and $\nu$ is
		\[
		\mathcal{W}_p(\mu,\nu)\coloneqq\left(\inf_{\pi\in\Pi(\mu,\nu)}
		\iint\limits_{\R\times\R}|x-y|^p\,\dd\pi(x,y)\right)^{1/p},
		\]
		where $\Pi(\mu,\nu)$ is the set of Borel measures $\pi$ on $\R^2$
		such that $\pi(A\times\R)=\mu(A)$ and $\pi(\R\times B)=\nu(B)$ for all
		Borel sets $A,B\subseteq\R$. This is the primal Kantorovich
		formulation of $\mathcal{W}_p$.
	\end{definition}
	
	Each of the following formulations is used below: the quantile
	formulation for empirical measures, and the last two items for $\SW_1$
	and $\RW_1$.
	
	\begin{proposition}[\textbf{Equivalent formulations of $\mathcal{W}_p$}]
		\label{prop:equiv_p}
		Let $p\in[1,\infty)$, $r>0$ and $\mu,\nu\in\mathcal{M}_r(\R)$. Then the
		following hold.
		\begin{enumerate}
			\item \emph{(Quantile and dual formulations.)}
			\[
			\mathcal{W}_p(\mu,\nu)
			=\left(r\int\limits_0^1
			|F_{\mu/r}^{-1}(t)-F_{\nu/r}^{-1}(t)|^p\,\dd t\right)^{\!1/p}
			=\left(\sup\left\{\int\limits_\R\varphi\,\dd\mu
			+\int\limits_\R\psi\,\dd\nu\right\}\right)^{\!1/p},
			\]
			where the supremum is taken over all pairs
			$(\varphi,\psi)\in L^1(\mu)\times L^1(\nu)$ such that
			$\varphi(x)+\psi(y)\le|x-y|^p$ for all $x,y\in\R$.
			
			\item \emph{(Distribution-function formulation, $p=1$.)}
			$\displaystyle
			\mathcal{W}_1(\mu,\nu)=\int\limits_{-\infty}^{\infty}
			|F_\mu(x)-F_\nu(x)|\,\dd x$, and the dual formulation reduces to the
			Kantorovich--Rubinstein formula
			$\displaystyle
			\mathcal{W}_1(\mu,\nu)=\sup\limits_{\Lip(f)\le1}\int\limits_\R f\,\dd(\mu-\nu)$.
			
			\item \emph{(Invariance, $p=1$.)} For all $s>0$ and
			$\gamma\in\mathcal{M}_s(\R)$, we have
			$\mathcal{W}_1(\mu+\gamma,\nu+\gamma)=\mathcal{W}_1(\mu,\nu)$.
		\end{enumerate}
	\end{proposition}
	
	\begin{proof}
		Items (1) and (2) are classical for probability measures (see
		\cite[Ch.~5]{Villani2009} and \cite[Ch.~2]{Santambrogio2015}), and they
		extend to $\mathcal{M}_r(\R)$ by applying them to $\mu/r$ and $\nu/r$.
		Item (3) follows from (2).
	\end{proof}

	\section{Projected and reflected Sliced Wasserstein kernels}
	\label{sec:kernel}
	
	We now combine the two ingredients of Section~\ref{sec:preliminares}.
	To define our distances, we project the points of the diagrams onto lines through
	the origin, turn each projection into an empirical measure on $\R$, and average the
	one-dimensional Wasserstein distances over all directions. In this way, we obtain
	two families of distances, namely $\SW_p$ and $\RW_p$, for $p\in[1,\infty)$.
	
	The distance $\SW_1$ is the one introduced by Carrière, Cuturi, and Oudot
	\cite{Carriere2017}, and Theorem~\ref{thm:dp_metrica} is classical
	\cite{CohenSteiner2007,Mileyko2011}. In contrast, the orders $p>1$, the reflected
	family $\RW_p$, and all results on $d_p^{r}$ and $\RW_1$ are new.
	
	The section is organized as follows. In Subsection~\ref{subsec:def_sw} we
	define $\SW_p$ and $\RW_p$ and show that, in the reflected case, the optimal
	transport in each direction is induced by a bijection between the sets $P$ and
	$Q$ of Definition~\ref{def:dp_r} (Lemma~\ref{lem:estructura_medidas_p}); this is
	the link between the sliced distances and $d_p^r$ that we use throughout the
	paper. In Subsection~\ref{subsec:dp_dpr} we compare $d_p^r$ with $d_p$ and
	determine for which $p$ the function $d_p^r$ is a metric. Finally, in
	Subsection~\ref{subsec:sw_rw_metricas} we prove that $\SW_1$ and $\RW_1$ are
	metrics; the case $p>1$ is treated in Section~\ref{sec:sw_rw_no_metricas}.

	\subsection{Definition of the Sliced Wasserstein distances}
	\label{subsec:def_sw}
	
	To compare two persistence diagrams $A$ and $B$ by means of the one-dimensional
	Wasserstein distance, we fix an angle $\theta\in[-\pi/2,\pi/2]$ and project
	each point of the diagrams onto the line spanned by the vector
	$u_\theta=(\cos\theta,\sin\theta)$; that is, we consider
	$\pi_\theta(b,d)\coloneqq\langle(b,d),u_\theta\rangle
	=b\cos\theta+d\sin\theta$. We also use the orthogonal projection onto the
	diagonal, $\pi_\Delta(b,d)\coloneqq\Bigl(\dfrac{b+d}{2},\dfrac{b+d}{2}\Bigr)$, and the
	reflection across it, $r_\Delta(b,d)\coloneqq(d,b)$.
	To match the total masses of the two projections, we add to the empirical measure
	of $A$ the projections onto the diagonal of the points of $B$ (or their reflections
	across it, depending on the distance), and vice versa. More precisely, we define
	the following measures.

	\begin{definition}[\textbf{Projected and reflected measures of a pair of diagrams}]
		\label{def:medidas_proyectadas}
		Let $A,B$ be two finite persistence diagrams. For each angle
		$\theta\in[-\pi/2,\pi/2]$, we define the discrete measures on
		$\R$
		\[
		\mu_{A,B}^{\theta}
		=
		\sum\limits_{x\in A}\delta_{\pi_\theta(x)}
		+
		\sum\limits_{y\in B}\delta_{\pi_\theta(\pi_\Delta(y))}, \ \ \ \ \ \ \ \
		\mu_{A,B}(\theta)
		=
		\sum\limits_{x\in A}\delta_{\pi_\theta(x)}
		+
		\sum\limits_{y\in B}\delta_{\pi_\theta(r_\Delta(y))}.
		\]
		
	\end{definition}

	Averaging the one-dimensional Wasserstein distances between these measures
	over all directions yields the two distances that we study.
	
	\begin{definition}[\textbf{Projected and reflected Sliced Wasserstein distances of order $p\ge1$}]
		\label{def:sw_p}
		Let $A,B$ be finite persistence diagrams and let $p\in[1,\infty)$. We define the
		following two distances.
		\begin{enumerate}
			\item The \emph{projected Sliced Wasserstein distance of
				order $p$} between $A$ and $B$ is
			\[
			\SW_p(A,B)
			=
			\left(
			\frac{1}{\pi}
			\int\limits_{-\pi/2}^{\pi/2}
			\mathcal{W}_p^p\bigl(\mu_{A,B}^{\theta},\mu_{B,A}^{\theta}\bigr)
			\dd\theta
			\right)^{\!1/p}.
			\]
			\item The \emph{reflected Sliced Wasserstein distance of
				order $p$} between $A$ and $B$ is
			\[
			\RW_p(A,B)
			=
			\left(
			\frac{1}{\pi}
			\int\limits_{-\pi/2}^{\pi/2}
			\mathcal{W}_p^p\bigl(\mu_{A,B}(\theta),\mu_{B,A}(\theta)\bigr)
			\dd\theta
			\right)^{\!1/p}.
			\]
		\end{enumerate}
		The integrals are finite, since $\mathcal{W}_p^p$ is a bounded, continuous
		function of $\theta$.
	\end{definition}
	
	In the reflected case, these measures are simply the projections of the sets
	$P$ and $Q$ of Definition~\ref{def:dp_r}, and the optimal transport between
	them is the matching that preserves the order of the projections.
	
	\begin{lemma}[\textbf{Structure of the reflected measures and an expression for $\mathcal{W}_p$, $p\ge1$}]
		\label{lem:estructura_medidas_p}
		Let $A,B\in\mathcal{D}_{\mathrm{fin}}$, $p\in[1,\infty)$, and
		$\theta\in[-\pi/2,\pi/2]$. Define the sets $P=A\cup r_\Delta(B)$ and
		$Q=B\cup r_\Delta(A)$. Then the following statements hold.
		
		\begin{enumerate}
			\item The reflected measures can be written as $
			\mu_{A,B}(\theta)=\sum\limits_{x\in P}\delta_{\pi_\theta(x)}$ and
			$\mu_{B,A}(\theta)=\sum\limits_{y\in Q}\delta_{\pi_\theta(y)}$.

			\item Let $r=|A|+|B|$. If $\alpha_1^\theta\le\cdots\le\alpha_r^\theta$ are the numbers
			$\pi_\theta(x)$ with $x\in P$ and $\beta_1^\theta\le\cdots\le\beta_r^\theta$ are the
			numbers $\pi_\theta(y)$ with $y\in Q$, both listed in nondecreasing order with repetitions,
			then $ \displaystyle
			\mathcal{W}_p\bigl(\mu_{A,B}(\theta),\mu_{B,A}(\theta)\bigr)
			=
			\left(\sum\limits_{i=1}^{r}|\alpha_i^\theta-\beta_i^\theta|^p\right)^{\!1/p}$.

			\item If $s^\theta:P\to Q$ is the bijection that matches the
			elements according to the increasing order of their projections,
			then $\displaystyle \mathcal{W}_p\bigl(\mu_{A,B}(\theta),\mu_{B,A}(\theta)\bigr)
			=
			\left(\sum\limits_{x\in P}\bigl|\langle x-s^\theta(x),u_\theta\rangle\bigr|^p\right)^{\!1/p}$.

			\item For every bijection $\displaystyle \sigma:P\to Q$, we have
			$\displaystyle \sum\limits_{x\in P}\bigl|\langle x-s^\theta(x),u_\theta\rangle\bigr|^p
			\le
			\sum\limits_{x\in P}\bigl|\langle x-\sigma(x),u_\theta\rangle\bigr|^p.
			$
		\end{enumerate}
	\end{lemma}
	
	\begin{proof}
		See Appendix~\ref{app:A}.
	\end{proof}

	\begin{remark}[\textbf{Version with the projection onto the diagonal}]
		\label{rem:version_proyeccion}
		The previous lemma is proved in exactly the same way if the
		reflection $r_\Delta$ is replaced by the orthogonal projection
		$\pi_\Delta$. Indeed, it suffices to replace $r_\Delta$ by
		$\pi_\Delta$ in the definitions of $P$ and $Q$ (taking copies with distinct
		indices in $\Delta\times\N$, as in the proof of
		Theorem~\ref{thm:equiv_p}) and in the measures, which then become the
		projected measures $\mu_{A,B}^{\theta}$ and $\mu_{B,A}^{\theta}$. In this way, the same properties hold.
	\end{remark}

	\subsection{The distances $d_p$ and $d_p^{r}$}
	\label{subsec:dp_dpr}
	
	We first recall that $d_p$ is a metric and then compare $d_p^{r}$ with $d_p$. This
	comparison shows that $d_1^{r}=2\,d_1$, so that the function $d_1^{r}$ is a
	metric, and it yields the stability of $d_p^{r}$ in
	Section~\ref{sec:equiv_metrica}. For $p>1$, in contrast, the function $d_p^{r}$ is not a
	metric.
	
	\begin{theorem}[{\textbf{The function $d_p$ is a metric for all $p\in[1,\infty]$}}]
		\label{thm:dp_metrica}
		For all $p\in[1,\infty]$, the function $d_p$ of
		Definition~\ref{def:dp} is a metric on $\mathcal{D}_{\mathrm{fin}}$.
	\end{theorem}
	
	\begin{proof}
		Symmetry follows by inverting partial bijections; definiteness follows from the
		fact that a point $(b,d)$ with $b<d$ lies at distance $(d-b)/2>0$ from the diagonal
		(Remark~\ref{rem:prop_diagonal}(1)); and the triangle inequality follows by
		composing partial bijections and applying the Minkowski inequality.
	\end{proof}

	\begin{theorem}[\textbf{Comparison of $d_p^{r}$ and $d_p$}]
		\label{thm:comparacion_dpr_dp}
		For all $p\in[1,\infty]$ and $A,B\in\mathcal{D}_{\mathrm{fin}}$, we have
		\[
		2^{1/p}\,d_p(A,B)\ \le\ d_p^{r}(A,B)\ \le\ 2\,d_p(A,B),
		\]
		with the convention $2^{1/\infty}=1$. Moreover, both constants are sharp.
	\end{theorem}
	
	\begin{proof}
		We write $P\coloneqq A\cup r_\Delta(B)$ and $Q\coloneqq B\cup r_\Delta(A)$, set
		$\delta(x)\coloneqq\|x-\pi_\Delta(x)\|_\infty$, and use the fact that
		$\delta\circ r_\Delta=\delta$. We first consider the case $p<\infty$, in which
		\[
		d_p^r(A,B)^p
		=
		\min_{\tau:P\to Q}\ \sum_{x\in P}\|x-\tau(x)\|_\infty^p ,
		\]
		where the minimum ranges over the bijections $\tau:P\to Q$, and, by
		Theorem~\ref{thm:dp_equivalencias}(2), we have
		\begin{equation}\label{eq:dp_descompuesta}
			d_p(A,B)^p
			=
			\min_{\substack{A_1\subseteq A,\ A_2\coloneqq A\setminus A_1\\
					B_1\subseteq B,\ B_2\coloneqq B\setminus B_1\\
					\gamma:A_1\to B_1\ \text{bijection}}}
			\Bigl(
			\sum_{a\in A_1}\|a-\gamma(a)\|_\infty^p
			+\sum_{a\in A_2}\delta(a)^p
			+\sum_{b\in B_2}\delta(b)^p
			\Bigr).
		\end{equation}
		
		\medskip
		\noindent\textit{Upper bound.}
		Fix $A_1,A_2,B_1,B_2$ and $\gamma$ attaining the minimum in
		\eqref{eq:dp_descompuesta}, and define $\sigma:P\to Q$ by
		\[
		\sigma(a)=\gamma(a),\quad
		\sigma(r_\Delta(\gamma(a)))=r_\Delta(a)
		\qquad(a\in A_1),
		\]
		\[
		\sigma(a)=r_\Delta(a)\quad(a\in A_2),
		\qquad
		\sigma(r_\Delta(b))=b\quad(b\in B_2).
		\]
		The domain blocks $A_1$, $A_2$, $r_\Delta(B_1)$, and $r_\Delta(B_2)$
		form a partition of $P$, and their images $B_1$, $r_\Delta(A_2)$,
		$r_\Delta(A_1)$, and $B_2$ form a partition of $Q$; hence the map $\sigma$
		is a bijection. Since $r_\Delta$ is an isometry, the pairs
		$a\mapsto\gamma(a)$ and $r_\Delta(\gamma(a))\mapsto r_\Delta(a)$ have the
		same cost, and, by
		Remark~\ref{rem:prop_diagonal}(4), we have
		$\|x-r_\Delta(x)\|_\infty=2\delta(x)$. Consequently,
		\begin{align*}
			d_p^r(A,B)^p
			&\le\sum_{x\in P}\|x-\sigma(x)\|_\infty^p
			=2\sum_{a\in A_1}\|a-\gamma(a)\|_\infty^p
			+2^p\Bigl(\sum_{a\in A_2}\delta(a)^p
			+\sum_{b\in B_2}\delta(b)^p\Bigr)\\
			&\le 2^p\Bigl(\sum_{a\in A_1}\|a-\gamma(a)\|_\infty^p
			+\sum_{a\in A_2}\delta(a)^p
			+\sum_{b\in B_2}\delta(b)^p\Bigr)
			=2^p\,d_p(A,B)^p,
		\end{align*}
		where the last equality follows from \eqref{eq:dp_descompuesta}. Then, taking
		$p$-th roots, we obtain $d_p^r(A,B)\le 2\,d_p(A,B)$.
		
		\medskip
		\noindent\textit{Lower bound.}
		Let $\tau:P\to Q$ be an optimal bijection for $d_p^r(A,B)$. Since $P=A\sqcup r_\Delta(B)$
		and $Q=B\sqcup r_\Delta(A)$, the sets
		\[
		\begin{aligned}
			P_1&\coloneqq A\cap\tau^{-1}(B), &
			P_2&\coloneqq A\cap\tau^{-1}(r_\Delta(A)),\\
			P_3&\coloneqq r_\Delta(B)\cap\tau^{-1}(B), &
			P_4&\coloneqq r_\Delta(B)\cap\tau^{-1}(r_\Delta(A))
		\end{aligned}
		\]
		form a partition of $P$, and we have $B=\tau(P_1)\sqcup\tau(P_3)$ and
		$r_\Delta(A)=\tau(P_2)\sqcup\tau(P_4)$. Define
		\[
		\begin{aligned}
			C&\coloneqq\sum_{x\in P_1}\|x-\tau(x)\|_\infty^p+\sum_{x\in P_2}\delta(x)^p
			+\sum_{x\in P_3}\delta(\tau(x))^p,\\
			C'&\coloneqq\sum_{x\in P_4}\|x-\tau(x)\|_\infty^p+\sum_{x\in P_2}\delta(\tau(x))^p
			+\sum_{x\in P_3}\delta(x)^p .
		\end{aligned}
		\]
		The sum $C$ is the quantity minimized in \eqref{eq:dp_descompuesta} for $A_1=P_1$ and
		$\gamma=\tau|_{P_1}$, and the sum $C'$ is the quantity minimized in
		\eqref{eq:dp_descompuesta} for
		$A_1=r_\Delta(\tau(P_4))$ and $\gamma(a)=r_\Delta(\tau^{-1}(r_\Delta(a)))$, since $r_\Delta$
		is an isometry and $\delta\circ r_\Delta=\delta$. Therefore, we have
		$C\ge d_p(A,B)^p$ and $C'\ge d_p(A,B)^p$. Moreover,
		\[
		C+C'=\sum_{x\in P_1\cup P_4}\|x-\tau(x)\|_\infty^p
		+\sum_{x\in P_2\cup P_3}\bigl(\delta(x)^p+\delta(\tau(x))^p\bigr).
		\]
		If $x\in P_2\cup P_3$, then the points $x$ and $\tau(x)$ lie on opposite sides of the
		diagonal, so Remark~\ref{rem:prop_diagonal}(3) and the
		inequality $s^p+t^p\le(s+t)^p$ give
		$\delta(x)^p+\delta(\tau(x))^p\le\|x-\tau(x)\|_\infty^p$. Therefore,
		\[
		2\,d_p(A,B)^p\le C+C'\le\sum_{x\in P}\|x-\tau(x)\|_\infty^p=d_p^r(A,B)^p .
		\]

		\medskip
		\noindent\textit{The case $p=\infty$.}
		Letting $p\to\infty$ in $2^{1/p}d_p(A,B)\le d_p^r(A,B)\le 2\,d_p(A,B)$
		and using Remarks~\ref{rem:p_infinito} and~\ref{rem:dpr_infinito}, we
		obtain $d_\infty(A,B)\le d_\infty^r(A,B)\le 2\,d_\infty(A,B)$.
		
		\medskip
		\noindent\textit{Sharpness of the constants.}
		For the upper constant, take $A=\{(0,1),(0,3)\}$ and
		$B=\{(0,3)\}$. The point $(0,3)$ is common to both diagrams, and the optimal
		partial bijection for $d_p$ matches it with itself and sends $(0,1)$ to its
		projection onto the diagonal, so that $d_p(A,B)=\delta((0,1))=1/2$. For $d_p^r$,
		in contrast, the point $(0,1)$ must be matched with an element of
		$Q=\{(0,3),(1,0),(3,0)\}$, and the closest one is $(1,0)$, at distance
		$1$. Therefore, every bijection has cost at least $1$, and the one that, in
		addition, matches $(0,3)$ with $(0,3)$ and $(3,0)$ with $(3,0)$ has cost exactly $1$;
		hence $d_p^r(A,B)=1=2\,d_p(A,B)$. For the lower constant, take
		$A=\{(0,L)\}$ and $B=\{(\varepsilon,L+\varepsilon)\}$ with
		$0<\varepsilon<L/2$, so that $d_p(A,B)=\varepsilon$. The only two
		bijections $P\to Q$ are the one that sends each point to its reflection, with
		cost $2^{1/p}L$, and the crossed one, with cost $2^{1/p}\varepsilon$; hence
		$d_p^r(A,B)=2^{1/p}\varepsilon=2^{1/p}d_p(A,B)$.
	\end{proof}
	
	\begin{corollary}[\textbf{The function $d_1^{r}$ is a metric}]
		\label{cor:d1r_metrica}
		For all $A,B\in\mathcal{D}_{\mathrm{fin}}$, we have
		$d_1^{r}(A,B)=2\,d_1(A,B)$. In particular, the function $d_1^{r}$ is a metric on
		$\mathcal{D}_{\mathrm{fin}}$. Moreover, for all $p\in[1,\infty]$, the
		function $d_p^{r}$ is symmetric and satisfies $d_p^{r}(A,B)=0$ if and
		only if $A=B$.
	\end{corollary}
	
	\begin{proof}
		For $p=1$, both constants in Theorem~\ref{thm:comparacion_dpr_dp}
		are equal to $2$. Furthermore, the symmetry of $d_p^{r}$ follows by inverting
		bijections, and its definiteness follows from $d_p\le d_p^{r}$ and
		Theorem~\ref{thm:dp_metrica}.
	\end{proof}
	
	\begin{proposition}[\textbf{The function $d_p^{r}$ is not a metric for $p>1$}]
		\label{prop:dpr_no_metrica}
		For all $p\in(1,\infty]$, the function $d_p^{r}$ does not satisfy
		the triangle inequality. More precisely, there exist single-point diagrams
		$A,B_\delta,C_\varepsilon\in\mathcal{D}_{\mathrm{fin}}$ such that
		\[
		d_p^{r}(A,B_\delta)\;>\;d_p^{r}(A,C_\varepsilon)+d_p^{r}(B_\delta,C_\varepsilon).
		\]
		
	\end{proposition}
	
	\begin{proof}
		Let $A=\{(0,1)\}$, $B_\delta=\{(0,\delta)\}$, and
		$C_\varepsilon=\{(\tfrac12,\tfrac12+\varepsilon)\}$ with $0<\delta,\varepsilon$ small.
		Since these are single-point diagrams, there are only two bijections $P\to Q$ in
		each case, and a direct computation gives
		\[
		d_p^{r}(A,B_\delta)=\min\bigl\{(1+\delta^p)^{1/p},\,2^{1/p}(1-\delta)\bigr\},\qquad
		d_p^{r}(B_\delta,C_\varepsilon)=(\delta^p+\varepsilon^p)^{1/p},\qquad
		d_p^{r}(A,C_\varepsilon)=2^{1/p-1}.
		\]
		As $\delta,\varepsilon\to0^+$, the first value tends to $1$, whereas the sum of the
		other two tends to $2^{1/p-1}<1$; hence the triangle inequality fails for
		sufficiently small $\delta,\varepsilon$. For $p=\infty$, the same computation, with
		maxima instead of sums, gives $1-\delta>\tfrac12+\max\{\delta,\varepsilon\}$.
	\end{proof}
	
	\subsection{The distances $\SW_1$ and $\RW_1$ are metrics}
	\label{subsec:sw_rw_metricas}
	
	Having settled for which $p$ the function $d_p^r$ is a metric, we turn to the
	sliced distances and prove that $\SW_1$ and $\RW_1$ are metrics. Nonnegativity,
	symmetry, and the triangle inequality follow from an $L^1$ representation
	(Lemma~\ref{lem:representacion_L1}), which is specific to $p=1$, since it relies
	on the distribution-function formulation of $\mathcal{W}_1$. Definiteness
	follows from the injectivity of the projections $\pi_\theta$ on finitely
	supported signed measures (Lemma~\ref{lem:slice}).
	
	\begin{lemma}[\textbf{Directional injectivity (Radon slice type)}]
		\label{lem:slice}
		Let $\tau=\sum\limits_{i=1}^{M}\tau_i\,\delta_{z_i}$ be a finitely
		supported signed measure on $\R^2$, with distinct points $z_i=(b_i,d_i)$.
		Suppose that there exists a set $E\subseteq[-\pi/2,\pi/2]$ of
		positive Lebesgue measure such that $(\pi_\theta)_\#\tau = 0$
		for all $\theta\in E$. Then $\tau=0$.
	\end{lemma}
	
	\begin{proof}
		Since $\tau$ is a finite combination of Dirac deltas, for each
		$\theta\in[-\pi/2,\pi/2]$ we have $
		(\pi_\theta)_\#\tau
		=
		\sum\limits_{i=1}^{M} \tau_i\,\delta_{\pi_\theta(z_i)}$. Moreover,
		for each pair $i\neq j$, we have $\pi_\theta(z_i)=\pi_\theta(z_j)$ if and only if
		$\langle z_i-z_j,u_\theta\rangle=0$, which happens for only finitely many values of
		$\theta\in[-\pi/2,\pi/2]$. Hence the set $F$ of angles at which
		two projections coincide is finite. Since $E$ has positive Lebesgue
		measure, we have $E\not\subseteq F$; therefore, there exists
		$\theta_0\in E\setminus F$ for which the points
		$\pi_{\theta_0}(z_1),\dots,\pi_{\theta_0}(z_M)$ are all distinct.
		By hypothesis, we have $(\pi_{\theta_0})_\#\tau=0$. Then, evaluating this signed
		measure at each singleton $\{\pi_{\theta_0}(z_i)\}$ gives $
		0=\bigl((\pi_{\theta_0})_\#\tau\bigr)\bigl(\{\pi_{\theta_0}(z_i)\}\bigr)=\tau_i$
		for all $i=1,\dots,M$. Therefore, we obtain $\tau=0$.
	\end{proof}

	\begin{lemma}[\textbf{An $L^1$ representation of $\SW_1$ and $\RW_1$}]
		\label{lem:representacion_L1}
		Let $A,B\in\mathcal{D}_{\mathrm{fin}}$. For each
		$X\in\mathcal{D}_{\mathrm{fin}}$, let
		$\nu_X\coloneqq\sum \limits_{x\in X}\delta_x$ be its counting measure.
		For $\theta\in[-\pi/2,\pi/2]$ and $t\in\R$, define
		\[
		g_X^{\theta}(t)
		\coloneqq
		\bigl[(\pi_\theta)_\#\nu_X-(\pi_\theta\circ\pi_\Delta)_\#\nu_X\bigr]\bigl((-\infty,t]\bigr), \  \  \
		h_X^{\theta}(t)
		\coloneqq
		\bigl[(\pi_\theta)_\#\nu_X-(\pi_\theta\circ r_\Delta)_\#\nu_X\bigr]\bigl((-\infty,t]\bigr).
		\]
		Then $\displaystyle
		\SW_1(A,B)
		=
		\frac{1}{\pi}
		\int\limits_{-\pi/2}^{\pi/2}
		\int\limits_{\R}
		\bigl|g_A^{\theta}(t)-g_B^{\theta}(t)\bigr|\,\dd t\,\dd\theta,
		\ \ \ \
		\RW_1(A,B)
		=
		\frac{1}{\pi}
		\int\limits_{-\pi/2}^{\pi/2}
		\int\limits_{\R}
		\bigl|h_A^{\theta}(t)-h_B^{\theta}(t)\bigr|\,\dd t\,\dd\theta.
		$
		In particular, the functions $\SW_1$ and $\RW_1$ are well defined, symmetric, and
		nonnegative.
	\end{lemma}
	
	\begin{proof}
		Fix $\theta\in[-\pi/2,\pi/2]$ and $t\in\R$. By the definition of the
		projected measures of a pair of diagrams, we have $
		\mu_{A,B}^{\theta}
		=
		(\pi_\theta)_\#\nu_A+(\pi_\theta\circ\pi_\Delta)_\#\nu_B $ and
		$\ \mu_{B,A}^{\theta}
		=
		(\pi_\theta)_\#\nu_B+(\pi_\theta\circ\pi_\Delta)_\#\nu_A$.
		Evaluating at $(-\infty,t]$ and subtracting, we obtain
		\[
		\begin{aligned}
			F_{\mu_{A,B}^{\theta}}(t)-F_{\mu_{B,A}^{\theta}}(t)
			&=
			\bigl[(\pi_\theta)_\#\nu_A\bigr]\bigl((-\infty,t]\bigr)
			+
			\bigl[(\pi_\theta\circ\pi_\Delta)_\#\nu_B\bigr]\bigl((-\infty,t]\bigr)
			\\
			&\qquad
			-
			\bigl[(\pi_\theta)_\#\nu_B\bigr]\bigl((-\infty,t]\bigr)
			-
			\bigl[(\pi_\theta\circ\pi_\Delta)_\#\nu_A\bigr]\bigl((-\infty,t]\bigr)
			\\
			&=	\bigl[(\pi_\theta)_\#\nu_A-(\pi_\theta\circ\pi_\Delta)_\#\nu_A\bigr]\bigl((-\infty,t]\bigr)-\bigl[(\pi_\theta)_\#\nu_B-(\pi_\theta\circ\pi_\Delta)_\#\nu_B\bigr]\bigl((-\infty,t]\bigr)\\
			&=
			g_A^{\theta}(t)-g_B^{\theta}(t).
		\end{aligned}
		\]
		By the formulation of $\mathcal{W}_1$ in terms of distribution functions
		(Proposition~\ref{prop:equiv_p}(2)), we have
		\[
		\mathcal{W}_1\bigl(\mu_{A,B}^{\theta},\mu_{B,A}^{\theta}\bigr)
		=
		\int\limits_{\R}
		\bigl|F_{\mu_{A,B}^{\theta}}(t)-F_{\mu_{B,A}^{\theta}}(t)\bigr|\,\dd t
		=
		\int\limits_{\R}
		\bigl|g_A^{\theta}(t)-g_B^{\theta}(t)\bigr|\,\dd t.
		\]
		Thus, integrating over $\theta$ and dividing by $\pi$ yields the formula for
		$\SW_1$. The case of $\RW_1$ is identical, with $\pi_\Delta$ replaced by
		$r_\Delta$.
		
		Symmetry is immediate from the absolute value. Moreover, the integrals are
		finite since, for a finite diagram $X$, the functions
		$g_X^\theta$ and $h_X^\theta$ are step functions with finitely many
		jumps and compact support in $t$, uniformly bounded in $\theta$.
	\end{proof}

	\begin{theorem}[\textbf{The distances $\SW_1$ and $\RW_1$ are metrics}]
		\label{thm:sw1_rw1_metricas}
		The functions $\SW_1$ and $\RW_1$ are metrics on $\mathcal{D}_{\mathrm{fin}}$.
	\end{theorem}
	
	\begin{proof}
		Let $\Lambda\coloneqq[-\pi/2,\pi/2]\times\R$ and, for
		$X\in\mathcal{D}_{\mathrm{fin}}$, let $G_X(\theta,t)\coloneqq g_X^\theta(t)/\pi$ and
		$H_X(\theta,t)\coloneqq h_X^\theta(t)/\pi$. By
		Lemma~\ref{lem:representacion_L1}, we have $\SW_1(A,B)=\|G_A-G_B\|_{L^1(\Lambda)}$ and
		$\RW_1(A,B)=\|H_A-H_B\|_{L^1(\Lambda)}$, whence nonnegativity,
		symmetry, and the triangle inequality follow from the corresponding properties
		of the norm.
		
		If $\SW_1(A,B)=0$, then, since the integrand
		$\mathcal{W}_1(\mu_{A,B}^\theta,\mu_{B,A}^\theta)$ in
		Definition~\ref{def:sw_p} is nonnegative, it vanishes for
		almost every $\theta$, and hence $\mu_{A,B}^\theta=\mu_{B,A}^\theta$ for
		almost every $\theta$. By the definition of these measures, this means that
		$(\pi_\theta)_\#\eta=0$, where
		$\eta\coloneqq\nu_A-(\pi_\Delta)_\#\nu_A-\nu_B+(\pi_\Delta)_\#\nu_B$, and
		Lemma~\ref{lem:slice} then gives $\eta=0$. Thus, we have
		$\nu_A+(\pi_\Delta)_\#\nu_B=\nu_B+(\pi_\Delta)_\#\nu_A$, and since the measures
		$(\pi_\Delta)_\#\nu_X$ are concentrated on the diagonal, it follows that
		$\nu_A=\nu_B$, that is, $A=B$. For $\RW_1$, the argument is the same, with
		$\pi_\Delta$ replaced by $r_\Delta$, since the measures $(r_\Delta)_\#\nu_X$ are
		concentrated below the diagonal, that is, outside $\Omega$.
	\end{proof}

	\section{Failure of the triangle inequality and of the CND property for $p>1$}
	\label{sec:sw_rw_no_metricas}
	
	Section~\ref{sec:kernel} showed that $\SW_1$ and $\RW_1$ are metrics
	(Theorem~\ref{thm:sw1_rw1_metricas}), and we will see in
	Section~\ref{sec:CND_sw1_rw1} that they are also CND
	(Lemma~\ref{lem:sw_nd}). In this section we show that both properties break
	down for larger orders: the distances $\SW_p$ and $\RW_p$ do not satisfy the
	triangle inequality for $p>1$, and $\SW_2$ and $\RW_2$ are not CND. All
	three results are new, since the behavior of
	$\SW_p$ and $\RW_p$ for $p>1$ has not been studied before, and the negative
	definiteness result of \cite{Kolouri2016} concerns measures of equal mass
	and therefore does not apply here.
	
	\begin{theorem}[\textbf{$\SW_p$ is not a metric for $p>1$}]
		\label{thm:sw_no_metrica}
		Let $p\in(1,\infty)$, let $Z=\{(b_z,d_z)\}$ be any single-point diagram, and
		set $t_1\coloneqq\tfrac{b_z+d_z}{2}$ and $t_*\coloneqq t_1-\tfrac{d_z-b_z}{4}$,
		so that $(t_1,t_1)$ is the projection of $Z$ onto the diagonal. For $w>0$,
		let $X_1\coloneqq\{(t_1,t_1+w)\}$ and $X_2\coloneqq\{(t_*,t_*+w)\}$ be the
		near-diagonal diagrams of persistence $w$ located at $t_1$ and $t_*$, respectively.
		Then, for every sufficiently small $w>0$, we have
		$\SW_p(X_1,Z)>\SW_p(X_1,X_2)+\SW_p(X_2,Z)$, so the distance $\SW_p$ is not a metric.
	\end{theorem}
	
	\begin{theorem}[\textbf{$\RW_p$ is not a metric for $p>1$}]
		\label{thm:rw_no_metrica}
		Let $p\in(1,\infty)$, let $Z=\{(b_z,d_z)\}$ be any single-point diagram, and
		set $t_1\coloneqq\tfrac{b_z+d_z}{2}$. For $T\in\R$ and $w>0$, let
		$X_1\coloneqq\{(T,T+w)\}$ and $X_2\coloneqq\{(t_1,t_1+w)\}$ be the
		near-diagonal diagrams of persistence $w$ located at $T$ and $t_1$, respectively.
		Then, for every sufficiently large $T$ and then every sufficiently small
		$w>0$, we have $\RW_p(X_1,Z)>\RW_p(X_1,X_2)+\RW_p(X_2,Z)$, so the distance
		$\RW_p$ is not a metric.
	\end{theorem}
	
	\begin{theorem}[\textbf{$\SW_2$ and $\RW_2$ are not CND}]
		\label{thm:sw2_no_CND}
		The distances $\SW_2$ and $\RW_2$ are not CND on
		$\mathcal{D}_{\mathrm{fin}}$. More precisely, consider the single-point
		diagrams $
		Z_1=\{(700,701)\},\
		Z_2=\{(500,501)\},\
		Z_3=\{(0,1000)\},\
		Z_4=\{(200,1200)\}$
		and the matrices $M,M'\in\R^{4\times 4}$ with entries
		$M_{ij}\coloneqq\SW_2(Z_i,Z_j)$ and $M'_{ij}\coloneqq\RW_2(Z_i,Z_j)$.
		Then the following hold.
		\begin{enumerate}
			\item The zero-sum vector $a=(-7,7,1,-1)^{\top}$ satisfies
			$a^{\top} M a \in [422.2423096,\,422.2423100]$.
			\item The zero-sum vector $a'=(-7,7,-1,1)^{\top}$ satisfies
			$a'^{\top} M' a' \in [396.4210427,\,396.4210428]$.
		\end{enumerate}
		In particular, we have $a^{\top} M a>0$ and $a'^{\top} M' a'>0$.
	\end{theorem}
	
	\paragraph{Strategy of the proofs.}
	All three counterexamples use only single-point diagrams. For these, the
	transport problem in each direction has just two candidate matchings, so
	$\SW_p$ and $\RW_p$ reduce to the angular average of the minimum of two
	explicit costs; we derive these closed forms in
	Subsection~\ref{subsec:formas_cerradas}. The proofs of
	Theorems~\ref{thm:sw_no_metrica} and~\ref{thm:rw_no_metrica} then rest on a
	single mechanism. As the persistence $w$ of a diagram $\{(t,t+w)\}$ tends to
	$0$, its distance to a fixed single-point diagram $z$ converges to a limit
	that depends on the position $t$ of the diagram along the diagonal. We call
	the values of this limit at the positions that we compare \emph{anchors}.
	For $p=1$ the limit does not depend on $t$, whereas for $p>1$, because of the
	strict convexity of $u\mapsto u^p$, there are positions with different
	anchors. Two near-diagonal diagrams placed at such positions are arbitrarily
	close to each other, but their distances to $z$ differ by a fixed amount,
	and this violates the triangle inequality. Subsection~\ref{subsec:anclas}
	computes the limits and compares the anchors, and
	Subsection~\ref{subsec:prueba_no_metricas} assembles the two proofs.
	Finally, Theorem~\ref{thm:sw2_no_CND} is proved in
	Subsection~\ref{subsec:certificado_sw2} by integrating the closed forms
	exactly for $p=2$ and enclosing the result with interval arithmetic.
	
	\subsection{Closed forms for single-point diagrams}
	\label{subsec:formas_cerradas}
	
	We first compute $\SW_p$ and $\RW_p$ between two single-point diagrams $x$
	and $y$. In each direction $\theta$, the two measures to be compared have two
	atoms each, so the optimal transport is the cheaper of two matchings: the
	\emph{parallel} matching, which pairs $x$ with $y$ and the two diagonal
	partners (projections or reflections) with each other, and the
	\emph{crossed} matching, which pairs each point with its own diagonal
	partner. The two lemmas below are used in all three proofs of this section.

	\begin{lemma}[\textbf{Closed form of $\SW_p$ for single-point diagrams}]
		\label{lem:forma_cerrada_sw}
		Let $x=(b,d)$, $y=(b',d')\in\Omega$ be single-point diagrams.
		Define $
		\Delta b \coloneqq b-b',
		\ \
		\Delta d \coloneqq d-d',
		\
		\Delta m \coloneqq \dfrac{(b+d)-(b'+d')}{2}$,
		$
		w\coloneqq d-b,
		\
		w'\coloneqq d'-b'.
		$
		Then $\displaystyle \SW_p(x,y)^p
		=
		\frac{1}{\pi}\int\limits_{-\pi/2}^{\pi/2}
		\min\{P,Q\}\,\dd\theta$, where $P$ and $Q$ are given by
		\[
		P
		=
		\bigl|\Delta b\,\cos(\theta)+\Delta d\,\sin(\theta)\bigr|^p
		+
		\bigl|\Delta m\,(\cos(\theta)+\sin(\theta))\bigr|^p, \ \
		Q
		=
		2^{-p}\bigl(w^p+(w')^p\bigr)\,
		\bigl|\cos(\theta)-\sin(\theta)\bigr|^p.
		\]
	\end{lemma}
	
	\begin{proof}
		Fix $\theta\in[-\pi/2,\pi/2]$. The projected measures of the pair $(x,y)$ are
		\[
		\mu_{x,y}^{\theta}
		=
		\delta_{\pi_\theta(x)}+\delta_{\pi_\theta(\pi_\Delta y)},
		\qquad
		\mu_{y,x}^{\theta}
		=
		\delta_{\pi_\theta(y)}+\delta_{\pi_\theta(\pi_\Delta x)}.
		\]
		Each measure has two atoms, so there are only two possible matchings,
		and, by Corollary~\ref{cor:empirico_p}, the quantity
		$\mathcal{W}_p^p(\mu_{x,y}^{\theta},\mu_{y,x}^{\theta})$ is the smaller of
		their two costs.
		
		\emph{$\star$ Parallel matching.} This matching consists of
		$x\leftrightarrow y$ and $\pi_\Delta y\leftrightarrow\pi_\Delta x$. We have
		$\pi_\theta(x)-\pi_\theta(y)=\Delta b\cos(\theta)+\Delta d\sin(\theta)$ and,
		since $\pi_\Delta y-\pi_\Delta x=(-\Delta m,-\Delta m)$, it follows that
		$\pi_\theta(\pi_\Delta y)-\pi_\theta(\pi_\Delta x)=-\Delta m(\cos(\theta)+\sin(\theta))$.
		Then, raising to the power $p$ and adding, we obtain $P$.
		
		\emph{$\star$ Crossed matching.} This matching consists of
		$x\leftrightarrow\pi_\Delta x$ and $y\leftrightarrow\pi_\Delta y$. Since
		$x-\pi_\Delta x=\bigl(\tfrac{b-d}{2},\tfrac{d-b}{2}\bigr)$, we have
		$\pi_\theta(x)-\pi_\theta(\pi_\Delta x)=\tfrac{b-d}{2}(\cos(\theta)-\sin(\theta))$,
		and similarly for $y$, with $w'$ in place of $w$. Thus, raising to the
		power $p$ and adding, we obtain $Q$.
		
		Therefore $\mathcal{W}_p^p(\mu_{x,y}^{\theta},\mu_{y,x}^{\theta})=\min\{P,Q\}$,
		and integrating over $\theta$, dividing by $\pi$, and applying
		Definition~\ref{def:sw_p} gives the stated formula.
	\end{proof}
	
	\begin{lemma}[\textbf{Closed form of $\RW_p$ for single-point diagrams}]
		\label{lem:forma_cerrada_rw}
		Let $x=(b,d)$, $y=(b',d')\in\Omega$ be single-point diagrams.
		Define $\Delta b\coloneqq b-b'$, $\Delta d\coloneqq d-d'$, and the
		persistences $w\coloneqq d-b$, $w'\coloneqq d'-b'$. Then $\displaystyle \RW_p(x,y)^p
		=
		\frac{1}{\pi}\int\limits_{-\pi/2}^{\pi/2}
		\min\{P,Q\}\,\dd\theta$,
		where $P$ and $Q$ are given by $
		P
		=
		\bigl|\Delta b\,\cos(\theta)+\Delta d\,\sin(\theta)\bigr|^p
		+
		\bigl|\Delta d\,\cos(\theta)+\Delta b\,\sin(\theta)\bigr|^p$
		and
		$Q
		=
		\bigl(w^p+(w')^p\bigr)\bigl|\cos(\theta)-\sin(\theta)\bigr|^p$.
	\end{lemma}
	
	\begin{proof}
		The argument is the same as in the proof of
		Lemma~\ref{lem:forma_cerrada_sw}, with $\pi_\Delta$ replaced by
		$r_\Delta$. In the parallel matching, given by $x\leftrightarrow y$ and
		$r_\Delta y\leftrightarrow r_\Delta x$, we use
		$r_\Delta y-r_\Delta x=-(\Delta d,\Delta b)$, which gives $P$. In the
		crossed matching, given by $x\leftrightarrow r_\Delta x$ and
		$y\leftrightarrow r_\Delta y$, we use $x-r_\Delta x=(b-d,\,d-b)$, so that
		$\pi_\theta(x)-\pi_\theta(r_\Delta x)=(b-d)(\cos(\theta)-\sin(\theta))$, and
		similarly for $y$, which gives $Q$.
	\end{proof}

	\subsection{Anchors}
	\label{subsec:anclas}
	
	In this subsection we study the distance from a fixed single-point
	diagram $z$ to a near-diagonal diagram $x_{t,w}$ in the limit $w\to0^+$.
	Lemmas~\ref{lem:limite_w0_sw} and~\ref{lem:limite_w0_rw} show that the
	limit exists, uniformly for $t$ in compact sets;
	Propositions~\ref{prop:anclas_sw} and~\ref{prop:anclas_rw} compare its
	values at different positions $t$; and
	Lemma~\ref{lem:distancia_casi_diagonales} shows that two near-diagonal
	diagrams of the same small persistence are close to each other, wherever
	they lie on the diagonal.
	
	We fix a single-point diagram $z=(b_z,d_z)$ with persistence
	$w_z=d_z-b_z$, set $
	t_1\coloneqq\dfrac{b_z+d_z}{2}$ and
	$x_{t,w}\coloneqq\{(t,t+w)\}$,
	and consider the limit functions
	\[
	K(t)\coloneqq\lim\limits_{w\to0^+}\SW_p(x_{t,w},z)^p,
	\qquad
	L(t)\coloneqq\lim\limits_{w\to0^+}\RW_p(x_{t,w},z)^p,
	\]
	as well as the constant $\displaystyle
	J
	\coloneqq
	\frac{1}{\pi}\int\limits_{-\pi/2}^{\pi/2}
	\bigl|\cos(\theta)-\sin(\theta)\bigr|^p\,\dd\theta
	\;>\;0$.
	The diagrams $x_{t,w}$ are ``near-diagonal'' perturbations. As
	$w\to0^+$, the diagram $x_{t,w}$ approaches a point of the diagonal, so the
	limit depends only on the position $t$ along the diagonal. This dependence
	is precisely what produces the gap between anchors, that is, between the
	values of $K$ (or $L$) at two different positions, and, ultimately, the
	violation of the triangle inequality. We will also use the
	following elementary consequence of the strict convexity of $u\mapsto u^p$
	on $[0,\infty)$: for $p>1$ and $s\in(0,1)$, we have
	\begin{equation}
		\label{eq:convexidad_estricta}
		s^p+(1-s)^p<1,
	\end{equation}
	whereas equality holds for all $s$ when $p=1$.

	\begin{lemma}[\textbf{Zero-persistence limit for $\SW_p$}]
		\label{lem:limite_w0_sw}
		For every $t\in\R$, the limit $K(t)$ exists and equals
		\[
		K(t)
		=
		\frac{1}{\pi}\int\limits_{-\pi/2}^{\pi/2}
		\min\bigl\{A(\theta),\,B(\theta)\bigr\}\,\dd\theta,
		\]
		where
		\[
		A(\theta)
		\coloneqq
		\bigl|t(\cos(\theta)+\sin(\theta))-(b_z\cos(\theta)+d_z\sin(\theta))\bigr|^p
		+
		\bigl|(t-t_1)(\cos(\theta)+\sin(\theta))\bigr|^p,
		\]
		\[
		B(\theta)
		\coloneqq
		\Bigl(\frac{w_z}{2}\Bigr)^p
		\bigl|\cos(\theta)-\sin(\theta)\bigr|^p.
		\]
		Moreover, the function $w\mapsto\SW_p(x_{t,w},z)^p$ converges to $K(t)$
		uniformly in $t$ on compact sets, with error $O(w)$.
	\end{lemma}
	
	\begin{proof}
		Fix a compact set $K_0\subseteq\R$, $t\in K_0$, and $0<w<1$. By
		Lemma~\ref{lem:forma_cerrada_sw} applied to the pair $(x_{t,w},z)$, with
		$\Delta b=t-b_z$, $\Delta d=t+w-d_z$, $\Delta m=t+\tfrac w2-t_1$, and
		persistences $w$ and $w_z$, we have
		\[
		\SW_p(x_{t,w},z)^p=\frac{1}{\pi}\int\limits_{-\pi/2}^{\pi/2}
		\min\{A_w(\theta),B_w(\theta)\}\,\dd\theta,
		\]
		where $A_w=|X_w|^p+|Y_w|^p$, with $X_w=(t-b_z)\cos\theta+(t+w-d_z)\sin\theta$ and
		$Y_w=(t+\tfrac w2-t_1)(\cos\theta+\sin\theta)$, and
		$B_w=2^{-p}(w^p+w_z^p)|\cos\theta-\sin\theta|^p$. For $w=0$, these
		expressions become $A=|X|^p+|Y|^p$ and the function $B$ of the statement,
		and
		\[
		X_w-X=w\sin\theta,\qquad
		Y_w-Y=\tfrac w2(\cos\theta+\sin\theta),\qquad
		0\le B_w-B=2^{-p}w^p|\cos\theta-\sin\theta|^p\le w^p .
		\]
		Since $X_w,X,Y_w,Y$ are bounded in absolute value by a constant
		$M_0=M_0(K_0)$, and
		$\bigl||a|^p-|b|^p\bigr|\le pM_0^{p-1}|a-b|$ for $|a|,|b|\le M_0$ (by the mean value
		theorem if $p>1$, and by the triangle inequality if $p=1$), we have
		$|A_w-A|\le pM_0^{p-1}\bigl(|X_w-X|+|Y_w-Y|\bigr)\le2pM_0^{p-1}w$. Then,
		using $|\min\{a,b\}-\min\{a',b'\}|\le\max\{|a-a'|,|b-b'|\}$ and $w^p\le w$,
		we obtain
		\[
		\Bigl|\SW_p(x_{t,w},z)^p-\frac{1}{\pi}\int\limits_{-\pi/2}^{\pi/2}
		\min\{A,B\}\,\dd\theta\Bigr|\le Cw,
		\qquad C\coloneqq\max\{2pM_0^{p-1},1\},
		\]
		for all $t\in K_0$ and $0<w<1$. This proves that $K(t)$ exists, that it
		has the stated form, and that the convergence is uniform on $K_0$ with
		error $O(w)$. Finally, since $|u^{1/p}-v^{1/p}|\le|u-v|^{1/p}$ for
		$u,v\ge0$, we also have $\SW_p(x_{t,w},z)\to K(t)^{1/p}$ uniformly on $K_0$.
	\end{proof}

	\begin{lemma}[\textbf{Zero-persistence limit for $\RW_p$}]
		\label{lem:limite_w0_rw}
		For every $t\in\R$, the limit $L(t)$ exists and equals
		\[
		L(t)
		=
		\frac{1}{\pi}\int\limits_{-\pi/2}^{\pi/2}
		\min\bigl\{A(\theta),\,
		w_z^p\bigl|\cos(\theta)-\sin(\theta)\bigr|^p\bigr\}\,\dd\theta,
		\]
		where
		\[
		A(\theta)
		\coloneqq
		\bigl|(t-b_z)\cos(\theta)+(t-d_z)\sin(\theta)\bigr|^p
		+
		\bigl|(d_z-t)\cos(\theta)+(b_z-t)\sin(\theta)\bigr|^p.
		\]
		Moreover, the function $w\mapsto\RW_p(x_{t,w},z)^p$ converges to $L(t)$
		uniformly in $t$ on compact sets, with error $O(w)$.
	\end{lemma}
	
	\begin{proof}
		The argument is the same as in the proof of Lemma~\ref{lem:limite_w0_sw},
		now using the closed form of Lemma~\ref{lem:forma_cerrada_rw} for the
		pair $(x_{t,w},z)$, with $\Delta b=t-b_z$ and $\Delta d=t+w-d_z$. Thus,
		the parallel cost is $|X_w|^p+|Y_w|^p$, with
		$X_w=(t-b_z)\cos\theta+(t+w-d_z)\sin\theta$ and
		$Y_w=(t+w-d_z)\cos\theta+(t-b_z)\sin\theta$, and the crossed cost is
		$(w^p+w_z^p)|\cos\theta-\sin\theta|^p$. For $w=0$, we obtain $A(\theta)$
		and $w_z^p|\cos\theta-\sin\theta|^p$, and the differences are
		\[
		X_w-X=w\sin\theta,\qquad Y_w-Y=w\cos\theta,\qquad
		0\le w^p|\cos\theta-\sin\theta|^p\le2^{p/2}w^p .
		\]
		As in Lemma~\ref{lem:limite_w0_sw}, we obtain
		$|\RW_p(x_{t,w},z)^p-L(t)|\le C'w$ for $t$ in a compact set $K_0$ and
		$0<w<1$, with $C'=\max\{2pM_0^{p-1},2^{p/2}\}$, where now $M_0=M_0(K_0)$ bounds
		$X_w$, $X$, $Y_w$, and $Y$, and therefore also
		$\RW_p(x_{t,w},z)\to L(t)^{1/p}$ uniformly on $K_0$.
	\end{proof}

	\begin{proposition}[\textbf{Anchors for $\SW_p$}]
		\label{prop:anclas_sw}
		With the notation of this subsection, let $t_1=\tfrac{b_z+d_z}{2}$, so
		that $\pi_\Delta(z)=(t_1,t_1)$. Then the following hold for every $p\ge1$.
		\begin{enumerate}
			\item $K(t_1)=\Bigl(\dfrac{w_z}{2}\Bigr)^p J$.
			\item If $p>1$, then there exists $t_*\in\R$ such that $K(t_*)<K(t_1)$.
			\item If $p=1$, then $K(t)=K(t_1)$ for all $t\in\R$.
		\end{enumerate}
	\end{proposition}
	
	\begin{proof}
		Since $t_1$ is the midpoint of the coordinates of $z$, we have
		$b_z=t_1-\tfrac{w_z}{2}$ and $d_z=t_1+\tfrac{w_z}{2}$. Then, writing
		$a\coloneqq t_1-t$, Lemma~\ref{lem:limite_w0_sw} gives
		\[
		K(t)=\frac{1}{\pi}\int\limits_{-\pi/2}^{\pi/2}\min\{A(\theta),B(\theta)\}\,\dd\theta,
		\qquad A=|u-v|^p+|u|^p,\qquad B=|v|^p,
		\]
		with $u\coloneqq a(\cos\theta+\sin\theta)$ and
		$v\coloneqq\tfrac{w_z}{2}(\cos\theta-\sin\theta)$. The term $B$ is the
		crossed cost and does not depend on $t$, so all the dependence on $t$
		lies in $A$.
		
		We first compute the value at $t=t_1$. In this case we have $a=0$, so
		$u=0$ and $A=B$, and therefore
		$K(t_1)=\tfrac1\pi\int B\,\dd\theta=\bigl(\tfrac{w_z}{2}\bigr)^pJ$.
		
		Now suppose that $p>1$; we show that $K$ takes a smaller value at
		$t_*\coloneqq t_1-\tfrac{w_z}{4}$. At this point we have
		$a=\tfrac{w_z}{4}$, so the condition $0<u<v$ is equivalent to
		$-1<\tan\theta<\tfrac13$, that is, to
		$\theta\in I\coloneqq\bigl(-\tfrac\pi4,\arctan\tfrac13\bigr)$. Thus, for
		$\theta\in I$, we can write $u=sv$ with $s\in(0,1)$, and
		inequality~\eqref{eq:convexidad_estricta} gives
		$A=v^p\bigl[(1-s)^p+s^p\bigr]<v^p=B$. Since $\min\{A,B\}\le B$ on all of
		$[-\pi/2,\pi/2]$, with strict inequality on $I$, which has positive
		measure, we obtain $K(t_*)<\tfrac1\pi\int B\,\dd\theta=K(t_1)$.
		
		Finally, suppose that $p=1$; we show that $K$ is constant. In this case,
		the triangle inequality gives $A=|u-v|+|u|\ge|v|=B$ for all $t$, so
		$\min\{A,B\}=B$, which does not depend on $t$, and therefore
		$K(t)=K(t_1)$ for all $t\in\R$.
	\end{proof}

	\begin{proposition}[\textbf{Anchors for $\RW_p$}]
		\label{prop:anclas_rw}
		With the notation of this subsection, let $t_1=\tfrac{b_z+d_z}{2}$. Then
		the following hold for every $p\ge1$.
		\begin{enumerate}
			\item $L(t_1)=2^{1-p}\,w_z^p\,J$.
			\item $\lim\limits_{t\to+\infty} L(t)=w_z^p\,J$.
			\item If $p=1$, then $L(t)=L(t_1)$ for all $t\in\R$.
		\end{enumerate}
		In particular, if $p>1$, then $L(t_1)<\lim\limits_{t\to+\infty}L(t)$.
	\end{proposition}
	
	\begin{proof}
		As in the proof of Proposition~\ref{prop:anclas_sw}, we write
		$b_z=t_1-\tfrac{w_z}{2}$ and $d_z=t_1+\tfrac{w_z}{2}$, but now we set
		$a\coloneqq t-t_1$. With $u\coloneqq a(\cos\theta+\sin\theta)$ and
		$v\coloneqq\tfrac{w_z}{2}(\cos\theta-\sin\theta)$,
		Lemma~\ref{lem:limite_w0_rw} gives
		\[
		L(t)=\frac{1}{\pi}\int\limits_{-\pi/2}^{\pi/2}\min\{A(\theta),B(\theta)\}\,\dd\theta,
		\qquad A=|v+u|^p+|v-u|^p,\qquad B=|2v|^p,
		\]
		where $B=w_z^p|\cos\theta-\sin\theta|^p$ does not depend on $t$ and
		$\tfrac1\pi\int B\,\dd\theta=w_z^pJ$.
		
		We first compute the value at $t=t_1$. In this case we have $u=0$, so
		$A=2|v|^p=2^{1-p}B\le B$; that is, the minimum is $A$, and therefore
		$L(t_1)=2^{1-p}w_z^pJ$.
		
		We now examine the behavior as $t\to+\infty$. For every
		$\theta\neq-\pi/4$, we have $\cos\theta+\sin\theta\neq0$, so
		$|u|\to\infty$, $A\to\infty$, and $\min\{A,B\}\to B$. Since
		$\min\{A,B\}\le B$ and $\{-\pi/4\}$ has measure zero, the dominated
		convergence theorem gives $\lim_{t\to+\infty}L(t)=w_z^pJ$. In particular,
		if $p>1$, then $L(t_1)=2^{1-p}w_z^pJ<w_z^pJ$.
		
		Finally, suppose that $p=1$; we show that $L$ is constant. In this case,
		the triangle inequality gives $A=|v+u|+|v-u|\ge|2v|=B$ for all $t$, so
		$\min\{A,B\}=B$, and therefore $L(t)=L(t_1)$ for all $t\in\R$.
	\end{proof}

	\begin{lemma}[\textbf{Distance between near-diagonal diagrams with equal persistence}]
		\label{lem:distancia_casi_diagonales}
		Let $s,t\in\R$, $w>0$, and $p\ge1$. Then
		\[
		\SW_p(x_{s,w},x_{t,w})^p\le 2w^p J,
		\qquad
		\RW_p(x_{s,w},x_{t,w})^p\le 2w^p J.
		\]
	\end{lemma}
	
	\begin{proof}
		Fix $\theta$. As in the proof of Lemma~\ref{lem:forma_cerrada_sw}, the
		projected measures of $(x_{s,w},x_{t,w})$ have two atoms each, so
		$\mathcal{W}_p^p=\min\{P,Q\}$, where the two arguments are now the costs
		of the parallel and crossed matchings. The parallel cost is
		$2|s-t|^p|\cos(\theta)+\sin(\theta)|^p$, and the crossed cost
		is $2^{1-p}w^p|\cos(\theta)-\sin(\theta)|^p$ for $\SW_p$ (matching
		each $x_{t,w}$ with $\pi_\Delta(x_{t,w})$) and
		$2w^p|\cos(\theta)-\sin(\theta)|^p$ for $\RW_p$ (matching with
		$r_\Delta(x_{t,w})$, which preserves the full persistence rather than
		half of it). Therefore, bounding $\mathcal{W}_p^p$ by the crossed cost and
		integrating, we obtain $\SW_p(x_{s,w},x_{t,w})^p\le2^{1-p}w^pJ\le2w^pJ$
		and $\RW_p(x_{s,w},x_{t,w})^p\le 2w^p J$.
	\end{proof}

	\subsection{Proofs of Theorems~\ref{thm:sw_no_metrica} and~\ref{thm:rw_no_metrica}}
	\label{subsec:prueba_no_metricas}
	
	The two proofs follow the same pattern. We take two positions at which
	the limit function ($K$ or $L$) takes different values, and we let
	$\delta>0$ be the gap between their $p$-th roots. By
	Lemmas~\ref{lem:limite_w0_sw} and~\ref{lem:limite_w0_rw}, for $w$ small
	enough the distances from the two near-diagonal diagrams at these
	positions to $z$ are within $\delta/4$ of the corresponding limits, and,
	by Lemma~\ref{lem:distancia_casi_diagonales}, the distance between the
	two near-diagonal diagrams is smaller than $\delta/4$. The positions are
	$t_1$ and $t_*$ for $\SW_p$ (Proposition~\ref{prop:anclas_sw}), and a
	sufficiently large $T$ and $t_1$ for $\RW_p$
	(Proposition~\ref{prop:anclas_rw}).
	
	\paragraph{Proof of Theorem~\ref{thm:sw_no_metrica}.}
	\mbox{}
	\begin{proof}
		Fix $p>1$ and a single-point diagram $z=(b_z,d_z)$ with
		$w_z=d_z-b_z>0$, and let $t_1=\tfrac{b_z+d_z}{2}$. By
		Proposition~\ref{prop:anclas_sw}, with $t_*=t_1-\tfrac{w_z}{4}$, we have
		$K(t_*)<K(t_1)$, and hence
		$\delta\coloneqq K(t_1)^{1/p}-K(t_*)^{1/p}>0$. The idea is to replace the
		limit values $K(t_1)$ and $K(t_*)$ by the distances to $z$ of two
		near-diagonal diagrams that are, in addition, very close to each other.
		
		We first choose the persistence $w$ of these diagrams. By
		Lemma~\ref{lem:limite_w0_sw}, we have $\SW_p(x_{t,w},z)\to K(t)^{1/p}$
		uniformly in $t\in\{t_1,t_*\}$ as $w\to0^+$. Moreover, by
		Lemma~\ref{lem:distancia_casi_diagonales}, we have
		$\SW_p(x_{t_1,w},x_{t_*,w})\le(2J)^{1/p}w\to0$. Therefore, for every
		sufficiently small $w>0$, we have
		\[
		\SW_p(x_{t_1,w},z)>K(t_1)^{1/p}-\tfrac{\delta}{4},
		\qquad
		\SW_p(x_{t_*,w},z)<K(t_*)^{1/p}+\tfrac{\delta}{4},
		\qquad
		\SW_p(x_{t_1,w},x_{t_*,w})<\tfrac{\delta}{4}.
		\]
		
		Fix such a $w$. The diagrams $X_1=x_{t_1,w}$, $X_2=x_{t_*,w}$, and $Z=z$
		are those of the statement, and the three inequalities above give
		\[
		\SW_p(X_1,Z)
		>K(t_1)^{1/p}-\tfrac{\delta}{4}
		=K(t_*)^{1/p}+\tfrac{3\delta}{4}
		>\SW_p(X_2,Z)+\tfrac{\delta}{2}
		>\SW_p(X_1,X_2)+\SW_p(X_2,Z)+\tfrac{\delta}{4},
		\]
		and in particular $\SW_p(X_1,Z)>\SW_p(X_1,X_2)+\SW_p(X_2,Z)$.
	\end{proof}

	\paragraph{Proof of Theorem~\ref{thm:rw_no_metrica}.}
	\mbox{}
	\begin{proof}
		Fix $p>1$ and a single-point diagram $z=(b_z,d_z)$ with
		$w_z=d_z-b_z>0$, and let $t_1=\tfrac{b_z+d_z}{2}$. By
		Proposition~\ref{prop:anclas_rw}, we have $L(t_1)<\lim_{t\to+\infty}L(t)$,
		so $L(T)>L(t_1)$ for every sufficiently large $T$. Fix such a $T$ and set
		$\delta\coloneqq L(T)^{1/p}-L(t_1)^{1/p}>0$.
		
		We first choose the persistence $w$. By Lemma~\ref{lem:limite_w0_rw}, we
		have $\RW_p(x_{t,w},z)\to L(t)^{1/p}$ for $t\in\{t_1,T\}$ as $w\to0^+$.
		Moreover, by Lemma~\ref{lem:distancia_casi_diagonales}, we have
		$\RW_p(x_{T,w},x_{t_1,w})\le(2J)^{1/p}w\to0$. Therefore, for every
		sufficiently small $w>0$, we have
		\[
		\RW_p(x_{T,w},z)>L(T)^{1/p}-\tfrac{\delta}{4},
		\qquad
		\RW_p(x_{t_1,w},z)<L(t_1)^{1/p}+\tfrac{\delta}{4},
		\qquad
		\RW_p(x_{T,w},x_{t_1,w})<\tfrac{\delta}{4}.
		\]
		
		Fix such a $w$. The diagrams $X_1=x_{T,w}$, $X_2=x_{t_1,w}$, and $Z=z$ are
		those of the statement, and the three inequalities above give
		\[
		\RW_p(X_1,Z)
		>L(T)^{1/p}-\tfrac{\delta}{4}
		=L(t_1)^{1/p}+\tfrac{3\delta}{4}
		>\RW_p(X_2,Z)+\tfrac{\delta}{2}
		>\RW_p(X_1,X_2)+\RW_p(X_2,Z)+\tfrac{\delta}{4},
		\]
		and in particular $\RW_p(X_1,Z)>\RW_p(X_1,X_2)+\RW_p(X_2,Z)$.
	\end{proof}
	\subsection{Certificates that $\SW_2$ and $\RW_2$ are not CND}
	\label{subsec:certificado_sw2}
	
	For $p=2$, the closed forms of Subsection~\ref{subsec:formas_cerradas}
	can be integrated exactly. After the substitution $t=\tan\theta$, the
	integrand becomes a piecewise rational function whose pieces have
	explicit antiderivatives, so no quadrature in $\theta$ is needed, and
	evaluating the resulting expressions in interval arithmetic gives
	rigorous enclosures of the entries of $M$ and $M'$, and hence of
	$a^\top Ma$ and $a'^\top M'a'$. The four diagrams of
	Theorem~\ref{thm:sw2_no_CND} have the same structure as those of
	Subsection~\ref{subsec:anclas}: $Z_1$ and $Z_2$ are near-diagonal
	diagrams of persistence $1$ located at the projections onto the diagonal
	of $Z_4$ and $Z_3$, respectively, which have persistence $1000$.
	
	\paragraph{Proof of Theorem~\ref{thm:sw2_no_CND}.}
	\mbox{}
	\begin{proof}
		\textit{Item~(1).} We first show that $\SW_2(x,y)$ can be computed in
		closed form when $x=(b,d)$ and $y=(b',d')$ are single-point diagrams with
		integer coordinates. By Lemma~\ref{lem:forma_cerrada_sw} with $p=2$ and
		the change of variable $t=\tan\theta$, which gives
		$\cos\theta\pm\sin\theta=(1\pm t)/\sqrt{1+t^2}$ and $\dd\theta=\dd t/(1+t^2)$,
		we have
		\[
		\SW_2(x,y)^2=\frac{1}{4\pi}\int\limits_{-\infty}^{\infty}
		\frac{\min\{q_1(t),q_2(t)\}}{(1+t^2)^2}\,\dd t,
		\]
		where $q_1(t)=4(\Delta b+\Delta d\,t)^2+4\Delta m^2(1+t)^2$ and
		$q_2(t)=(w^2+w'^2)(1-t)^2$ are polynomials of degree at most $2$ with
		integer coefficients, since $4\Delta m^2=\bigl((b+d)-(b'+d')\bigr)^2$.
		Since $q_1-q_2$ has degree at most $2$, either it vanishes
		identically or it has at most two real roots, which divide $\R$ into at most three
		intervals. In both cases, on each interval the minimum coincides with either $q_1$ or
		$q_2$. Since an antiderivative of
		$(\alpha t^2+\beta t+\gamma)/(1+t^2)^2$ is
		\[
		F(t)=\frac12\Bigl[\alpha\Bigl(\arctan t-\frac{t}{1+t^2}\Bigr)
		-\frac{\beta}{1+t^2}
		+\gamma\Bigl(\arctan t+\frac{t}{1+t^2}\Bigr)\Bigr],
		\qquad
		\lim_{t\to\pm\infty}F(t)=\pm\frac{\pi}{4}(\alpha+\gamma),
		\]
		each $\SW_2(x,y)^2$ is a finite sum of values of these antiderivatives at
		the roots of $q_1-q_2$ and at $\pm\infty$.
		
		We then apply this formula to the six pairs $(Z_i,Z_j)$ with $i<j$.
		Evaluating the above expressions with interval
		arithmetic\footnote{Interval arithmetic replaces each real number
			by a closed interval containing it and each operation by the
			corresponding operation on intervals, so that the result is
			an interval that rigorously contains the exact value.} and taking
		square roots, we obtain the intervals in
		Table~\ref{tab:intervalos_M} for $M_{ij}=\SW_2(Z_i,Z_j)$.
		
		Finally, since $M$ is symmetric with zero diagonal and
		$a=(-7,7,1,-1)^{\top}$, we have
		\[
		a^{\top}Ma=2\bigl[-49M_{12}-7M_{13}+7M_{14}+7M_{23}-7M_{24}-M_{34}\bigr],
		\]
		and evaluating this expression with the intervals in the table gives
		\[
		a^{\top}Ma\in[422.2423096,\,422.2423100]\subseteq(0,\infty).
		\]
		Since $a$ is a zero-sum vector, it follows that $\SW_2$ is not CND.
		
		\textit{Item~(2).} We proceed in the same way, now using
		Lemma~\ref{lem:forma_cerrada_rw}. With $p=2$ and $t=\tan\theta$, we have
		\[
		\RW_2(x,y)^2=\frac{1}{\pi}\int\limits_{-\infty}^{\infty}
		\frac{\min\{r_1(t),r_2(t)\}}{(1+t^2)^2}\,\dd t,
		\]
		where $r_1(t)=(\Delta b+\Delta d\,t)^2+(\Delta d+\Delta b\,t)^2$ and
		$r_2(t)=(w^2+w'^2)(1-t)^2$ are again polynomials of degree at most $2$
		with integer coefficients. Hence each $\RW_2(x,y)^2$ is a finite sum of
		values of the antiderivative $F$ at the roots of $r_1-r_2$ and at
		$\pm\infty$, and evaluating these sums with interval arithmetic gives
		the intervals for $M'_{ij}=\RW_2(Z_i,Z_j)$ in
		Table~\ref{tab:intervalos_M}. Since $a'=(-7,7,-1,1)^{\top}$, we have
		\[
		a'^{\top}M'a'=2\bigl[-49M'_{12}+7M'_{13}-7M'_{14}-7M'_{23}+7M'_{24}-M'_{34}\bigr]
		\in[396.4210427,\,396.4210428]\subseteq(0,\infty),
		\]
		and since $a'$ is a zero-sum vector, it follows that $\RW_2$ is not CND.
		The scripts that carry out both computations are included in the
		supplementary material.
	\end{proof}
	
	\begin{table}[htbp]
		\centering
		\footnotesize
		\begin{tabular}{ccc}
			\toprule
			$(i,j)$ & $M_{ij}=\SW_2(Z_i,Z_j)$ & $M'_{ij}=\RW_2(Z_i,Z_j)$ \\
			\midrule
			$(1,2)$ & [0.7063561210, 0.7063561211] & [1.4112093470, 1.4112093471] \\
			$(1,3)$ & [465.2199364403, 465.2199364404] & [744.0797523169, 744.0797523170] \\
			$(1,4)$ & [499.4661849391, 499.4661849392] & [706.4000280124, 706.4000280125] \\
			$(2,3)$ & [499.4661849391, 499.4661849392] & [706.4000280124, 706.4000280125] \\
			$(2,4)$ & [465.2831642245, 465.2831642246] & [743.7901486617, 743.7901486618] \\
			$(3,4)$ & [233.2722796707, 233.2722796708] & [258.1291353155, 258.1291353156] \\
			\bottomrule
		\end{tabular}
		\caption{\footnotesize Rigorous intervals for the entries of the matrices
			$M$ and $M'$ in Theorem~\ref{thm:sw2_no_CND}.}
		\label{tab:intervalos_M}
	\end{table}
	
	\section{Metric equivalence and stability}
	\label{sec:equiv_metrica}
	
	By Section~\ref{sec:sw_rw_no_metricas}, for $p>1$ the distances $\SW_p$
	and $\RW_p$ lack two of the properties of $\SW_1$ and $\RW_1$. We now
	establish a property that holds for every order $p\ge1$.
	In this section we prove that, on the space of persistence diagrams of
	bounded cardinality, the Sliced Wasserstein distances $\SW_p$ and $\RW_p$
	are metrically equivalent to the corresponding diagram distances $d_p$ and
	$d_p^{r}$, respectively, and we deduce the stability of $d_\infty^{r}$. For
	$p=1$, the metric equivalence between $\SW_1$ and $d_1$ is due to
	\cite{Carriere2017}, although our constants improve on theirs. The case of general
	$p$, all the statements for $\RW_p$ and $d_p^{r}$, and
	Corollary~\ref{cor:estabilidad_dpr} are new; the latter relies on the
	bottleneck stability theorem of \cite{CohenSteiner2007}.
	
	\begin{theorem}[\textbf{Metric equivalence between $\SW_p$ and $d_p$ for $p\ge1$}]
		\label{thm:equiv_p}
		Let $N\in\N$ and $X=\mathcal{D}_{\le N}$, and set $\Lambda_N=1+2N(2N-1)$ and
		$I_p=\int \limits_{0}^{\pi/2}\sin^p t\,\dd t$. For every $p\in[1,\infty)$
		and $A,B\in X$, the following hold.
		\begin{enumerate}
			\item $\mathcal{W}_p(\mu_{A,B}^{\theta},\mu_{B,A}^{\theta})\le2^{1/p}\sqrt2\,d_p(A,B)$
			for all $\theta\in[-\pi/2,\pi/2]$.
			\item $
			C_{p,N}\,d_p(A,B)
			\le
			\SW_p(A,B)
			\le
			\kappa_p\,d_p(A,B),$
			where $
			C_{p,N}
			=
			\frac{2^{1/p}}{\pi^{1/p}\Lambda_N}\,I_p^{1/p}$
			and
			$\kappa_p
			=
			\sqrt2\left(\frac{4}{\pi}\,I_p\right)^{1/p}$.
			
		\end{enumerate}
		The constant $\kappa_p$ cannot be improved. Moreover, we have
		$\kappa_1=4\sqrt2/\pi$, and the inequality $\kappa_p\le\sqrt2$ holds if and only if
		$p\ge2$.
	\end{theorem}
	
	The same argument, with the reflection in place of the projection onto the
	diagonal, yields the reflected counterpart.
	
	\begin{theorem}[\textbf{Metric equivalence between $ \RW_p$ and $d_p^{r}$ for $p\ge1$}]
		\label{thm:equiv_p_rw}
		Let $N\in\N$ and $X=\mathcal{D}_{\le N}$. For every
		$p\in[1,\infty)$ and $A,B\in X$, the following hold.
		\begin{enumerate}
			\item $\mathcal{W}_p(\mu_{A,B}(\theta),\mu_{B,A}(\theta))\le\sqrt2\,d_p^{r}(A,B)$
			for all $\theta\in[-\pi/2,\pi/2]$.
			\item With $\Lambda_N$, $I_p$, and $C_{p,N}$ as in Theorem~\ref{thm:equiv_p}, we have $
			C_{p,N}\,d_p^{r}(A,B)
			\le
			\RW_p(A,B)
			\le
			\kappa_p^{r}\,d_p^{r}(A,B),$
			where $\kappa_p^{r}=\sqrt2\left(\frac{2}{\pi}\,I_p\right)^{1/p}=2^{-1/p}\kappa_p$.
		\end{enumerate}
		The constant $\kappa_p^{r}$ cannot be improved. Moreover, we have
		$\kappa_1^{r}=2\sqrt2/\pi$, $\kappa_2^{r}=1$, and $\kappa_p^{r}<\sqrt2$ for every
		$p\in[1,\infty)$.
	\end{theorem}
	
	\begin{remark}[\textbf{On the lower constant}]
		\label{rem:constante_inferior}
		Since $\Lambda_N\sim4N^2$, the lower constant $C_{p,N}$ decays like
		$1/N^2$, which is the same order as in \cite{Carriere2017}, although the
		experiments in Section~\ref{sec:experimentos} show that it is far
		from tight for typical diagrams. Nevertheless, Carri\`ere and Bauer
		\cite{CarriereBauer2019} showed that, for several stable embeddings of
		diagrams into infinite-dimensional Hilbert spaces, any lower bound must
		depend on the cardinality of the diagrams. Whether the order $1/N^2$ can be
		improved for $\SW_p$ and $\RW_p$ remains an open question.
	\end{remark}
	
	As an application, the comparison in
	Theorem~\ref{thm:comparacion_dpr_dp} transfers the bottleneck stability
	theorem to $d_\infty^{r}$.
	
	\begin{corollary}[\textbf{Stability of the reflected bottleneck distance}]
		\label{cor:estabilidad_dpr}
		Let $X$ be a triangulable space, let $f,g:X\to\R$ be continuous tame functions in the
		sense of \cite{CohenSteiner2007}, let $k\ge0$, and let
		$A,B\in\mathcal{D}_{\mathrm{fin}}$ be the finite parts of the
		persistence diagrams of $f$ and $g$ in degree $k$. Then
		$d_\infty^{r}(A,B)\le 2\,\|f-g\|_\infty$.
	\end{corollary}
	
	\begin{proof}
		By Theorem~\ref{thm:comparacion_dpr_dp} with $p=\infty$, we have
		$d_\infty^{r}(A,B)\le 2\,d_\infty(A,B)$. Moreover, by the bottleneck stability
		theorem \cite{CohenSteiner2007}, the bottleneck distance
		between the full diagrams is at most $\|f-g\|_\infty$. Since a
		finite-cost matching pairs essential points with essential points, the
		restriction of such a matching to the finite parts gives
		$d_\infty(A,B)\le\|f-g\|_\infty$. The constant $2$ in the first
		inequality cannot be improved, since for $A=\{(0,1),(0,3)\}$ and
		$B=\{(0,3)\}$ we have $d_\infty(A,B)=1/2$ and $d_\infty^{r}(A,B)=1$.
	\end{proof}
	
	\paragraph{Strategy of the proofs.}
	The proofs of Theorems~\ref{thm:equiv_p} and~\ref{thm:equiv_p_rw} rely on
	the description of the optimal transport given by
	Lemma~\ref{lem:estructura_medidas_p} and
	Remark~\ref{rem:version_proyeccion}: in each direction $\theta$, the
	measures are the projections of two finite sets $P$ and $Q$, and the
	optimal transport is induced by the bijection $s^\theta:P\to Q$ that matches
	the projections in increasing order. For the upper bounds, we replace
	$s^\theta$ by a single bijection $\sigma:P\to Q$, independent of $\theta$
	and built from an optimal matching for $d_p$ or $d_p^r$; averaging
	$|\langle v,u_\theta\rangle|^p$ over $\theta$ then gives the constants
	$\kappa_p$ and $\kappa_p^r$, and a pair of single-point diagrams shows that
	they are sharp. For the lower bounds, the order of the projections changes
	only at finitely many critical angles, so $s^\theta$ is piecewise constant
	in $\theta$. On each piece we bound the integral of
	$|\cos(\theta-\varphi)|^p$ from below in terms of the length of the piece,
	and we add up the pieces by means of Hölder's inequality. Since the number
	of pieces is at most $\Lambda_N$, the lower constant $C_{p,N}$ depends on
	$N$. Subsection~\ref{subsec:aux_equiv} contains the auxiliary lemmas, and
	Subsection~\ref{subsec:prueba_equiv} contains the proofs.
	
	\subsection{Auxiliary lemmas}
	\label{subsec:aux_equiv}
	
	The first two lemmas bound the integral of $|\cos(\theta-\varphi)|^p$
	over an interval from below by a quantity that depends only on the
	length of the interval. The third bounds the number of intervals on which
	the optimal bijection $s^\theta$ is constant, and the fourth combines the
	contributions of these intervals.

	\begin{lemma}[\textbf{Exact lower bound for the integral of $|\cos|^p$, $p\ge1$}]
		\label{lem:integral_coseno_exacta_p}
		Let $\alpha,\beta\in[-\pi/2,\pi/2]$ with $\alpha<\beta$, let
		$\varphi\in[-\pi,\pi]$ be fixed, and let $p\ge1$. Then $\displaystyle 	\int\limits_{\alpha}^{\beta} |\cos(\theta-\varphi)|^p\,\dd\theta
		\;\ge\;
		2\int\limits_{0}^{L/2}\sin^p t\,\dd t$, where
		$L=\beta-\alpha$.
		The inequality is sharp, since equality is attained for $
		\varphi=\frac{\alpha+\beta}{2}-\frac{\pi}{2}$.
	\end{lemma}
	
	\begin{proof}
		Let $L=\beta-\alpha\in(0,\pi]$. The function $f(u)=|\cos u|^p$ is $\pi$-periodic, symmetric
		about $\pi/2$, and increasing in the distance from $\pi/2$ on $[0,\pi]$. Therefore, among all
		intervals of length $L$, the integral of $f$ is smallest over the one centered at
		$\pi/2$, and its value is
		$\int_{\frac{\pi}{2}-\frac{L}{2}}^{\frac{\pi}{2}+\frac{L}{2}}|\cos u|^p\,\dd u
		=2\int_{0}^{L/2}\sin^p t\,\dd t$.
		Finally, under the substitution $u=\theta-\varphi$, this interval corresponds to $\varphi=\frac{\alpha+\beta}{2}-\frac{\pi}{2}$.
	\end{proof}

	\begin{lemma}[\textbf{Sharp polynomial lower bound for the integral of $|\cos|^p$}]
		\label{lem:integral_coseno_p_p}
		Under the same hypotheses as in
		Lemma~\ref{lem:integral_coseno_exacta_p}, we have $\displaystyle
		\int\limits_{\alpha}^{\beta} |\cos(\theta-\varphi)|^p\,\dd\theta
		\;\ge\;
		\frac{2}{\pi^{p+1}}
		\left(\int\limits_{0}^{\pi/2}\sin^p t\,\dd t\right)
		(\beta-\alpha)^{p+1}$.
		The constant is sharp.
	\end{lemma}
	
	\begin{proof}
		By Lemma~\ref{lem:integral_coseno_exacta_p}, it suffices to show that the function $H(L)=\frac{2}{L^{p+1}}\int_{0}^{L/2}\sin^p t\,\dd t$ is decreasing on $(0,\pi]$, since its value at $L=\pi$ is $H(\pi)=\frac{2}{\pi^{p+1}}\int_0^{\pi/2}\sin^p t\,\dd t$. Indeed, the substitution $t=\frac{L}{2}s$ gives
		$H(L)=2^{-p}\int_0^1\bigl(\tfrac{\sin(Ls/2)}{Ls/2}\bigr)^p s^p\,\dd s$, which is decreasing
		because $\sin u/u$ is decreasing on $(0,\pi/2]$. Equality is attained for $\beta-\alpha=\pi$ and
		$\varphi$ as in Lemma~\ref{lem:integral_coseno_exacta_p}.
	\end{proof}

	\begin{lemma}[\textbf{Partition of the interval by critical angles}]
		\label{lem:particion_p}
		Let $A,B\in\mathcal{D}_{\le N}$ and $p\ge1$, and let $P$ and $Q$ be the sets of
		Lemma~\ref{lem:estructura_medidas_p}, or those of the proof of
		Theorem~\ref{thm:equiv_p} in the projected case. For each
		$\theta\in[-\pi/2,\pi/2]$, let $s^\theta:P\to Q$ be the optimal bijection given by
		Lemma~\ref{lem:estructura_medidas_p}. Then there exists a
		partition $-\frac{\pi}{2}=\Theta_0<\Theta_1<\cdots<\Theta_K=\frac{\pi}{2}$,
		with $K\le 1+2N(2N-1)$,
		such that $s^\theta$ is constant on each open subinterval
		$(\Theta_i,\Theta_{i+1})$.
	\end{lemma}
	
	\begin{proof}
		The bijection $s^\theta$ is determined by the increasing order of the
		projections of $P$ and of $Q$. This order can change only at an angle
		$\theta$ where $\langle x-y,u_\theta\rangle=0$ for some pair
		$x,y$ of points lying both in $P$ or both in $Q$; moreover, each such pair contributes at
		most one solution in $(-\pi/2,\pi/2)$. Since $|P|=|Q|\le2N$,
		there are at most $2\binom{2N}{2}=2N(2N-1)$ such angles, and
		adding the two endpoints yields a partition into at most
		$1+2N(2N-1)$ subintervals on which $s^\theta$ is constant.
	\end{proof}

	\begin{lemma}[\textbf{Lower bound for sums of powers of the lengths of a partition}]
		\label{lem:holder_particion}
		Let $q\ge1$ and let $-\frac{\pi}{2}=\Phi_0<\Phi_1<\cdots<\Phi_m=\frac{\pi}{2}$ be a partition.
		Then $
		\sum\limits_{i=1}^{m}(\Delta\Phi_i)^q
		\ge
		\dfrac{\pi^q}{m^{q-1}}$,
		where $\Delta\Phi_i=\Phi_i-\Phi_{i-1}$. If $q>1$, equality holds if and only if the partition is uniform.
	\end{lemma}
	
	\begin{proof}
		For $q>1$, Hölder's inequality applied to
		$(\Delta\Phi_1,\dots,\Delta\Phi_m)$ and $(1,\dots,1)$ gives $
		\pi=\sum\limits_i\Delta\Phi_i
		\le
		\left(\sum\limits_i(\Delta\Phi_i)^q\right)^{1/q}m^{1-1/q}$.
		Raising both sides to the power $q$ and solving for
		$\sum \limits_i(\Delta\Phi_i)^q$ yields the bound. The case $q=1$ is
		immediate.
	\end{proof}
	
	\subsection{Proofs of Theorems~\ref{thm:equiv_p} and~\ref{thm:equiv_p_rw}}
	\label{subsec:prueba_equiv}
	
	We give the proof in detail for $\SW_p$. For $\RW_p$ we only indicate
	the changes, the main one being that an optimal bijection for $d_p^r$
	already matches $P$ with $Q$, which removes a factor $2$ from the upper
	bound.
	
	\paragraph{Proof of Theorem~\ref{thm:equiv_p}.}
	\mbox{}
	\begin{proof}
		We fix distinct indices $n_a$ ($a\in A$) and $m_b$ ($b\in B$) and
		consider the sets of diagonal copies
		\[
		A_0\coloneqq\{(\pi_\Delta(a),n_a):a\in A\},\qquad
		B_0\coloneqq\{(\pi_\Delta(b),m_b):b\in B\},
		\]
		which are contained in $\Delta\times\N$ and satisfy $|A_0|=|A|$ and $|B_0|=|B|$. Let
		$P\coloneqq A\cup B_0$ and $Q\coloneqq B\cup A_0$, each of cardinality
		$|A|+|B|\le2N$. Since the norm involves only the first two coordinates,
		we write $\pi_\Delta(a)$ and $\pi_\Delta(b)$ instead of
		$(\pi_\Delta(a),n_a)$ and $(\pi_\Delta(b),m_b)$. With this notation, we have
		$\mu_{A,B}^\theta=\sum_{x\in P}\delta_{\pi_\theta(x)}$ and
		$\mu_{B,A}^\theta=\sum_{y\in Q}\delta_{\pi_\theta(y)}$, so
		Lemma~\ref{lem:estructura_medidas_p}, in its version with projection onto
		the diagonal (Remark~\ref{rem:version_proyeccion}), applies to
		$P$ and $Q$. We write $\delta(x)\coloneqq\|x-\pi_\Delta(x)\|_\infty$.
		
		\textbf{$\blacktriangleright$ Step 1.} Upper bounds.
		
		By Theorem~\ref{thm:dp_equivalencias}(2), there exist
		$A_1\subseteq A$, $B_1\subseteq B$, and a bijection $\gamma:A_1\to B_1$
		such that, setting $A_2\coloneqq A\setminus A_1$ and
		$B_2\coloneqq B\setminus B_1$,
		\[
		d_p(A,B)^p
		=
		\sum_{a\in A_1}\|a-\gamma(a)\|_\infty^p
		+\sum_{a\in A_2}\delta(a)^p
		+\sum_{b\in B_2}\delta(b)^p .
		\]
		We define $\sigma:P\to Q$ by
		\[
		\sigma(a)=\gamma(a),\quad
		\sigma(\pi_\Delta(\gamma(a)))=\pi_\Delta(a)
		\qquad(a\in A_1),
		\]
		\[
		\sigma(a)=\pi_\Delta(a)\quad(a\in A_2),
		\qquad
		\sigma(\pi_\Delta(b))=b\quad(b\in B_2).
		\]
		The domain blocks $A_1$, $A_2$, $\pi_\Delta(B_1)$, and
		$\pi_\Delta(B_2)$ form a partition of $P$, and their respective images $B_1$,
		$\pi_\Delta(A_2)$, $\pi_\Delta(A_1)$, and $B_2$ form a partition of $Q$,
		so $\sigma$ is a bijection. Then, by
		Lemma~\ref{lem:estructura_medidas_p}(3)--(4), for all $\theta$,
		\begin{equation}\label{eq:cota_sigma}
			\mathcal{W}_p^p(\mu_{A,B}^{\theta},\mu_{B,A}^{\theta})
			\le
			\sum_{x\in P}|\langle x-\sigma(x),u_\theta\rangle|^p .
		\end{equation}
		Each $a\in A_1$ contributes two terms to the sum
		$\sum_{x\in P}\|x-\sigma(x)\|_\infty^p$, corresponding to
		$a-\gamma(a)$ and to $\pi_\Delta(\gamma(a))-\pi_\Delta(a)$, and by
		Remark~\ref{rem:prop_diagonal}(2) the norm of the second one is
		at most $\|a-\gamma(a)\|_\infty$. Moreover, the points of $A_2$ and of $B_2$
		contribute $\delta(a)^p$ and $\delta(b)^p$, respectively. Therefore,
		\begin{equation}\label{eq:suma_sigma}
			\sum_{x\in P}\|x-\sigma(x)\|_\infty^p
			\le
			2\sum_{a\in A_1}\|a-\gamma(a)\|_\infty^p
			+\sum_{a\in A_2}\delta(a)^p
			+\sum_{b\in B_2}\delta(b)^p
			\le 2\,d_p(A,B)^p .
		\end{equation}
		
		\emph{Item 1.} By the Cauchy--Schwarz inequality, we have
		$|\langle v,u_\theta\rangle|\le\|v\|_2\le\sqrt2\|v\|_\infty$ for all
		$v\in\R^2$, whence
		$|\langle x-\sigma(x),u_\theta\rangle|^p\le(\sqrt2)^p\|x-\sigma(x)\|_\infty^p$.
		Therefore, \eqref{eq:cota_sigma} and \eqref{eq:suma_sigma} give
		$\mathcal{W}_p^p(\mu_{A,B}^{\theta},\mu_{B,A}^{\theta})\le2\,(\sqrt2)^p\,d_p(A,B)^p$.
		
		\emph{Upper bound in item 2.} For $v\in\R^2$, we write
		$v=\|v\|_2(\cos\varphi,\sin\varphi)$, so that
		$\langle v,u_\theta\rangle=\|v\|_2\cos(\theta-\varphi)$. Since
		$|\cos|^p$ is $\pi$-periodic, we have
		\begin{equation}\label{eq:promedio_angular}
			\frac1\pi\int\limits_{-\pi/2}^{\pi/2}|\langle v,u_\theta\rangle|^p\,\dd\theta
			=
			\frac{\|v\|_2^p}{\pi}\int\limits_{-\pi/2}^{\pi/2}|\cos t|^p\,\dd t
			=
			\frac{2I_p}{\pi}\,\|v\|_2^p .
		\end{equation}
		Then, since $\sigma$ does not depend on $\theta$, we can integrate \eqref{eq:cota_sigma}
		term by term. Using \eqref{eq:promedio_angular},
		$\|v\|_2\le\sqrt2\|v\|_\infty$, and \eqref{eq:suma_sigma}, we obtain
		\[
		\SW_p(A,B)^p
		\le
		\frac{2I_p}{\pi}\,(\sqrt2)^p\cdot2\,d_p(A,B)^p
		=
		\kappa_p^p\,d_p(A,B)^p .
		\]
		
		\emph{Sharpness of $\kappa_p$.} Let $A=\{(0,L)\}$ and
		$B=\{(\varepsilon,L+\varepsilon)\}$ with $0<\varepsilon<L/2$, so that
		$d_p(A,B)=\varepsilon$. There are only two bijections $P\to Q$, so, by
		Lemma~\ref{lem:estructura_medidas_p}(3)--(4),
		\[
		\mathcal{W}_p^p(\mu_{A,B}^{\theta},\mu_{B,A}^{\theta})
		=
		\min\Bigl\{2\,\varepsilon^p|\cos\theta+\sin\theta|^p,\;
		2\,(L/2)^p|\cos\theta-\sin\theta|^p\Bigr\}.
		\]
		As $L\to\infty$, the minimum equals the first term for all
		$\theta\neq\pi/4$; hence, by dominated convergence and
		\eqref{eq:promedio_angular} with $v=(1,1)$, we obtain
		$\SW_p(A,B)^p/\varepsilon^p\to2\cdot(\sqrt2)^p\,\frac{2I_p}{\pi}=\kappa_p^p$.
		Finally, since $I_1=1$, we have $\kappa_1=4\sqrt2/\pi$. Moreover, since $I_p$ is decreasing in
		$p$ and $I_2=\pi/4$, we have $\kappa_p\le\sqrt2$ if and only if $p\ge2$.
		
		\textbf{$\blacktriangleright$ Step 2.} Lower bound.
		
		By Lemma~\ref{lem:particion_p}, there exists a partition
		$-\frac{\pi}{2}=\Theta_0<\cdots<\Theta_K=\frac{\pi}{2}$ with
		$K\le\Lambda_N$ such that $s^\theta=s_i$ on each subinterval
		$(\Theta_i,\Theta_{i+1})$. For each $x\in P$ with $x\ne s_i(x)$,
		we write $x-s_i(x)=\|x-s_i(x)\|_2(\cos\varphi_x,\sin\varphi_x)$, so
		that, by Lemma~\ref{lem:estructura_medidas_p}(3), for
		$\theta\in(\Theta_i,\Theta_{i+1})$ we have
		\[
		\mathcal{W}_p^p(\mu_{A,B}^{\theta},\mu_{B,A}^{\theta})
		=\sum_{x\in P}|\langle x-s_i(x),u_\theta\rangle|^p
		=\sum_{\substack{x\in P\\ x\ne s_i(x)}}\|x-s_i(x)\|_2^p\,
		|\cos(\theta-\varphi_x)|^p .
		\]
		Then, integrating over $(\Theta_i,\Theta_{i+1})$, whose length we denote by
		$\Delta\Theta_i\coloneqq\Theta_{i+1}-\Theta_i$, and applying
		Lemma~\ref{lem:integral_coseno_p_p} to each term, we obtain
		\[
		\int\limits_{\Theta_i}^{\Theta_{i+1}}\mathcal{W}_p^p\,\dd\theta
		\ge
		\frac{2I_p}{\pi^{p+1}}\,(\Delta\Theta_i)^{p+1}
		\sum_{x\in P}\|x-s_i(x)\|_2^p .
		\]
		On the other hand, the map $s_i:A\cup B_0\to A_0\cup B$ is a bijection with
		$A_0,B_0\subseteq\Delta\times\N$, $|A_0|=|A|$, and $|B_0|=|B|$, so, by
		Theorem~\ref{thm:dp_equivalencias}(1) and the inequality
		$\|\cdot\|_\infty\le\|\cdot\|_2$,
		\[
		d_p(A,B)^p
		\le
		\sum_{x\in P}\|x-s_i(x)\|_\infty^p
		\le
		\sum_{x\in P}\|x-s_i(x)\|_2^p .
		\]
		Therefore, we have
		\[
		\int\limits_{\Theta_i}^{\Theta_{i+1}}\mathcal{W}_p^p\,\dd\theta
		\ge
		\frac{2I_p}{\pi^{p+1}}\,(\Delta\Theta_i)^{p+1}\,d_p(A,B)^p .
		\]
		Summing over $i$ and applying Lemma~\ref{lem:holder_particion} with
		$q=p+1$, which gives $\sum_i(\Delta\Theta_i)^{p+1}\ge\pi^{p+1}K^{-p}$, we obtain
		\[
		\SW_p(A,B)^p
		=\frac{1}{\pi}\int\limits_{-\pi/2}^{\pi/2}\mathcal{W}_p^p\,\dd\theta
		\ge
		\frac{2I_p}{\pi}\,K^{-p}\,d_p(A,B)^p .
		\]
		Finally, taking $p$-th roots and using $K\le\Lambda_N$, we conclude that
		\[
		\SW_p(A,B)
		\ge
		\frac{1}{\Lambda_N}\Bigl(\frac{2I_p}{\pi}\Bigr)^{1/p}d_p(A,B)
		=
		C_{p,N}\,d_p(A,B).
		\]
	\end{proof}
	
	\paragraph{Proof of Theorem~\ref{thm:equiv_p_rw}.}
	\mbox{}
	\begin{proof}
		
		The lower bound follows by repeating Step 2 of the proof of
		Theorem~\ref{thm:equiv_p} with $P=A\sqcup r_\Delta(B)$ and
		$Q=B\sqcup r_\Delta(A)$, with the bijection $s^\theta$ given by
		Lemma~\ref{lem:estructura_medidas_p} in the reflected case, and with
		$\pi_\Delta$, $d_p$, and $\mu_{A,B}^{\theta}$ replaced throughout the argument by
		$r_\Delta$, $d_p^{r}$, and $\mu_{A,B}(\theta)$, respectively. For the upper bounds, in contrast,
		we take $\sigma$ to be an optimal bijection in Definition~\ref{def:dp_r};
		since it already matches $P$ with $Q$, no pair is counted
		twice. Thus, by Lemma~\ref{lem:estructura_medidas_p}(4), for all $\theta$ we have
		\begin{equation}\label{eq:cota_sigma_r}
			\mathcal{W}_p^p(\mu_{A,B}(\theta),\mu_{B,A}(\theta))
			\le
			\sum\limits_{x\in P}|\langle x-\sigma(x),u_\theta\rangle|^p
			,\qquad
			\sum\limits_{x\in P}\|x-\sigma(x)\|_\infty^p=d_p^{r}(A,B)^p .
		\end{equation}
		
		\emph{Item 1.} By \eqref{eq:cota_sigma_r} and the Cauchy--Schwarz inequality
		$|\langle v,u_\theta\rangle|\le\sqrt2\|v\|_\infty$, we have
		\[
		\mathcal{W}_p^p(\mu_{A,B}(\theta),\mu_{B,A}(\theta))\le(\sqrt2)^p\,d_p^{r}(A,B)^p .
		\]
		
		\emph{Upper bound in item 2.} Since $\sigma$ does not depend on $\theta$, we can
		integrate \eqref{eq:cota_sigma_r} term by term. Using
		\eqref{eq:promedio_angular} and $\|v\|_2\le\sqrt2\|v\|_\infty$, we obtain
		\[
		\RW_p(A,B)^p
		\le
		\frac{2I_p}{\pi}\sum\limits_{x\in P}\|x-\sigma(x)\|_2^p
		\le
		\frac{2I_p}{\pi}\,(\sqrt2)^p\,d_p^{r}(A,B)^p
		=
		(\kappa_p^{r})^p\,d_p^{r}(A,B)^p .
		\]
		In contrast to Theorem~\ref{thm:equiv_p}, the factor $2$ in \eqref{eq:suma_sigma}
		does not appear; hence $\kappa_p^{r}=2^{-1/p}\kappa_p$.
		
		\emph{Sharpness of $\kappa_p^{r}$.} Let $A=\{(0,L)\}$ and
		$B=\{(\varepsilon,L+\varepsilon)\}$ with $0<\varepsilon<L$. Then
		$P=\{(0,L),(L+\varepsilon,\varepsilon)\}$ and
		$Q=\{(\varepsilon,L+\varepsilon),(L,0)\}$, and there are only two
		bijections $P\to Q$. The first one sends $(0,L)$ to
		$(\varepsilon,L+\varepsilon)$ and $(L+\varepsilon,\varepsilon)$ to $(L,0)$,
		with displacements $\pm(\varepsilon,\varepsilon)$, while the second one
		has displacements $\pm(L,-L)$. Hence
		$d_p^{r}(A,B)=2^{1/p}\varepsilon$, and, by
		Lemma~\ref{lem:estructura_medidas_p}(3)--(4),
		\[
		\mathcal{W}_p^p(\mu_{A,B}(\theta),\mu_{B,A}(\theta))
		=
		\min\Bigl\{2\,\varepsilon^p|\cos\theta+\sin\theta|^p,\;
		2\,L^p|\cos\theta-\sin\theta|^p\Bigr\}.
		\]
		As $L\to\infty$, the minimum equals the first term for all
		$\theta\neq\pi/4$; hence, by dominated convergence and
		\eqref{eq:promedio_angular} with $v=(1,1)$, we obtain
		$\RW_p(A,B)^p/d_p^{r}(A,B)^p\to(\sqrt2)^p\,\frac{2I_p}{\pi}=(\kappa_p^{r})^p$.
		Finally, since $I_1=1$ and $I_2=\pi/4$, we have $\kappa_1^{r}=2\sqrt2/\pi$ and
		$\kappa_2^{r}=1$. Moreover, since $2I_p/\pi<1$, we have $\kappa_p^{r}<\sqrt2$.
	\end{proof}

	\section{Conditional negative definiteness of $\SW_1$ and $\RW_1$}
	\label{sec:CND_sw1_rw1}
	
	The metric equivalence of Section~\ref{sec:equiv_metrica} holds for every
	$p$, but the positive definiteness of the exponential kernels requires the
	CND property, which fails for $p=2$ by Section~\ref{sec:sw_rw_no_metricas}.
	In this section we prove that the distances $\SW_1$ and $\RW_1$ are CND, so
	that, by Schoenberg's theorem, the associated exponential kernels are
	positive definite, and we then examine what these kernels guarantee when
	they are used in learning (Proposition~\ref{prop:kernels_aprendizaje} and
	Remark~\ref{rem:kernels_aprendizaje}). Schoenberg's theorem (Theorem~\ref{thm:schoenberg}) is classical
	\cite{Schoenberg1938a,Schoenberg1938b,Berg1984}, and the conditional
	negative definiteness of $\SW_1$, together with the kernel $k_{\SW_1}$, is
	due to \cite{Carriere2017}. In contrast, the corresponding statements for
	$\RW_1$, and hence the kernel $k_{\RW_1}$, are new.
	
	We first recall the two notions on which the construction is based.
	
	\begin{definition}[\textbf{Conditionally negative definite (CND) function}]
		\label{def:nd}
		Let $f:X\times X\to\R$ be a function. We say that $f$ is conditionally
		negative definite (CND) if it is symmetric, if $f(x,x)=0$ for all $x\in X$,
		and if $\sum\limits_{i,j=1}^{n}a_i a_j f(x_i,x_j)\le0$ for all $n\ge1$,
		$x_1,\dots,x_n\in X$, and
		$a_1,\dots,a_n\in\R$ with $\sum\limits_{i=1}^{n} a_i=0$.
		
	\end{definition}
	
	\begin{theorem}[\textbf{Schoenberg, 1938}]\label{thm:schoenberg}
		Let $f:X\times X\to\R$ be a symmetric function with
		$f(x,x)=0$ and $f\ge0$. Then $f$ is CND if and only if, for all
		$t>0$, the kernel $K_t(x,y)=\exp\bigl(-t f(x,y)\bigr)$ is positive
		definite on $X$. Moreover, the function $f$ is CND if and only if there
		exist a Hilbert space $H$ and a map $\Phi:X\to H$ such that
		$f(x,y)=\|\Phi(x)-\Phi(y)\|_H^2$.
	\end{theorem}
	
	\begin{proof}
		See \cite{Schoenberg1938a,Schoenberg1938b} and \cite[Ch.~3]{Berg1984}.
	\end{proof}
	
	\begin{lemma}[\textbf{$\SW_1$ and $\RW_1$ are CND}]
		\label{lem:sw_nd}
		The distances $\SW_1$ and $\RW_1$ are conditionally negative
		definite functions on $\mathcal{D}_{\mathrm{fin}}$.
	\end{lemma}
	
	\begin{proof}
		For $\SW_1$, this is the result of Carrière et al.\ \cite{Carriere2017},
		whose proof uses the facts that $\mathcal{W}_1$ is CND on $\mathcal{M}_r(\R)$ and that,
		for each $\theta$, a common measure can be added to the measures of
		$A_1,\dots,A_n$ without changing $\mathcal{W}_1$
		(Proposition~\ref{prop:equiv_p}(3)). For $\RW_1$, the argument is analogous,
		with $\pi_\Delta$ replaced by $r_\Delta$.
	\end{proof}
	
	Combining the two previous results yields the kernels that we test in
	Subsection~\ref{subsec:orbitas}.
	
	\begin{definition}[\textbf{Sliced Wasserstein kernels}]
		\label{def:kernel_sw}
		For $\sigma>0$, we define the kernels
		\[
		k_{\SW_1}(A,B)=\exp\left(-\frac{\SW_1(A,B)}{2\sigma^2}\right),
		\qquad
		k_{\RW_1}(A,B)=\exp\left(-\frac{\RW_1(A,B)}{2\sigma^2}\right).
		\]
		By Lemma~\ref{lem:sw_nd} and Theorem~\ref{thm:schoenberg}, both kernels
		are positive definite.
	\end{definition}
	
	\begin{proposition}[\textbf{The kernels $k_{\SW_1}$ and $k_{\RW_1}$ in learning}]
		\label{prop:kernels_aprendizaje}
		Let $d$ be $\SW_1$ or $\RW_1$, let $\sigma>0$, and let
		$k(A,B)=\exp\bigl(-d(A,B)/(2\sigma^2)\bigr)$. Then the following hold.
		\begin{enumerate}
			\item The kernel $k$ is positive definite on
			$\mathcal{D}_{\mathrm{fin}}$; that is,
			$\sum\limits_{i,j=1}^{n}a_i a_j\,k(A_i,A_j)\ge0$ for all
			$A_1,\dots,A_n\in\mathcal{D}_{\mathrm{fin}}$ and $a\in\R^n$.
			\item If $X\subseteq\mathcal{D}_{\mathrm{fin}}$ is compact with respect
			to $d$, then $k$ is universal on $X$; that is, $k$ is continuous on
			$X\times X$ and its reproducing kernel Hilbert space $H_k$ is dense in
			$C(X)$ with respect to the supremum norm. Here $H_k$ is the completion
			of the linear span of the functions $k(A,\bullet)$, with $A\in X$, with
			respect to the inner product given by
			$\langle k(A,\bullet),k(B,\bullet)\rangle_{H_k}=k(A,B)$, and its
			elements are continuous functions on $X$
			\cite[Ch.~4]{SteinwartChristmann2008}. This
			applies, for instance, to the set $X_{N,L}$
			of diagrams $A$ with $|A|\le N$ all of whose points lie in $[0,L]^2$.
			\item For all $A_1,\dots,A_n\in\mathcal{D}_{\mathrm{fin}}$ and
			$\beta_1,\dots,\beta_n,b\in\R$, the function
			$f=\sum\limits_{i=1}^{n}\beta_i\,k(A_i,\bullet)+b$ satisfies
			\[
			|f(A)-f(B)|\le\frac{c_d}{2\sigma^2}\Bigl(\sum_{i=1}^{n}|\beta_i|\Bigr)\,d_1(A,B)
			\qquad(A,B\in\mathcal{D}_{\mathrm{fin}}),
			\]
			where $c_{\SW_1}=c_{\RW_1}=4\sqrt2/\pi$.
		\end{enumerate}
	\end{proposition}
	
	\begin{proof}
		Throughout the proof, we use the bound
		\begin{equation}\label{eq:cota_d_d1}
			d(A,B)\le c_d\,d_1(A,B)\quad\text{for all }A,B \in\mathcal{D}_{\text{fin}}.
		\end{equation}
		For $d=\SW_1$, this is the upper bound $\SW_1\le\kappa_1 d_1$ of
		Theorem~\ref{thm:equiv_p}, with $\kappa_1=4\sqrt2/\pi$. For $d=\RW_1$, it
		follows from the upper bound $\RW_1\le\kappa_1^{r}\,d_1^{r}$ of
		Theorem~\ref{thm:equiv_p_rw}, with $\kappa_1^{r}=2\sqrt2/\pi$, together with
		the identity $d_1^{r}=2d_1$ of Corollary~\ref{cor:d1r_metrica}. These
		theorems are stated on $\mathcal{D}_{\le N}$, but the constants in their
		upper bounds do not depend on $N$, so~\eqref{eq:cota_d_d1} holds on all of
		$\mathcal{D}_{\mathrm{fin}}$.
		
		Item~(1) follows from Lemma~\ref{lem:sw_nd} and
		Theorem~\ref{thm:schoenberg} with $t=1/(2\sigma^2)$. For~(2), the
		function $d$ is continuous on $X\times X$, as is every metric, and hence
		so is $k$. Moreover, Lemma~\ref{lem:sw_nd} and Theorem~\ref{thm:schoenberg}
		provide a Hilbert space $H$ and a map $\Phi\colon\mathcal{D}_{\mathrm{fin}}\to H$
		with $d(A,B)=\|\Phi(A)-\Phi(B)\|_H^2$, so that
		\[
		k(A,B)=\exp\Bigl(-\frac{\|\Phi(A)-\Phi(B)\|_H^2}{2\sigma^2}\Bigr).
		\]
		Since $d$ is a metric (Theorem~\ref{thm:sw1_rw1_metricas}), so is
		$d^{1/2}$, and both define the same topology on $X$. The map $\Phi$ is
		then an isometry from $(X,d^{1/2})$ into $H$; hence it is injective and
		continuous on $(X,d)$. Since $X$ is a compact metric space, it is
		separable, and therefore the closed linear span $H_X$ of $\Phi(X)$ is a
		separable Hilbert space. Theorem~2.2(iii) of \cite{ChristmannSteinwart2010},
		applied to the continuous injective map $\Phi\colon X\to H_X$, states that
		the kernel $\exp\bigl(-\gamma\|\Phi(A)-\Phi(B)\|_H^2\bigr)$ is universal on
		$X$ for every $\gamma>0$. In particular, for $\gamma=1/(2\sigma^2)$, this
		kernel is $k$.
		
		It remains to check that $X_{N,L}$ is compact with respect to $d$. Let
		$T=\{(b,e)\in[0,L]^2:b\le e\}$. The map that sends
		$(x_1,\dots,x_m)\in T^m$, with $0\le m\le N$, to the diagram in which each
		point of $T\setminus\Delta$ appears as many times as it occurs among
		$x_1,\dots,x_m$ is continuous with respect to $d_1$, and it maps the compact
		set $\bigcup\limits_{m=0}^{N}T^m$ onto $X_{N,L}$. Hence
		$X_{N,L}$ is compact with respect to $d_1$ and, by~\eqref{eq:cota_d_d1},
		also with respect to $d$.
		
		For~(3), the function $s\mapsto e^{-s/(2\sigma^2)}$ is
		$\dfrac{1}{2\sigma^2}$-Lipschitz on $[0,\infty)$. Combined with the
		triangle inequality for $d$ (Theorem~\ref{thm:sw1_rw1_metricas})
		and~\eqref{eq:cota_d_d1}, this gives, for every $i$,
		\[
		|k(A_i,A)-k(A_i,B)|
		\le\frac{|d(A_i,A)-d(A_i,B)|}{2\sigma^2}
		\le\frac{d(A,B)}{2\sigma^2}
		\le\frac{c_d}{2\sigma^2}\,d_1(A,B).
		\]
		Multiplying by $|\beta_i|$ and summing over $i$ yields~(3).
	\end{proof}
	
	\begin{remark}[\textbf{Interpretation in learning}]
		\label{rem:kernels_aprendizaje}
		Let us recall how a support vector machine (SVM) uses a kernel to
		separate two classes of diagrams. Suppose that the training diagrams
		$A_1,\dots,A_n$ belong to two classes, and set
		\[
		y_i=\begin{cases}
			\phantom{-}1 & \text{if } A_i \text{ is in the first class},\\
			-1 & \text{if } A_i \text{ is in the second class}.
		\end{cases}
		\]
		The SVM assigns a diagram $A\in\mathcal{D}_{\mathrm{fin}}$ to a class
		according to the sign of a function $f$,
		\[
		\mathrm{SVM}(A)=\begin{cases}
			\text{first class} & \text{if } f(A)\ge0,\\
			\text{second class} & \text{if } f(A)<0,
		\end{cases}
		\qquad
		f=\sum_{i=1}^{n}v_i\,k(A_i,\bullet)+b,
		\qquad v_i=\alpha_iy_i,
		\]
		where $\alpha=(\alpha_1,\dots,\alpha_n)$ maximizes the function
		$L(\alpha)=\sum\limits_{i=1}^{n}\alpha_i-\frac12\,v^{\top}Kv$, with
		$v=(v_1,\dots,v_n)^{\top}$ and
		$K=\bigl(k(A_i,A_j)\bigr)_{i,j=1}^{n}$,
		subject to $0\le\alpha_i\le C$ and $\sum_i\alpha_i y_i=0$. Here $C>0$ is
		a regularization parameter, and $b$ is determined by the optimal
		$\alpha$. Each item of Proposition~\ref{prop:kernels_aprendizaje} says
		something concrete about this procedure.
		
		By item~(1), the Gram matrix $K$ is positive semidefinite, so we have
		$u^{\top}Ku\ge0$ for every $u\in\R^n$, which makes $L$ concave.
		Indeed, let $\alpha$ and $\alpha'$ satisfy the constraints, let $v$ and
		$v'$ be the corresponding vectors, and let $t\in[0,1]$. Since $v$
		depends linearly on $\alpha$, the linear part of $L$ cancels, and
		expanding the quadratic form gives
		\[
		L\bigl(t\alpha+(1-t)\alpha'\bigr)-tL(\alpha)-(1-t)L(\alpha')
		=\frac{t(1-t)}{2}\,(v-v')^{\top}K(v-v')\ge0 .
		\]
		Since the constraints define a convex set, every local maximum of
		$L$ on this set is a global maximum. To see this, note that if
		$L(\alpha')>L(\alpha)$, then the inequality above with $t=1-s$ gives
		\[
		L\bigl(\alpha+s(\alpha'-\alpha)\bigr)\ge L(\alpha)+s\bigl(L(\alpha')-L(\alpha)\bigr)>L(\alpha)
		\qquad(0<s\le1),
		\]
		so $L$ takes larger values at points arbitrarily close to $\alpha$, and
		thus $\alpha$ is not a local maximum. Hence an algorithm that finds a local
		maximum of $L$ has found an optimal $\alpha$, whatever the values of
		$\sigma$ and $C$. In contrast, if the kernel is not positive definite, the Gram matrix $K$ may have
		a vector $u$ with $u^{\top}Ku<0$, and in that case the right-hand side of the identity
		above can be negative. In that case, the function $L$ may fail to
		be concave, and therefore the algorithm may stop at a point that is not
		optimal.
		
		Item~(2) describes the kinds of separation that an SVM can learn. The
		function $f$ is a linear combination of the functions $k(A_i,\bullet)$ plus
		a constant, and universality means that combinations of this kind
		approximate every continuous function on $X$ uniformly. That is, for every
		continuous $g\colon X\to\R$ and every $\varepsilon>0$, there exist diagrams
		$B_1,\dots,B_m\in X$ and coefficients $c_1,\dots,c_m\in\R$ such that
		\[
		\sup_{A\in X}\Bigl|\,g(A)-\sum_{j=1}^{m}c_j\,k(B_j,A)\Bigr|<\varepsilon .
		\]
		In particular, this approximation has the same sign as $g$ at every
		diagram satisfying $|g(A)|\ge\varepsilon$. Hence, if the two classes are
		separated by the sign of a continuous function $g$, then a combination of
		the functions $k(B_j,\bullet)$ reproduces this separation everywhere
		except in the thin region where $|g|<\varepsilon$. However, not every
		positive definite kernel has this property. For example, if
		$k(A,A')=\langle\varphi(A),\varphi(A')\rangle_{\R^D}$ for some map
		$\varphi\colon X\to\R^D$, then
		$\sum_j c_j\,k(B_j,\bullet)=\langle z,\varphi(\bullet)\rangle$ with
		$z=\sum_j c_j\,\varphi(B_j)$, so all these combinations lie in a space of
		dimension at most $D$, which cannot approximate every continuous function
		on an infinite set $X$. Moreover, universality is the key hypothesis in the
		results guaranteeing that, as the number of training diagrams grows and
		$C$ is chosen suitably, the error of the SVM approaches the smallest error
		achievable by any classifier \cite{Steinwart2005}; see also
		\cite{SteinwartChristmann2008,ChristmannSteinwart2010}.
		
		Item~(3) says that diagrams that are close with respect to $d_1$ are
		assigned close values by $f$. Since $|v_i|=\alpha_i\le C$, item~(3) gives
		\[
		|f(A)-f(B)|\le\frac{c_d\,nC}{2\sigma^2}\,d_1(A,B).
		\]
		
		Carri\`ere et al.~\cite{Carriere2017} observe, using \cite{Kwitt2015},
		that $\exp(k_{\SW_1})$ is universal. Item~(2) shows that the kernel
		$k_{\SW_1}$ itself is already universal on compact sets, because the
		distance $\SW_1$ is both CND and a metric.
		
		For $p>1$, the situation is different. The kernel
		$k_{\SW_2}=\exp\bigl(-\SW_2/(2\sigma^2)\bigr)$ is not positive definite
		for any $\sigma>0$, as the diagrams of Theorem~\ref{thm:sw2_no_CND}
		show explicitly. Indeed, for $\lambda>0$ and $A\in\mathcal{D}_{\mathrm{fin}}$,
		let $\lambda A$ be the diagram obtained by multiplying the first two
		coordinates of every triple of $A$ by $\lambda$. This multiplies the
		atoms of all the projected measures by $\lambda$, and hence also every
		Wasserstein distance between them, so that
		$\SW_2(\lambda A,\lambda B)=\lambda\,\SW_2(A,B)$. Now fix $\sigma>0$,
		take $Z_1,\dots,Z_4$, $M$, and $a$ as in Theorem~\ref{thm:sw2_no_CND}(1),
		and let
		\[
		K_\lambda=\Bigl(\exp\bigl(-\SW_2(\lambda Z_i,\lambda Z_j)/(2\sigma^2)\bigr)\Bigr)_{i,j=1}^{4}
		=\Bigl(\exp\bigl(-\lambda M_{ij}/(2\sigma^2)\bigr)\Bigr)_{i,j=1}^{4}
		\]
		be the Gram matrix of the rescaled diagrams. Since $\sum_i a_i=0$,
		expanding the exponential gives
		\[
		a^{\top}K_\lambda a=-\frac{\lambda}{2\sigma^2}\,a^{\top}Ma+O(\lambda^2)
		\qquad(\lambda\to0),
		\]
		which is negative for sufficiently small $\lambda$, because $a^{\top}Ma>0$.
		Hence $K_\lambda$ is not positive semidefinite for such $\lambda$, and
		this holds whatever the value of $\sigma$. Since we also have
		$\RW_2(\lambda A,\lambda B)=\lambda\,\RW_2(A,B)$, the same argument, with
		$M'$ and $a'$ from Theorem~\ref{thm:sw2_no_CND}(2), shows that
		$k_{\RW_2}=\exp\bigl(-\RW_2/(2\sigma^2)\bigr)$ is not positive definite
		for any $\sigma>0$. Without positive definiteness, there is no
		reproducing kernel Hilbert space either, so item~(2) is not even
		meaningful for $k_{\SW_2}$ and $k_{\RW_2}$. Moreover, the proof of
		item~(3) does not carry over, since it uses the triangle inequality,
		which fails for $p>1$ (Theorems~\ref{thm:sw_no_metrica}
		and~\ref{thm:rw_no_metrica}). This does not make $k_{\SW_2}$
		useless in practice because an SVM can still be trained with a kernel that
		is not positive definite, and in the orbit benchmark of
		Subsection~\ref{subsec:orbitas} the kernel $k_{\SW_2}$ was even slightly
		more accurate than $k_{\SW_1}$, although its training Gram matrix had
		negative eigenvalues in all ten runs. What is lost for $p>1$ are the
		guarantees: the concavity of the training problem, the consistency of the
		SVM, and our stability bound.
	\end{remark}
	
	\section{Numerical experiments}
	\label{sec:experimentos}
	
	The previous sections leave two questions that can be explored
	numerically: how far from sharp the constants of the metric equivalence
	are on typical diagrams, and for which orders $p$ between $1$ and $2$, or
	above $2$, the CND property fails. They also raise a practical one, namely
	how the kernels $k_{\SW_1}$ and $k_{\RW_1}$ compare with other kernels, and
	with $k_{\SW_2}$ and $k_{\RW_2}$, on real classification data.
	In this section we test the results of the paper numerically by means of the three
	experiments summarized in Figure~\ref{fig:resumen_experimentos}.
	Subsections~\ref{subsec:exp_equivalencia} and~\ref{subsec:exp_csdn} use synthetic diagrams to check the metric equivalence and the CND property, while
	Subsection~\ref{subsec:orbitas} applies the kernels \(k_{\SW_1}\) and \(k_{\RW_1}\) to a
	classification task. The distances \(d_p\) and \(d_p^{r}\), which are used only in
	Subsection~\ref{subsec:exp_equivalencia}, are computed exactly by solving an
	assignment problem on the augmented sets of Definitions~\ref{def:biyeccion_parcial}
	and~\ref{def:dp_r}. In all three experiments, the integrals over \(\theta\) that define
	\(\SW_p\) and \(\RW_p\) are approximated by the midpoint rule with \(M\) subintervals. The Python code that reproduces all
	tables and figures is provided in the supplementary material.

	We build each random diagram from a birth interval \([0,B]\) and two persistence
	intervals, one of small values and one of large values. These intervals, together
	with the number of points in each diagram, are given in
	Figure~\ref{fig:resumen_experimentos}(b). Each point \((b,d)\) is drawn independently
	of the others, as follows.
	\begin{enumerate}
		\item The birth \(b\) is drawn from the uniform distribution on \([0,B]\).
		\item With probability \(1/2\), the point has small persistence; otherwise, it has
		large persistence.
		\item The persistence \(d-b\) is drawn from the uniform distribution on the
		corresponding interval.
	\end{enumerate}
	Thus, two points of the same diagram generally have different persistences. Points
	of small persistence lie close to the diagonal, and points of large persistence lie far
	from it (Figure~\ref{fig:resumen_experimentos}(a)). We use both kinds of points
	because, in all our distances, a point can be matched with a point of the other
	diagram or with the diagonal (or with its reflection, in the case of \(d_p^r\) and
	\(\RW_p\)). Matching a point with the diagonal or with its reflection is cheap if its
	persistence is small and expensive if it is large. In this way, pairs of random
	diagrams involve both kinds of matching. Moreover, in
	Subsection~\ref{subsec:exp_csdn} the two intervals are on the scale of
	Theorem~\ref{thm:sw2_no_CND}, whose diagrams have persistences \(1\) and \(1000\).
	
	\begin{figure}[!htb]
		\centering
		\begin{tikzpicture}[x=0.9cm,y=0.62cm]
			\fill[green!18] (0,0.15)--(4,4.15)--(4,4.8)--(0,0.8)--cycle;
			\fill[orange!22] (0,0.8)--(4,4.8)--(4,6.2)--(0,2.2)--cycle;
			\draw[dashed, gray] (0,0)--(4.4,4.4); \node[font=\scriptsize, gray, rotate=34] at (2.9,2.35) {\(d=b\)};
			\draw[-{Stealth[length=1.8mm]}] (0,0)--(4.6,0) node[right, font=\footnotesize] {\(b\)};
			\draw[-{Stealth[length=1.8mm]}] (0,0)--(0,6.5) node[above, font=\footnotesize] {\(d\)};
			\draw[line width=1.2pt, blue!70!black] (0,-0.12)--(4,-0.12);
			\node[font=\scriptsize, blue!70!black, below] at (2,-0.12) {births uniform in \([0,B]\)};
			\foreach \b/\w in {0.4/0.4,1.3/0.6,2.1/0.3,3.3/0.5}
			\fill[green!50!black] (\b,\b+\w) circle (1.6pt);
			\foreach \b/\w in {0.6/1.6,1.7/1.1,2.6/1.8,3.5/1.3}
			\fill[orange!80!black] (\b,\b+\w) circle (1.6pt);
			\draw[{Stealth[length=1.2mm]}-{Stealth[length=1.2mm]}] (4.25,4.15)--(4.25,4.8);
			\node[font=\scriptsize, anchor=west, green!40!black] at (4.35,4.3) {small persistence};
			\draw[{Stealth[length=1.2mm]}-{Stealth[length=1.2mm]}] (4.25,4.8)--(4.25,6.2);
			\node[font=\scriptsize, anchor=west, orange!80!black] at (4.35,5.6) {large persistence};
			\node[font=\footnotesize] at (2.3,-1.35) {(a) a random diagram};
			\node[anchor=north west, font=\scriptsize] (tab) at (7.2,5.4) {
				\begin{tabular}{lcc}
					\toprule
					& Subsection~\ref{subsec:exp_equivalencia}
					& Subsection~\ref{subsec:exp_csdn} \\
					\midrule
					births & \([0,10]\) & \([0,1000]\) \\
					\textcolor{green!45!black}{small persistence} & \([0.05,0.5]\) & \([0.5,2]\) \\
					\textcolor{orange!80!black}{large persistence} & \([0.5,5]\) & \([200,1000]\) \\
					points per diagram & \(1\)--\(3\) & \(1\)--\(3\) \\
					\bottomrule
			\end{tabular}};
			\node[font=\footnotesize, below=3mm of tab] {(b) parameters of the random diagrams};
		\end{tikzpicture}
		
		\vspace{4mm}
		\begin{tikzpicture}[
			card/.style={draw, rounded corners=3pt, align=left, font=\scriptsize,
				text width=2.62cm, minimum height=3.35cm, inner sep=4pt, line width=0.6pt,
				anchor=north west}]
			\node[card, fill=green!8] (e2) at (0.00,0) {
				{\footnotesize\textbf{\ref{subsec:exp_equivalencia}\ Metric}}\\{\footnotesize\textbf{equivalence}}\\[2pt]
				\(500\) pairs in \(\mathcal{D}_{\le3}\)\\\(p\in\{1,1.5,2,3\}\)\\\(\SW_p/d_p\), \(\RW_p/d_p^r\)\\[2pt]\(M=2\,000\)};
			\node[card, fill=red!8] (e3) at (3.18,0) {
				{\footnotesize\textbf{\ref{subsec:exp_csdn}\ CND}}\\{\footnotesize\textbf{property}}\\[2pt]
				\(150\) sets of \(6\)\\search over \(4\)\\one-point diagrams\\\(p\in[1,4]\)\\[2pt]\(M=1\,200\); \(300\) and\\\(30\,000\) in the search};
			\node[card, fill=orange!10] (e4) at (6.36,0) {
				{\footnotesize\textbf{\ref{subsec:orbitas}\ Orbit}}\\{\footnotesize\textbf{benchmark}}\\[2pt]
				\(500\) point clouds\\Vietoris--Rips\\diagrams\\six kernels, SVM\\[2pt]\(M=200\)};
			\node[font=\footnotesize] at (4.63,-3.85) {(c) the three experiments and the number \(M\) of subintervals};
		\end{tikzpicture}
		\caption{\footnotesize Overview of the numerical experiments. (a) A random
			diagram. Each point has a uniformly distributed birth \(b\) and, with
			probability \(1/2\) each, a small or a large persistence \(d-b\), drawn
			uniformly from the corresponding interval. (b) The intervals used; the scale
			in Subsection~\ref{subsec:exp_csdn} is that of Theorem~\ref{thm:sw2_no_CND}.
			(c) The quantity computed in each experiment, and the number \(M\) of
			subintervals in the midpoint rule.}
		\label{fig:resumen_experimentos}
	\end{figure}
	
	\subsection{Metric equivalence}
	\label{subsec:exp_equivalencia}
	
\begin{figure}[!htb]
		\centering
		\definecolor{kSW}{HTML}{DD8452}\definecolor{kRW}{HTML}{55A868}
		\begin{tikzpicture}[x=5cm, y=0.92cm,
			band/.style={fill=#1!20},
			obs/.style={fill=#1},
			mn/.style={circle, fill=white, draw=#1, line width=0.6pt, inner sep=0pt, minimum size=4pt}]
			\foreach \v in {0,0.25,...,1.75} \draw[gray!18] (\v,0.45)--(\v,-4.25);
			\draw[line width=0.5pt] (0,-4.25)--(1.85,-4.25);
			\foreach \v in {0,0.5,1,1.5} {
				\draw (\v,-4.25)--(\v,-4.32) node[below, font=\scriptsize] {\(\v\)};}
			\node[font=\footnotesize] at (0.92,-5.0) {ratio};
			\foreach \p/\C/\K/\Kr/\sl/\sh/\sm/\rl/\rh/\rmn/\y in {
				1/0.0205/1.801/0.900/0.642/1.084/0.826/0.469/0.900/0.742/0,
				1.5/0.0218/1.519/0.957/0.550/0.951/0.834/0.480/0.946/0.758/-1.1,
				2/0.0228/1.414/1.000/0.546/0.994/0.855/0.490/0.988/0.779/-2.2,
				3/0.0242/1.339/1.063/0.559/1.057/0.899/0.509/1.050/0.820/-3.3} {
				\node[font=\footnotesize, anchor=east] at (-0.33,\y-0.2) {\(p=\p\)};
				\node[font=\scriptsize, anchor=east, kSW!70!black] at (-0.015,\y) {\(\SW_p/d_p\)};
				\node[font=\scriptsize, anchor=east, kRW!60!black] at (-0.015,\y-0.4) {\(\RW_p/d_p^r\)};
				\fill[band=kSW] (\C,\y-0.09) rectangle (\K,\y+0.09);
				\fill[obs=kSW] (\sl,\y-0.09) rectangle (\sh,\y+0.09);
				\draw[kSW!70!black, line width=1pt] (\K,\y-0.15)--(\K,\y+0.15);
				\node[mn=kSW!70!black] at (\sm,\y) {};
				\node[font=\tiny, anchor=west, kSW!70!black] at (\K+0.008,\y) {\(\K\)};
				\fill[band=kRW] (\C,\y-0.49) rectangle (\Kr,\y-0.31);
				\fill[obs=kRW] (\rl,\y-0.49) rectangle (\rh,\y-0.31);
				\draw[kRW!60!black, line width=1pt] (\Kr,\y-0.55)--(\Kr,\y-0.25);
				\node[mn=kRW!60!black] at (\rmn,\y-0.4) {};
				\node[font=\tiny, anchor=west, kRW!60!black] at (\Kr+0.008,\y-0.4) {\(\Kr\)};
				\draw[gray!60!black, line width=1pt] (\C,\y-0.55)--(\C,\y+0.15);
			}
			\node[font=\scriptsize, gray!60!black, anchor=south] at (0.0205,0.15) {\(C_{p,N}\approx0.02\)};
			\node[font=\scriptsize, kSW!70!black, anchor=south] at (1.801,0.15) {\(\kappa_p\)};
			\node[font=\scriptsize, kRW!60!black, anchor=south] at (0.900,0.15) {\(\kappa_p^r\)};
			\begin{scope}[shift={(0,1.05)}]
				\fill[gray!22] (0.12,-0.09) rectangle (0.24,0.09);
				\node[font=\scriptsize, anchor=west] at (0.25,0) {band of the theorem};
				\fill[gray!65] (0.85,-0.09) rectangle (0.97,0.09);
				\node[font=\scriptsize, anchor=west] at (0.98,0) {observed, min to max};
				\node[mn=black] at (1.66,0) {};
				\node[font=\scriptsize, anchor=west] at (1.675,0) {mean};
			\end{scope}
		\end{tikzpicture}
		\caption{\footnotesize Observed ratios \(\SW_p/d_p\) (orange) and
			\(\RW_p/d_p^{r}\) (green) over the \(500\) pairs in \(\mathcal{D}_{\le3}\)
			(dark bars, from the smallest to the largest ratio, with the mean marked),
			shown inside the bands of Theorems~\ref{thm:equiv_p}
			and~\ref{thm:equiv_p_rw} (light bars). Each band extends from \(C_{p,N}\)
			(gray mark) to \(\kappa_p\) or \(\kappa_p^{r}\) (colored mark, labeled with
			its value). The dark green bars come close to \(\kappa_p^{r}\), whereas all bars stay
			far from \(C_{p,N}\).}
		\label{fig:equivalencia}
	\end{figure}
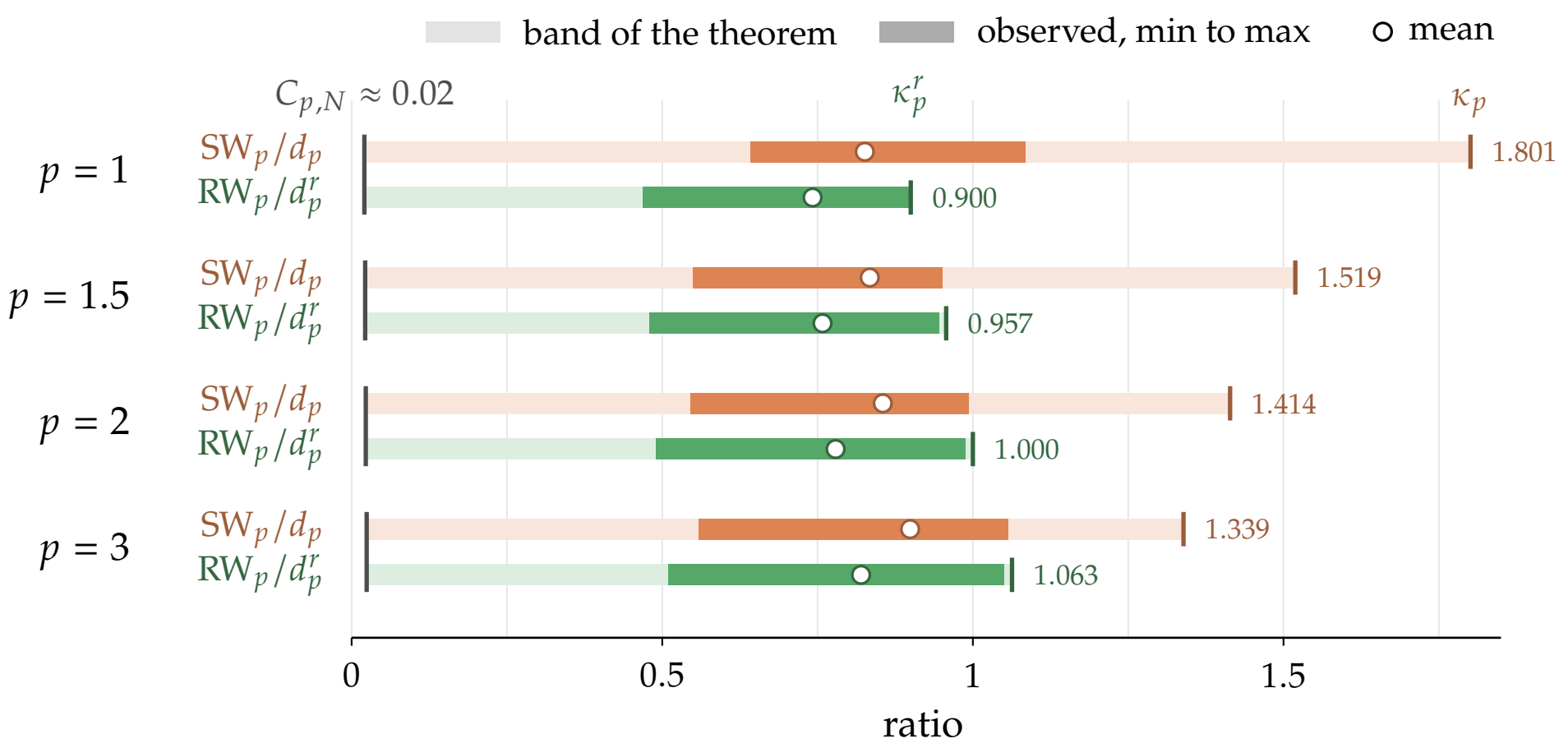

	We fix \(N=3\) and, for \(p\in\{1,1.5,2,3\}\), compare the ratios \(\SW_p/d_p\) and
	\(\RW_p/d_p^r\) on \(500\) random pairs of diagrams in \(\mathcal{D}_{\le3}\) with
	the bands \([C_{p,N},\kappa_p]\) and \([C_{p,N},\kappa_p^{r}]\) of
	Theorems~\ref{thm:equiv_p} and~\ref{thm:equiv_p_rw}, respectively.
	Table~\ref{tab:exp_equivalencia} lists the constants, and
	Figure~\ref{fig:equivalencia} shows where the ratios fall. All ratios lie inside
	their bands, and two facts stand out. First, the upper constant \(\kappa_p^{r}\) is
	almost attained: for \(p=1\), the largest ratio \(\RW_1/d_1^{r}\) is \(0.8995\),
	compared with \(\kappa_1^{r}=0.9003\), in agreement with the sharpness of this
	constant. Second, by contrast, the ratios never drop below \(0.46\), which is far
	above \(C_{p,N}\approx0.02\); so the lower bound is valid but does not seem to be
	sharp (Remark~\ref{rem:constante_inferior}).
	
	\begin{table}[!htb]
		\centering
		\footnotesize
		\begin{tabular}{cccccc}
			\toprule
			\(p\) & \(C_{p,N}\) & \(\kappa_p\) & \(\kappa_p^{r}\) & range of \(\SW_p/d_p\) & range of \(\RW_p/d_p^{r}\) \\
			\midrule
			\(1\)   & \(0.0205\) & \(1.801\) & \(0.900\) & \([0.642,\,1.084]\) & \([0.469,\,0.900]\) \\
			\(1.5\) & \(0.0218\) & \(1.519\) & \(0.957\) & \([0.550,\,0.951]\) & \([0.480,\,0.946]\) \\
			\(2\)   & \(0.0228\) & \(1.414\) & \(1.000\) & \([0.546,\,0.994]\) & \([0.490,\,0.988]\) \\
			\(3\)   & \(0.0242\) & \(1.339\) & \(1.063\) & \([0.559,\,1.057]\) & \([0.509,\,1.050]\) \\
			\bottomrule
		\end{tabular}
		\caption{\footnotesize Constants of Theorems~\ref{thm:equiv_p}
			and~\ref{thm:equiv_p_rw} for \(N=3\), together with the smallest and largest
			ratios observed over the \(500\) pairs. Every ratio lies inside its band.}
		\label{tab:exp_equivalencia}
	\end{table}
	
	\subsection{Numerical failure of the CND property}
	\label{subsec:exp_csdn}
	
	In this subsection, we look for violations of the CND property of \(\SW_p\)
	and \(\RW_p\) on two kinds of collections of diagrams. The values of \(p\)
	are chosen in light of the theory. For \(p=1\), both distances are CND
	(Lemma~\ref{lem:sw_nd}), so no violation should appear; this case thus serves as a
	check of the test. For \(p=2\), however, they are not CND
	(Theorem~\ref{thm:sw2_no_CND}), and for the other values of \(p>1\) the question
	remains open. The two kinds of collections are as follows.
	
	\begin{itemize}
		\item The first kind consists of \emph{random collections}, which show how often
		a violation appears, for \(p\in\{1,1.5,2,3\}\), on diagrams that were not built
		for that purpose.
		
		\item The second kind is produced by an \emph{adversarial search}, which looks,
		on a fine grid of \(p\in[1,4]\), for the collection of \(4\) single-point
		diagrams that most strongly violates the CND property, in order to locate the
		smallest \(p\) at which the property fails.
	\end{itemize}
	
	Both rely on the following matrix test.
	
	\begin{remark}[\textbf{Matrix reformulation of the CND condition}]
		\label{rem:cnd_centrado}
		Let \(H_n\coloneqq I_n-\tfrac1n\mathbf{1}\mathbf{1}^\top\) be the orthogonal
		projection onto \(V_n=\{a\in\R^n:\sum_i a_i=0\}\). Since
		\(v^\top H_nMH_nv=(H_nv)^\top M(H_nv)\) and \(H_nv\) ranges over all of \(V_n\), a
		symmetric function \(f\) with \(f(x,x)=0\) is CND if and only if, for every
		\(n\ge1\) and all \(x_1,\dots,x_n\in X\), the matrix \(H_nMH_n\) with
		\(M_{ij}=f(x_i,x_j)\) is negative semidefinite.
	\end{remark}
	
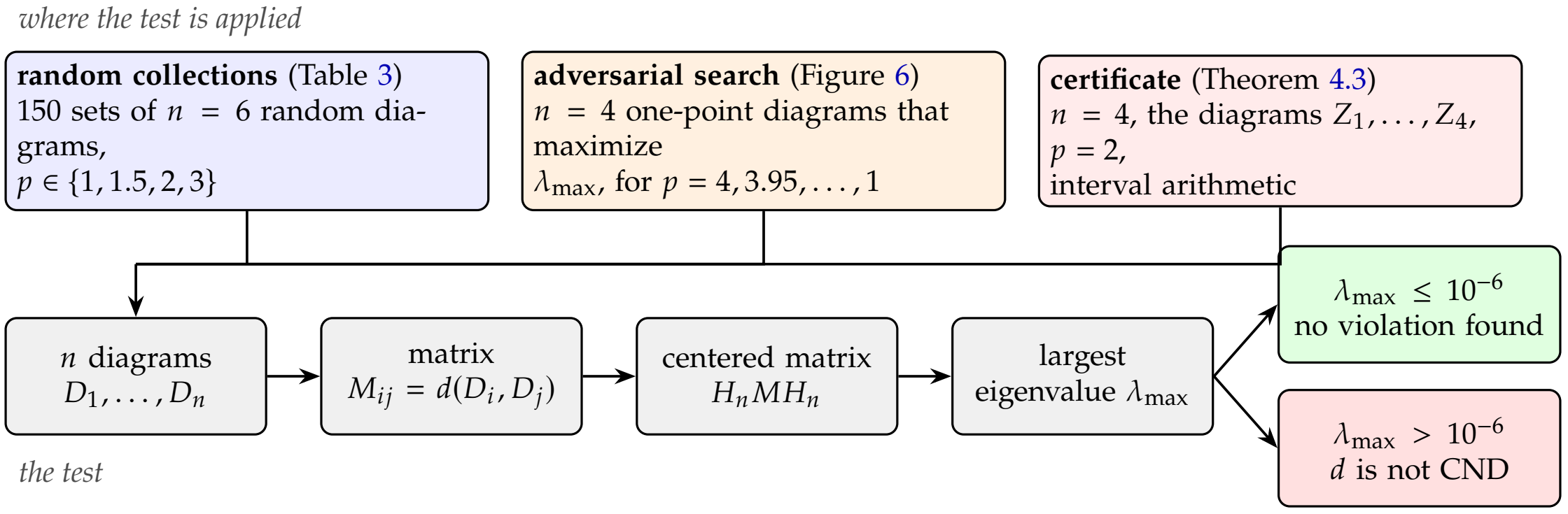
\begin{figure}[!htb]
		\centering
		\begin{tikzpicture}[
			box/.style={draw, rounded corners=3pt, align=center, font=\scriptsize,
				minimum height=1.1cm, inner sep=3pt, line width=0.6pt, text width=2.25cm},
			src/.style={draw, rounded corners=3pt, align=left, font=\scriptsize,
				inner sep=3pt, line width=0.6pt, text width=4.35cm, minimum height=1.25cm},
			arr/.style={-{Stealth[length=2mm]}, line width=0.7pt}]
			\node[src, fill=blue!8] (s1) at (0,0)
			{\textbf{random collections} (Table~\ref{tab:exp_csdn})\\
				\(150\) sets of \(n=6\) random diagrams,\\\(p\in\{1,1.5,2,3\}\)};
			\node[src, fill=orange!12, right=0.3cm of s1] (s2)
			{\textbf{adversarial search} (Figure~\ref{fig:csdn_vs_p})\\
				\(n=4\) one-point diagrams that maximize\\\(\lambda_{\max}\), for \(p=4,3.95,\dots,1\)};
			\node[src, fill=red!8, right=0.3cm of s2] (s3)
			{\textbf{certificate} (Theorem~\ref{thm:sw2_no_CND})\\
				\(n=4\), the diagrams \(Z_1,\dots,Z_4\), \(p=2\),\\interval arithmetic};
			\node[box, fill=gray!12, below=1.0cm of s1.south west, anchor=north west] (D) {\(n\) diagrams\\\(D_1,\dots,D_n\)};
			\node[box, fill=gray!12, right=0.5cm of D] (M) {matrix\\\(M_{ij}=d(D_i,D_j)\)};
			\node[box, fill=gray!12, right=0.5cm of M] (H) {centered matrix\\\(H_nMH_n\)};
			\node[box, fill=gray!12, right=0.5cm of H] (L) {largest\\eigenvalue \(\lambda_{\max}\)};
			\node[box, fill=green!12, text width=2.45cm, right=0.6cm of L, yshift=0.68cm] (ok)
			{\(\lambda_{\max}\le10^{-6}\)\\no violation found};
			\node[box, fill=red!12, text width=2.45cm, right=0.6cm of L, yshift=-0.68cm] (no)
			{\(\lambda_{\max}>10^{-6}\)\\\(d\) is not CND};
			\draw[arr] (D)--(M); \draw[arr] (M)--(H); \draw[arr] (H)--(L);
			\draw[arr] (L.east)--(ok.west); \draw[arr] (L.east)--(no.west);
			\coordinate (j) at ($(D.north)+(0,0.5)$);
			\draw[line width=0.7pt] (s1.south |- j) -- (s1.south);
			\draw[line width=0.7pt] (s2.south) |- (j);
			\draw[line width=0.7pt] (s3.south) |- (j);
			\draw[arr] (j) -- (D.north);
			\node[font=\scriptsize\itshape, gray!60!black, anchor=south west] at (s1.north west) {where the test is applied};
			\node[font=\scriptsize\itshape, gray!60!black, anchor=north west] at ([yshift=-1mm]D.south west) {the test};
		\end{tikzpicture}
		\caption{\footnotesize The CND test of Remark~\ref{rem:cnd_centrado} as an
			algorithm (bottom row) and the three cases in which we apply it (top
			row). A positive \(\lambda_{\max}\) yields a zero-sum vector \(a\) with
			\(a^\top Ma>0\). For the certificate, the computation is carried out in
			interval arithmetic, so that neither rounding errors nor the quadrature in
			\(\theta\) can produce a spurious violation.}
		\label{fig:test_cnd}
	\end{figure}

	Since \(H_n\mathbf{1}=0\), the largest eigenvalue \(\lambda_{\max}\) of
	\(H_nMH_n\) is \(0\) when \(f\) is CND on \(x_1,\dots,x_n\), and positive
	otherwise. We therefore compute \(\lambda_{\max}\) for many collections of
	diagrams and record a violation when \(\lambda_{\max}>10^{-6}\), a threshold above
	the level of rounding errors. Figure~\ref{fig:test_cnd} shows this test as an
	algorithm (bottom row) together with the three cases in which we apply it (top
	row), namely the two kinds of collections and the certificate of
	Theorem~\ref{thm:sw2_no_CND}.
	
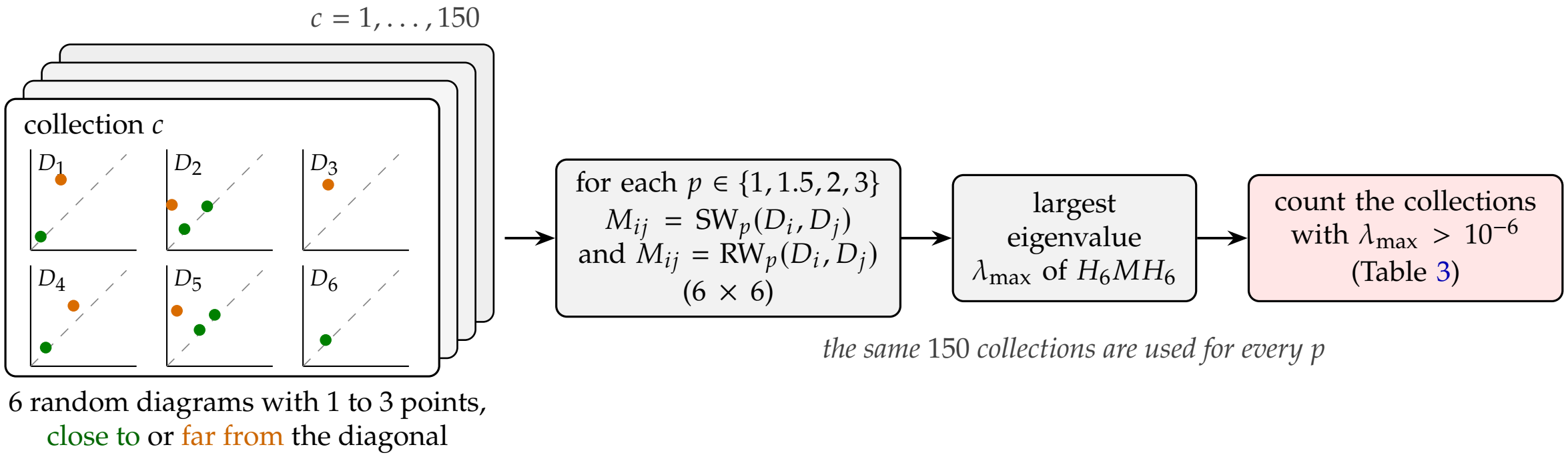
\begin{figure}[!htb]
		\centering
		\begin{tikzpicture}[
			x=1cm,y=1cm,
			box/.style={draw, rounded corners=3pt, align=center, font=\scriptsize,
				minimum height=1.25cm, inner sep=3pt, line width=0.6pt},
			arr/.style={-{Stealth[length=2mm]}, line width=0.7pt},
			near/.style={circle, fill=green!50!black, inner sep=0pt, minimum size=3.3pt},
			far/.style={circle, fill=orange!85!black, inner sep=0pt, minimum size=3.3pt}]
			\foreach \k in {3,2,1}
			\draw[rounded corners=3pt, fill=gray!\the\numexpr 4+3*\k\relax, line width=0.5pt]
			(0.18*\k,0.18*\k) rectangle ++(4.3,2.75);
			\draw[rounded corners=3pt, fill=white, line width=0.6pt] (0,0) rectangle (4.3,2.75);
			\node[font=\scriptsize, anchor=north west] at (0.05,2.75) {collection \(c\)};
			\node[font=\scriptsize, anchor=south east, gray!50!black] at (4.84,3.29) {\(c=1,\dots,150\)};
			\foreach \i/\x/\y/\pts in {
				1/0.25/1.25/{n/0.2/0.27,f/0.6/1.4},
				2/1.6/1.25/{n/0.8/0.87,f/0.1/0.9,n/0.35/0.42},
				3/2.95/1.25/{f/0.5/1.3},
				4/0.25/0.1/{n/0.3/0.37,f/0.85/1.2},
				5/1.6/0.1/{n/0.65/0.72,f/0.2/1.1,n/0.95/1.02},
				6/2.95/0.1/{n/0.45/0.52}}{
				\begin{scope}[shift={(\x,\y)}, x=0.5cm, y=0.5cm]
					\draw[line width=0.4pt] (0,0)--(2.1,0); \draw[line width=0.4pt] (0,0)--(0,2.0);
					\draw[dashed, gray, line width=0.3pt] (0,0)--(1.9,1.9);
					\node[font=\tiny, anchor=north west, inner sep=1pt] at (0.05,2.0) {\(D_{\i}\)};
					\foreach \t/\b/\d in \pts {
						\if n\t \node[near] at (\b,\d) {}; \else \node[far] at (\b,\d) {}; \fi}
			\end{scope}}
			\node[box, fill=gray!10, text width=3.2cm, anchor=west] (A1) at (5.45,1.37)
			{for each \(p\in\{1,1.5,2,3\}\)\\[1pt]
				\(M_{ij}=\SW_p(D_i,D_j)\)\\ and \(M_{ij}=\RW_p(D_i,D_j)\)\\[1pt] (\(6\times6\))};
			\node[box, fill=gray!10, text width=2.2cm, right=0.5cm of A1] (A2)
			{largest eigenvalue\\\(\lambda_{\max}\) of \(H_6MH_6\)};
			\node[box, fill=red!10, text width=2.9cm, right=0.5cm of A2] (A3)
			{count the collections\\with \(\lambda_{\max}>10^{-6}\)\\[1pt]
				(Table~\ref{tab:exp_csdn})};
			\draw[arr] (4.95,1.37)--(A1.west);
			\draw[arr] (A1)--(A2); \draw[arr] (A2)--(A3);
			\node[font=\scriptsize, align=center] at (2.4,-0.45)
			{\(6\) random diagrams with \(1\) to \(3\) points,\\
				\textcolor{green!40!black}{close to} or \textcolor{orange!80!black}{far from} the diagonal};
			\node[font=\scriptsize\itshape, gray!50!black, align=center] at (10.6,0.25)
			{the same \(150\) collections are used for every \(p\)};
			
		\end{tikzpicture}
		\caption{\footnotesize How the random collections are built. We draw
			\(150\) collections of \(6\) random diagrams with the parameters of
			Figure~\ref{fig:resumen_experimentos}(b) and apply the test of
			Figure~\ref{fig:test_cnd} to each of them, for each \(p\) and for both
			distances.}
		\label{fig:cnd_aleatorias}
	\end{figure}

	\paragraph{Random collections.}
	We build \(150\) collections of \(6\) random diagrams
	(Figure~\ref{fig:resumen_experimentos}) and, for each \(p\in\{1,1.5,2,3\}\),
	compute \(\lambda_{\max}\) for \(\SW_p\) and for \(\RW_p\)
	(Figure~\ref{fig:cnd_aleatorias}). Table~\ref{tab:exp_csdn} shows the results.
	For \(p=1\), the eigenvalue never exceeds \(10^{-12}\), in agreement with
	Lemma~\ref{lem:sw_nd}. For \(p>1\), by contrast, violations do occur; they become
	more numerous as \(p\) grows and are much more frequent for \(\RW_p\). For
	\(p=2\), these violations are numerical evidence whose rigorous counterpart is
	Theorem~\ref{thm:sw2_no_CND}. For \(p=1.5\), however, the distance \(\SW_p\) fails
	on just one collection; therefore, random diagrams say little about values of
	\(p\) close to \(1\).
	
	\begin{table}[!htb]
		\centering
		\footnotesize
		\begin{tabular}{ccccc}
			\toprule
			& \multicolumn{2}{c}{\(\SW_p\)} & \multicolumn{2}{c}{\(\RW_p\)} \\
			\cmidrule(lr){2-3}\cmidrule(lr){4-5}
			\(p\) & violations & largest \(\lambda_{\max}\) & violations & largest \(\lambda_{\max}\) \\
			\midrule
			\(1\)   & \(0\)  & \(<10^{-12}\) & \(0\)  & \(<10^{-12}\) \\
			\(1.5\) & \(1\)  & \(2.66\)  & \(43\) & \(13.50\) \\
			\(2\)   & \(17\) & \(8.32\)  & \(65\) & \(33.02\) \\
			\(3\)   & \(25\) & \(16.07\) & \(74\) & \(57.96\) \\
			\bottomrule
		\end{tabular}
		\caption{\footnotesize Number of collections, out of \(150\), with
			\(\lambda_{\max}>10^{-6}\), and the largest \(\lambda_{\max}\) found. The
			entries of the matrices \(M\) reach several hundred, since the
			persistences go up to \(1000\).}
		\label{tab:exp_csdn}
	\end{table}
	
\begin{figure}[!htb]
		\centering
		\begin{tikzpicture}[
			box/.style={draw, rounded corners=3pt, align=center, font=\scriptsize,
				minimum height=1.25cm, inner sep=3pt, line width=0.6pt},
			st/.style={draw, rounded corners=2pt, align=center, font=\scriptsize,
				inner sep=2pt, line width=0.5pt, text width=2.9cm, minimum height=0.62cm},
			arr/.style={-{Stealth[length=2mm]}, line width=0.7pt}]
			\node[st, fill=red!8, anchor=west] (s1) at (0,2.0) {(1) \(\{Z_1,\dots,Z_4\}\) of\\Theorem~\ref{thm:sw2_no_CND}};
			\node[st, fill=blue!8, below=0.18cm of s1] (s2) {(2) \(4\) random\\collections};
			\node[st, fill=green!12, below=0.18cm of s2] (s3) {(3) the best one\\for \(p+0.05\)\\(none if \(p=4\))};
			\node[draw, dashed, rounded corners=3pt, gray, fit=(s1)(s3), inner sep=3pt] (starts) {};
			\node[font=\scriptsize, align=center, above=0pt of starts] {initial collections, \(6\) in total\\(\(5\) if \(p=4\)), each with\\\(4\) one-point diagrams};
			\node[box, fill=gray!10, text width=3.0cm, right=0.5cm of starts] (B1)
			{Nelder--Mead from\\each initial collection\\maximizes \(\lambda_{\max}\)\\of \(H_4MH_4\)};
			\node[box, fill=gray!10, text width=3.0cm, right=0.45cm of B1] (B2)
			{choose, among the\\results, the collection\\with the largest \(\lambda_{\max}\)};
			\node[box, fill=orange!12, text width=3.3cm, right=0.45cm of B2] (B3)
			{output for \(p\)\\[2pt] \(\lambda_{\max}(p)\), a point of\\Figure~\ref{fig:csdn_vs_p}, and the\\collection attaining it};
			\draw[arr] (starts)--(B1); \draw[arr] (B1)--(B2); \draw[arr] (B2)--(B3);
			\draw[arr, green!40!black] (B3.south) -- ++(0,-1.05) -| node[pos=0.25, below, font=\scriptsize, text=green!30!black]
			{this collection is the initial collection (3) for \(p-0.05\)} (s3.south);
			\node[font=\scriptsize\itshape, gray!50!black, anchor=south] at ($(B1.north)!0.5!(B2.north)+(0,0.25)$)
			{\(p=4,\,3.95,\,\dots,\,1\), for \(\SW_p\) and for \(\RW_p\)};
		\end{tikzpicture}
		\caption{\footnotesize How the adversarial search works. A collection of \(4\)
			single-point diagrams in the search region is determined by \(8\) numbers.
			For each \(p\), the Nelder--Mead method adjusts these numbers to make
			\(\lambda_{\max}\) as large as possible, starting from each initial
			collection. The values of \(p\) are visited from \(4\) down to \(1\), so that
			the best collection for one value of \(p\) is one of the initial collections
			for the next.}
		\label{fig:cnd_busqueda}
	\end{figure}
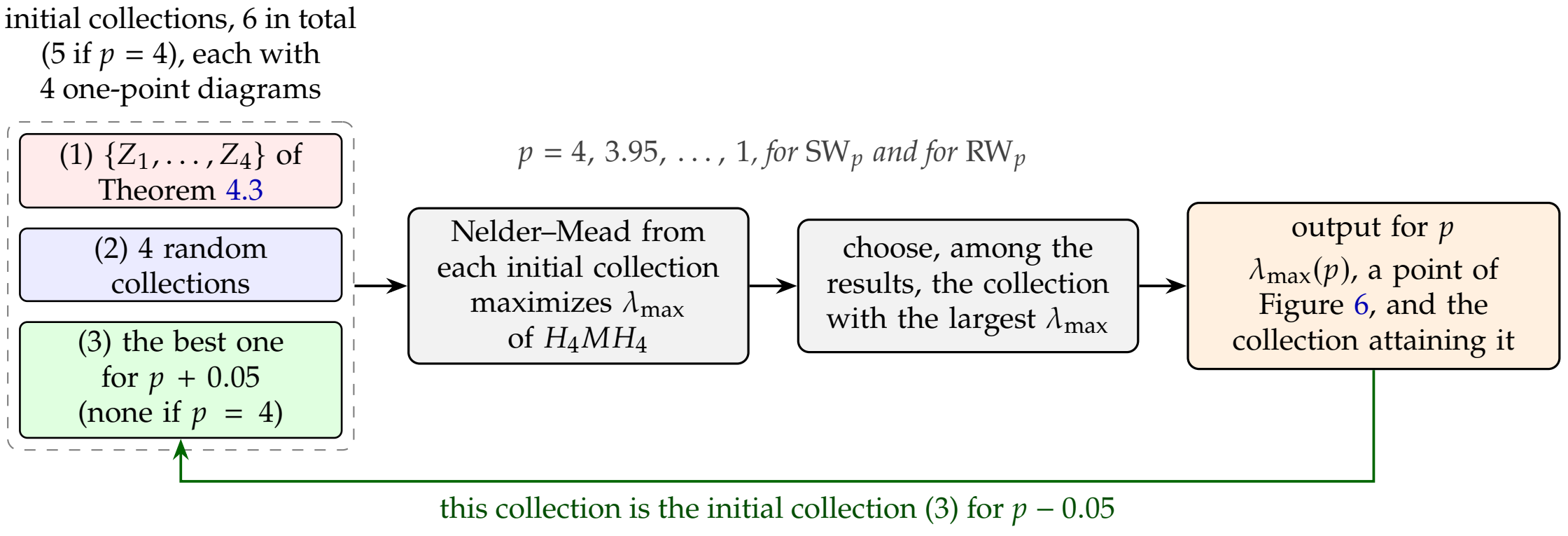

\begin{figure}[!htb]
		\centering
		\includegraphics[width=0.8\textwidth]{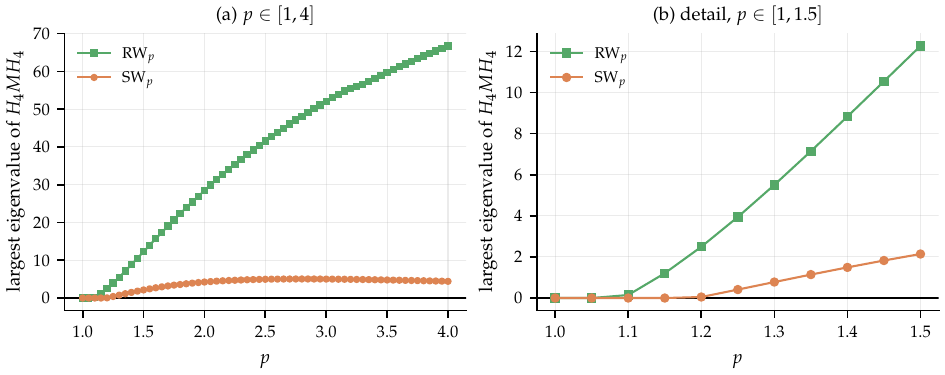}
		\caption{\footnotesize The largest eigenvalue \(\lambda_{\max}(p)\) found by the
			adversarial search for \(\SW_p\) and \(\RW_p\), for
			\(p=1.00,1.05,\dots,4.00\). A value greater than \(10^{-6}\) means that the
			search found a collection of \(4\) diagrams on which the CND property fails.
			Panel (a) shows the full range, and panel (b) shows a close-up near the
			threshold.}
		\label{fig:csdn_vs_p}
	\end{figure}

	\paragraph{Adversarial search.}
	To study values of \(p\) close to \(1\), we search for violations directly
	(Figure~\ref{fig:cnd_busqueda}). We work with collections
	\(\{Y_1,Y_2,Y_3,Y_4\}\) of \(4\) single-point diagrams,
	\[
	Y_i=\{(b_i,\,b_i+w_i)\},\qquad b_i\in[0,1000],\quad w_i\in[0.5,1000],
	\]
	so that each collection is determined by the \(8\) numbers \(b_1,\dots,b_4\) and
	\(w_1,\dots,w_4\). For each \(p\) on the grid \(4,3.95,\dots,1\) and for each of the
	distances \(\SW_p\) and \(\RW_p\), we use the Nelder--Mead method to maximize the
	value \(\lambda_{\max}\) of the collection, viewed as a function of these \(8\)
	numbers. Since this method finds only local maxima, we run it from each of the
	following initial collections.
	\begin{enumerate}
		\item The collection \(\{Z_1,Z_2,Z_3,Z_4\}\) of Theorem~\ref{thm:sw2_no_CND}.
		\item Four collections chosen at random from the search region.
		\item The best collection found for the previous value of \(p\), that is, for
		\(p+0.05\). For \(p=4\), the first value on the grid, this collection does
		not exist.
	\end{enumerate}
	Thus, for each \(p\) we start from \(1+4+1=6\) collections, except for \(p=4\),
	where we start from \(1+4=5\), and among the collections returned by the
	Nelder--Mead method we choose the one with the largest \(\lambda_{\max}\). For each
	\(p\), this procedure yields two outputs.
	\begin{itemize}
		\item The number \(\lambda_{\max}(p)\), which is the largest eigenvalue found for
		that \(p\). Since the search may miss the global maximum, the number
		\(\lambda_{\max}(p)\) is only a lower bound for the supremum of
		\(\lambda_{\max}\) over all collections of \(4\) single-point diagrams in the
		search region. Therefore, if \(\lambda_{\max}(p)>10^{-6}\), then the collection
		found yields a violation, and the distance is not CND for that \(p\), up to
		numerical error. If, instead, \(\lambda_{\max}(p)\le10^{-6}\), then no violation
		has been found, but this does not prove that the distance is CND, since
		violations could occur on other collections.
		\item The collection that attains \(\lambda_{\max}(p)\), which serves as initial
		collection~(3) for the next value of \(p\), that is, for \(p-0.05\).
	\end{itemize}
	Regarding numerical precision, during the search the integrals over \(\theta\) are
	approximated with \(M=300\) subintervals, so that each evaluation of
	\(\lambda_{\max}\) is fast, whereas the value \(\lambda_{\max}(p)\) of the chosen
	collection is computed with \(M=30\,000\) (Figure~\ref{fig:resumen_experimentos}(c)).
	Figure~\ref{fig:csdn_vs_p} shows \(\lambda_{\max}(p)\) as a function of \(p\), and
	Table~\ref{tab:busqueda} summarizes its main values. The results suggest that the
	CND property fails for every \(p\) above a threshold \(p^*\), with
	\(p^*\in(1.05,1.10]\) for \(\RW_p\) and \(p^*\in(1.15,1.20]\) for \(\SW_p\).
	
	\begin{table}[!htb]
		\centering
		\footnotesize
		\begin{tabular}{lcccc}
			\toprule
			& smallest \(p\) with \(\lambda_{\max}(p)>10^{-6}\) & \(\lambda_{\max}(p)\) at that \(p\) & \(\lambda_{\max}(2)\) & maximum of \(\lambda_{\max}(p)\) \\
			\midrule
			\(\SW_p\) & \(1.20\) & \(0.05\) & \(4.23\)  & \(5.05\) at \(p=2.8\) \\
			\(\RW_p\) & \(1.10\) & \(0.15\) & \(28.45\) & \(66.7\) at \(p=4\) \\
			\bottomrule
		\end{tabular}
		\caption{\footnotesize Main values of the adversarial search shown in
			Figure~\ref{fig:csdn_vs_p}.}
		\label{tab:busqueda}
	\end{table}
	
	\subsection{Orbit recognition benchmark}
	\label{subsec:orbitas}
	
	In this subsection we apply \(k_{\SW_1}\) and \(k_{\RW_1}\) to the orbit
	recognition benchmark of \cite{AdamsPersistenceImages2017,Carriere2017}, compare
	them with other kernels, and analyze which part of the results depends on the
	kernel being positive definite. Five of the six kernels we compare have the form
	\(e^{-\delta/(2\sigma^2)}\) for some function \(\delta\) between diagrams, so we
	begin with a result on their Gram matrices, which we use below to explain the
	negative eigenvalues of the kernels that are not positive definite.
	\begin{proposition}[\textbf{Negative eigenvalues and bandwidth}]
		\label{prop:ancho_banda}
		Let \(n\ge2\), let \(A_1,\dots,A_n\) be diagrams, let \(\delta\) be a
		symmetric function with \(\delta\ge0\) and \(\delta(A,A)=0\), and let
		\(D=\bigl(\delta(A_i,A_j)\bigr)_{i,j}\) be the corresponding matrix. Let
		\(K_\sigma=\bigl(e^{-D_{ij}/(2\sigma^2)}\bigr)_{i,j}\), and write
		\(M_D=\max\limits_{i,j}D_{ij}\) and
		\(\ell=\lambda_{\max}(H_nDH_n)\ge0\). Then the following hold.
		\begin{enumerate}
			\item If \(\delta\) is CND on the collection, that is, if \(\ell=0\), then the
			matrix \(K_\sigma\) is positive semidefinite for every \(\sigma>0\).
			\item If \(\ell>0\), then for every \(\sigma>0\)
			\[
			\lambda_{\min}(K_\sigma)\le-\frac{\ell}{2\sigma^2}+\frac{nM_D^2}{8\sigma^4},
			\]
			so the matrix \(K_\sigma\) has a negative eigenvalue whenever
			\(\sigma^2>nM_D^2/(4\ell)\).
			\item If \(\ell>0\) and \(\sigma^2\ge M_D\), then
			\[
			\Bigl|\lambda_{\min}(K_\sigma)+\frac{\ell}{2\sigma^2}\Bigr|\le\frac{5nM_D^2}{8\sigma^4}
			\qquad\text{and}\qquad
			e^{-1/2}\,n\le\lambda_{\max}(K_\sigma)\le n .
			\]
		\end{enumerate}
	\end{proposition}

	Item~(1) states that the Gram matrix is positive semidefinite for every
	\(\sigma\) when \(\delta\) is CND. By contrast, items~(2) and~(3) state that,
	when \(\delta\) is not CND, negative eigenvalues always appear for large
	\(\sigma\), and that they are small, of order \(\ell/(2\sigma^2)\), compared
	with the largest eigenvalue, which is of order \(n\). The proof uses only
	Schoenberg's theorem for a fixed collection of points and elementary linear
	algebra; we give it in Appendix~\ref{app:ancho_banda}.
	
	The experiment has four steps, namely the data, the persistence diagrams, the
	kernels with their parameters, and the classification.
	
\begin{figure}[!htb]
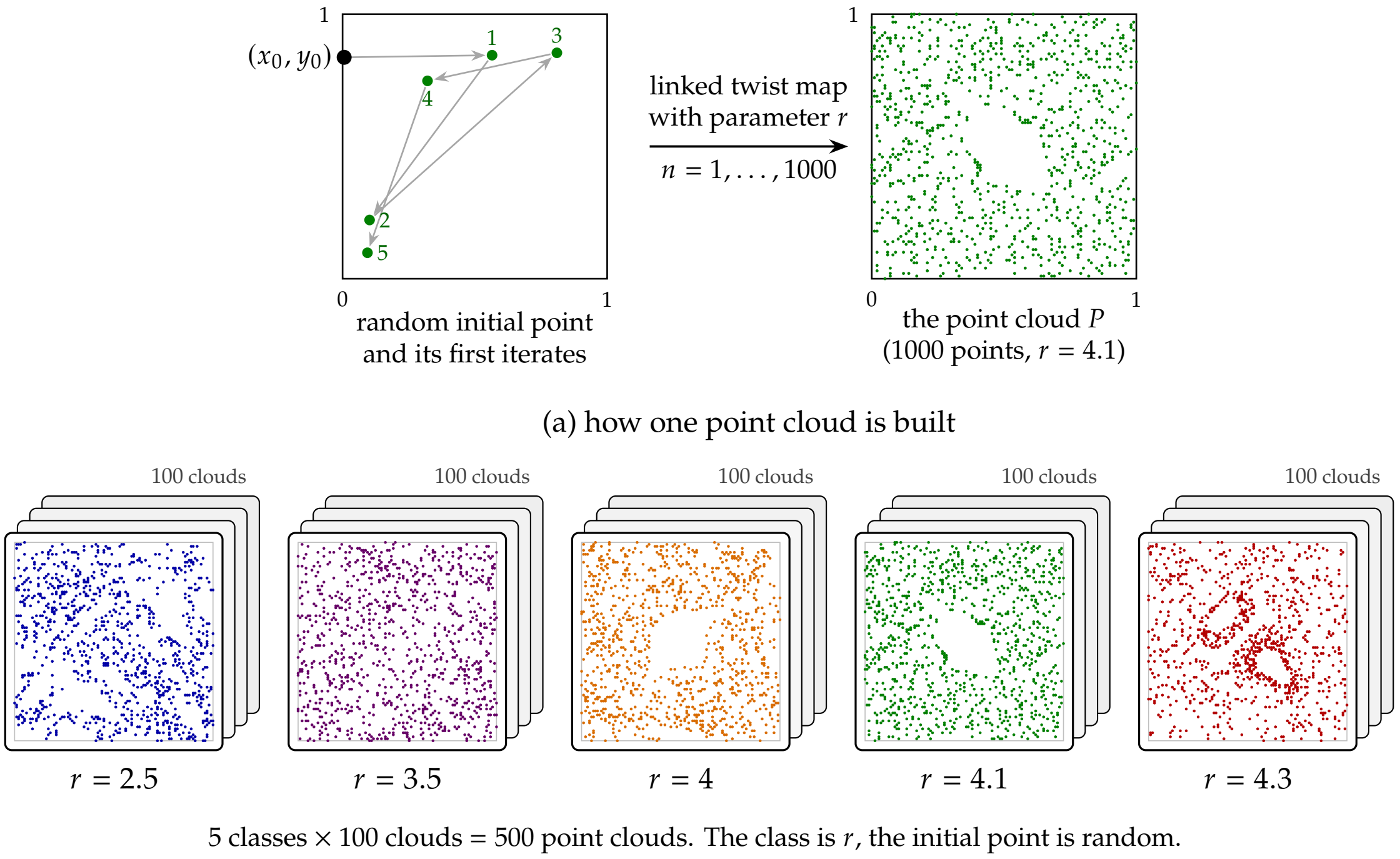

		\centering

		\caption{\footnotesize Data of the orbit benchmark. (a) A random initial
			point \((x_0,y_0)\), its first five iterates under the linked twist map,
			and the cloud formed by the first \(1000\) iterates. (b) For each value of
			\(r\) there are \(100\) clouds, one for each random initial point; the
			front card shows one of them (for \(r=4.1\), the cloud of (a)). Clouds of
			the same class can look different, so the classifier has to recognize
			\(r\) regardless of the initial point.}
		\label{fig:orbitas_datos}
	\end{figure}

	\paragraph{Step 1. Data.}
	For a parameter \(r>0\), the linked twist map acts on the unit square
	\([0,1]^2\) by
	\[
	x_{n+1}=x_n+r\,y_n(1-y_n)\ \ \mathrm{mod}\ 1,\qquad
	y_{n+1}=y_n+r\,x_{n+1}(1-x_{n+1})\ \ \mathrm{mod}\ 1 .
	\]
	For each \(r\in\{2.5,3.5,4,4.1,4.3\}\) we choose \(100\) initial points
	\((x_0,y_0)\) uniformly at random in \([0,1]^2\), and each of them yields the
	cloud \(P\) formed by the first \(1000\) iterates
	(Figure~\ref{fig:orbitas_datos}). The task is to recognize \(r\) from the
	persistence diagrams of \(P\), as in Table~2 of \cite{Carriere2017}.
	
	\paragraph{Step 2. Persistence diagrams.}
	We use the diagrams \(D_0(P)\) and \(D_1(P)\) of the Vietoris--Rips filtration,
	in which a simplex enters at half its diameter. We compute them with Ripser
	\cite{Bauer2021}, which indexes the filtration by the diameter, and we divide its
	output by \(2\). Every diagram \(D_0(P)\) has \(999\) points, all of the form
	\((0,d)\), and the diagrams \(D_1(P)\) have between \(1\) and \(281\) points,
	with median \(244.5\). The empty regions of a cloud produce the points of
	\(D_1(P)\) far from the diagonal.
	
	\paragraph{Step 3. Kernels and their parameters.}
	For \(d\in\{\SW_1,\RW_1,\SW_2,\RW_2\}\) and \(\tau>0\), let
	\(k_d(A,B)=\exp\bigl(-d(A,B)/\tau\bigr)\). For \(d=\SW_1\) and \(d=\RW_1\) these
	are the kernels of Definition~\ref{def:kernel_sw} with \(\tau=2\sigma^2\), and
	therefore they are positive definite for every \(\tau>0\). In contrast, by
	Theorems~\ref{thm:sw2_no_CND} and~\ref{thm:schoenberg}, there are values of
	\(\tau\) for which \(k_{\SW_2}\) and \(k_{\RW_2}\) are not positive definite. The
	other two kernels are the positive definite kernels used for comparison in
	\cite{Carriere2017}, as implemented in Gudhi \cite{Gudhi}. Both are built from
	the Gaussian kernel \(g_s(x,y)=\exp\bigl(-\|x-y\|^2/(2s^2)\bigr)\) on \(\R^2\),
	with \(s>0\) and the Euclidean norm \cite{Reininghaus2015,Kusano2018}, and we write
	them without their constant factors, which play no role in the protocol below.
	The Euclidean norm is needed here, since with the maximum norm of the previous
	sections \(g_s\) is not positive definite on \(\R^2\). For example, if
	\(x_1=(0,5)\), \(x_2=(1,6)\), \(x_3=(1,4)\), \(x_4=(2,5)\) and \(u=(1,-1,-1,1)\),
	the maximum norm gives \(\sum_{i,j}u_iu_j\,g_s(x_i,x_j)=4(1-2a+a^4)\), with
	\(a=e^{-1/(2s^2)}\), which is negative if \(s>0.91\).
	\begin{itemize}
		\item The \emph{persistence scale space kernel} \cite{Reininghaus2015} is
		\[
		k_{\mathrm{PSS}}(A,B)=\sum_{x\in A}\sum_{y\in B}
		\Bigl(g_s(x,y)-g_s\bigl(x,r_\Delta(y)\bigr)\Bigr).
		\]
		\item The \emph{persistence weighted Gaussian kernel} \cite{Kusano2016,Kusano2018}
		is \(k_{\mathrm{PWG}}(A,B)=\exp\bigl(-D_{\mathrm{PWG}}(A,B)/\tau\bigr)\), where
		\[
		\begin{aligned}
			D_{\mathrm{PWG}}(A,B)={}&\sum_{x,x'\in A}w(x)w(x')\,g_s(x,x')
			+\sum_{y,y'\in B}w(y)w(y')\,g_s(y,y')\\
			&-2\sum_{x\in A}\sum_{y\in B}w(x)w(y)\,g_s(x,y)
		\end{aligned}
		\]
		and \(w(x)=\arctan\bigl(c\,\pers{x}^4\bigr)\) is a weight depending on a
		parameter \(c>0\). The number \(D_{\mathrm{PWG}}(A,B)\) is the squared
		distance between the functions \(\sum_{x\in A}w(x)\,g_s(x,\cdot)\) and
		\(\sum_{y\in B}w(y)\,g_s(y,\cdot)\) in the Hilbert space associated with
		\(g_s\). The exponent \(4\) is the one used for this benchmark in
		\cite{Carriere2017}, and it is the smallest integer satisfying the condition
		\(p>3\) of the stability theorem of \cite{Kusano2018} for point clouds in the
		plane.
	\end{itemize}
	
	Each parameter, and the parameter \(C\) of the SVM, ranges over the grid given in
	Table~\ref{tab:orbitas_detalles}, and its value is selected in Step~4. For the five kernels of the form \(\exp(-d/\tau)\), that is, our four
	kernels and \(k_{\mathrm{PWG}}\) with \(d=D_{\mathrm{PWG}}\), the grid is
	\[
	\tau\in\{2^{j}m:\ j=-4,\dots,4\},\qquad
	m=\operatorname{med}\bigl(d(A_i,A_j):\ 1\le i<j\le n\bigr),
	\]
	where \(A_1,\dots,A_n\) are the training diagrams. The median \(m\) is computed
	without the labels. If \(d\) is multiplied by a constant, the whole grid is
	multiplied by the same constant, so the Gram matrices that are tested do not
	change, and a given position in the grid has the same meaning for the five
	kernels, although their distances take values of different magnitudes (for
	\(D_1(P)\), \(m\) is about \(0.18\) for \(\SW_1\) and about \(0.029\) for
	\(\SW_2\)). The grids of \(s\) and \(c\) are built around the values \(s_0\) and
	\(c_0=P_0^{-4}\) proposed in \cite[Section~3.4]{Kusano2018}, where \(s_0\) is the
	median, over the diagrams of the collection with at least two points, of the
	median distance between two points of a diagram, and \(P_0\) is the median, over the diagrams, of the median
	persistence of a diagram. For \(D_1(P)\) we obtain \(s_0\approx0.010\) and
	\(P_0\approx0.0032\). The integrals over \(\theta\) in \(\SW_p\) and \(\RW_p\) are
	computed with the midpoint rule with \(M=200\). For two diagrams \(D_1(P)\), the
	double sums in \(k_{\mathrm{PSS}}\) and \(D_{\mathrm{PWG}}\) have about
	\(6\cdot10^4\) terms, and, as in \cite{Carriere2017}, we approximate
	\(g_s(x,y)\) by an inner product of \(R=2^{14}\) random Fourier features
	\cite[Algorithm~1]{RahimiRecht2007}. Since the terms of these sums partly cancel,
	the error bound of \cite[Claim~1]{RahimiRecht2007} does not control the relative
	error of the matrix entries, so we measure it on \(30\) diagrams
	(Table~\ref{tab:orbitas_detalles}).
	\begin{table}[!htb]
		\centering
		\footnotesize
		\begin{tabular}{@{}>{\raggedright\arraybackslash}p{2.3cm}>{\raggedright\arraybackslash}p{3.1cm}>{\raggedright\arraybackslash}p{3.2cm}>{\raggedright\arraybackslash}p{2.3cm}>{\raggedright\arraybackslash}p{3.1cm}@{}}
			\toprule
			kernel & parameters and grids & computation & relative error & time for two diagrams \\
			\midrule
			\(k_{\SW_p}\), \(k_{\RW_p}\)\newline(\(p=1,2\)) & \(\tau=2^{j}m\), \(j=-4,\dots,4\)
			& midpoint rule with \(M=200\), one distance matrix for all \(\tau\)
			& \(<0.05\%\) in the distances
			& \(6.4\) ms for the four distances (\(0.16\) ms with \(M=6\)) \\[2pt]
			\(k_{\mathrm{PSS}}\) & \(s=s_0\,2^{j}\), \(j=-6,\dots,6\)
			& \(2^{14}\) Fourier features, matrix divided by the mean of its diagonal
			& \(<1.3\%\) in the matrix (\(2.5\%\) if \(s=s_0\,2^{-5}\), \(5.1\%\) if \(s=s_0\,2^{-6}\))
			& \(3.8\) ms for each \(s\), with the exact computation \\[2pt]
			\(k_{\mathrm{PWG}}\) & \(\tau\) as above,\newline \(s=s_0\,4^{j}\), \(c=c_0\,10^{j'}\), \(j,j'=-2,\dots,2\)
			& \(2^{14}\) Fourier features, one matrix \(D_{\mathrm{PWG}}\) for each \((s,c)\) for all \(\tau\)
			& \(<1.3\%\) in the matrix
			& \(4.5\) ms for each \((s,c)\), with the exact computation \\
			\midrule
			all kernels & \multicolumn{4}{l}{\(C=10^{j}\), \(j=-3,\dots,3\)} \\
			pairs \((D_0,D_1)\) & \multicolumn{4}{l}{\(\lambda\in\{0,1/4,1/2,3/4,1\}\)} \\
			\bottomrule
		\end{tabular}
		\caption{\footnotesize Parameters, computation and cost of each kernel in the
			orbit benchmark. The grids are defined in Step~3, where \(m\), \(s_0\) and
			\(c_0\) are computed from the diagrams; the selected values of \(\tau\),
			\(s\), \(c\) and \(C\) are discussed in the paragraph \emph{The
				parameters}, and the weight \(\lambda\) is defined in Step~4. The column \emph{relative error} refers to the quantities
			computed approximately, measured on \(30\) diagrams: the sliced distances,
			compared with the midpoint rule with \(M=4000\), and the matrices of
			\(k_{\mathrm{PSS}}\) and \(k_{\mathrm{PWG}}\), compared with the exactly
			computed matrices up to a constant factor that does not affect the
			protocol. The two smallest widths of \(k_{\mathrm{PSS}}\), \(s=s_0\,2^{-6}\) and
			\(s=s_0\,2^{-5}\), are never selected. The column \emph{time} is measured on a single core for two
			diagrams \(D_1(P)\); for \(k_{\mathrm{PSS}}\) and \(k_{\mathrm{PWG}}\), it
			corresponds to the exact computation of the double sums, which is the
			computation compared in the item \emph{Cost} of the results.}
		\label{tab:orbitas_detalles}
	\end{table}
	
	\paragraph{Step 4. Classification and accuracy.}
	We follow the protocol of \cite{Carriere2017}. In each of \(10\) runs, the
	\(500\) clouds are split at random into \(350\) training clouds and \(150\) test
	clouds, that is, \(70\) and \(30\) from each class, and the same splits are used
	for every kernel. The parameters are chosen by their mean accuracy over \(10\)
	random halves of the training set, training on one half and evaluating on the
	other, and the SVM is then trained on the \(350\) training clouds. We use the
	SVM of scikit-learn, which trains one binary SVM, as in
	Remark~\ref{rem:kernels_aprendizaje}, for each of the \(10\) pairs of classes and
	assigns each cloud to the class that wins the most of these comparisons. The
	accuracy of a run is the percentage of the \(150\) test clouds whose class is
	recognized correctly, and we report its mean and standard deviation over the
	\(10\) runs.
	
	Since the results of \cite{Carriere2017} use the diagrams of all degrees, we also
	represent each cloud \(P\) by the pair \(X=\bigl(D_0(P),D_1(P)\bigr)\), with the
	distance
	\[
	d^{\lambda}(X,X')=\lambda\,d(A_0,A_0')+(1-\lambda)\,d(A_1,A_1')
	\]
	between \(X=(A_0,A_1)\) and \(X'=(A_0',A_1')\), and the kernel
	\(\exp(-d^{\lambda}/\tau)\), where \(m\) is now the median of \(d^{\lambda}\) and
	the weight \(\lambda\in\{0,1/4,1/2,3/4,1\}\) is selected with the other
	parameters. For \(k_{\mathrm{PSS}}\), the two Gram matrices are combined in the
	same way. If \(d\) is \(\SW_1\) or \(\RW_1\), then \(d^{\lambda}\) is CND on pairs of
	diagrams, since it is a combination with nonnegative coefficients of two CND
	functions (Lemma~\ref{lem:sw_nd}), and Theorem~\ref{thm:schoenberg} shows that the
	kernel on pairs is still positive definite. The supplementary material contains
	the code that reproduces every number in this subsection.
	\begin{table}[!htb]
		\centering
		\footnotesize
		\begin{tabular}{lcccccc}
			\toprule
			& & & \multicolumn{2}{c}{\(D_1\)} & & \\
			\cmidrule(lr){4-5}
			kernel & positive definite & \(D_0\) & Rips & alpha & \((D_0,D_1)\) & reported in \cite{Carriere2017} \\
			\midrule
			\(k_{\SW_1}\)        & yes & \(50.0\pm2.5\) & \(85.1\pm2.9\) & \(84.1\pm3.2\) & \(85.1\pm2.9\) & \(83.7\pm0.5\) \\
			\(k_{\RW_1}\)        & yes & \(49.5\pm2.2\) & \(85.6\pm2.7\) & \(84.9\pm3.9\) & \(85.3\pm2.6\) & -- \\
			\(k_{\SW_2}\)        & no  & \(50.5\pm3.1\) & \(86.3\pm2.7\) & \(85.5\pm2.7\) & \(86.3\pm2.7\) & -- \\
			\(k_{\RW_2}\)        & no  & \(50.8\pm3.8\) & \(86.6\pm3.5\) & \(86.1\pm3.1\) & \(87.0\pm2.9\) & -- \\
			\(k_{\mathrm{PSS}}\) & yes & \(48.9\pm2.2\) & \(82.6\pm2.9\) & \(81.7\pm2.8\) & \(82.6\pm2.9\) & \(63.6\pm1.2\) \\
			\(k_{\mathrm{PWG}}\) & yes & \(49.1\pm3.8\) & \(84.9\pm2.8\) & \(79.8\pm4.0\) & \(84.9\pm2.8\) & \(77.7\pm1.2\) \\
			\(k_{\mathrm{PWG}}^{*}\) & yes & \(49.1\pm3.8\) & \(86.1\pm1.8\) & \(86.4\pm2.4\) & \(86.3\pm2.2\) & -- \\
			\bottomrule
		\end{tabular}
		\caption{\footnotesize Test accuracy (\(\%\), mean \(\pm\) standard deviation
			over \(10\) runs) on the orbit benchmark with the diagrams \(D_0(P)\), with
			the diagrams \(D_1(P)\) of the Vietoris--Rips filtration (Ripser) and of
			the alpha complex (Gudhi), and with the pairs \(\bigl(D_0(P),D_1(P)\bigr)\)
			of the Vietoris--Rips filtration. Here, the kernel \(k_{\mathrm{PWG}}^{*}\) is
			\(k_{\mathrm{PWG}}\) with the grid of \(c\) extended down to \(10^{-7}\)
			times \(c_0\). The last column reproduces Table~2 of
			\cite{Carriere2017}, which uses the alpha diagrams of all degrees, that
			is, \(D_0\) and \(D_1\) together, and \(6\) directions for \(k_{\SW_1}\).}
		\label{tab:orbitas}
	\end{table}
	
	\paragraph{Results.}
	Table~\ref{tab:orbitas} lists the accuracies. The main observations are as
	follows.
	\begin{itemize}
		\item \underline{The information is in \(D_1(P)\):} The diagrams \(D_0(P)\)
		alone give an accuracy of about \(50\%\), well above the \(20\%\) of random
		guessing among five classes, but far below the \(85\%\) obtained with
		\(D_1(P)\). Moreover, with the pairs, the selected \(\lambda\) is \(0\) or
		\(1/4\) in all \(60\) runs of the six kernels.
		\item \underline{Comparison with the reference kernels:} With the diagrams
		\(D_1(P)\), the kernels \(k_{\SW_1}\) and \(k_{\RW_1}\) reach \(85.1\%\) and
		\(85.6\%\), respectively. If we compare the runs one by one, on the same
		splits, then the kernel \(k_{\SW_1}\) outperforms \(k_{\mathrm{PSS}}\) in \(9\)
		of the \(10\) runs, with a mean difference of \(2.5\) points, and the kernel
		\(k_{\RW_1}\) outperforms it in \(8\) runs, with a mean difference of \(3.0\)
		points. In contrast, the mean differences with respect to \(k_{\mathrm{PWG}}\)
		are small, namely \(0.2\) and \(0.7\) points.
		\item \underline{Comparison with \cite{Carriere2017}:} That work uses the alpha
		diagrams of all degrees. With the alpha diagrams \(D_1(P)\), the kernel
		\(k_{\SW_1}\) reaches \(84.1\%\), and with the Vietoris--Rips pairs
		\(\bigl(D_0(P),D_1(P)\bigr)\), it reaches \(85.1\%\), in line with the
		\(83.7\%\) reported there.
		\item \underline{Alpha diagrams:} The diagrams \(D_0(P)\) of the two filtrations
		coincide, but the alpha diagrams \(D_1(P)\) have many more points close to the
		diagonal (median \(918\), against \(244.5\) for Vietoris--Rips). Indeed, the
		Vietoris--Rips complex fills a small triangle as soon as its longest edge appears,
		whereas the alpha complex, which has the homotopy type of the union of the balls of
		radius \(t\) \cite{EdelsbrunnerHarer2010}, may fill it later, and in the meantime
		its edges form a short-lived cycle. With the alpha diagrams and the same
		splits, the accuracy of the sliced kernels and of \(k_{\mathrm{PSS}}\) decreases
		by \(0.5\) to \(1\) point, and their ranking does not change, whereas
		\(k_{\mathrm{PWG}}\) loses \(5\) points, for a reason explained below.
		\item \underline{Errors:} Between \(50\%\) and \(59\%\) of the errors made by
		the sliced kernels confuse \(r=2.5\) with \(r=3.5\).
		\item \underline{Number of directions:} For \(M\in\{6,12,25,50,100,200\}\), the
		mean accuracy of each sliced kernel with \(D_1(P)\) varies by less than \(0.8\)
		points. In particular, with \(M=6\), as in \cite{Carriere2017}, it changes by
		at most \(0.6\) points with respect to \(M=200\).
		\item \underline{Cost:} The four sliced distances between two diagrams take
		\(6.4\) ms with \(M=200\) and \(0.16\) ms with \(M=6\), and they are computed
		only once, since the Gram matrix for each of the \(9\) values of \(\tau\) is
		obtained from the distance matrix by applying \(\exp(-\,\cdot/\tau)\)
		entrywise. In contrast, an exact evaluation of \(k_{\mathrm{PSS}}\) between two
		diagrams takes \(3.8\) ms and must be repeated for each of the \(13\) values of
		\(s\), that is, about \(49\) ms in total. Likewise, an exact evaluation of
		\(D_{\mathrm{PWG}}\) takes \(4.5\) ms and must be repeated for each of the
		\(25\) pairs \((s,c)\), that is, about \(112\) ms in total
		(Table~\ref{tab:orbitas_detalles}).
	\end{itemize}
	
	\paragraph{The parameters.}
	Table~\ref{tab:seleccion} summarizes the values selected in the \(10\) runs of
	Step~4.
	Three of them, \(\tau=16m\), \(c=10^{-2}P_0^{-4}\) and \(C=10^3\), lie at an end
	of their grids, so we checked that the grids do not limit the results.
	\begin{itemize}
		\item \underline{Lower end of \(\tau\):} If \(\tau\le m/16\), the Gram matrix is
		practically the identity. For instance, for \(k_{\SW_1}\), in each training set
		at least \(95\%\) of its off-diagonal entries are below \(4\cdot10^{-5}\). As
		\(\tau\to0\), the decision function of Remark~\ref{rem:kernels_aprendizaje}
		tends to a constant, so extending the grid downward only adds classifiers that
		are increasingly close to a trivial one.
		\item \underline{Upper ends of \(\tau\) and \(C\):} We repeated the experiment
		with \(\tau\) up to \(2^{8}m\) and \(C\) up to \(10^{5}\). The accuracy of the
		four sliced kernels changed by less than \(0.2\) points, that of
		\(k_{\mathrm{PSS}}\) rose from \(82.6\%\) to \(84.3\%\), and that of
		\(k_{\mathrm{PWG}}\) from \(84.9\%\) to \(85.6\%\).
		\item \underline{The weight parameter \(c\):} If we write \(c=t_0^{-4}\), the
		weight \(w(x)=\arctan\bigl((\pers{x}/t_0)^4\bigr)\) equals \(\pi/4\) when
		\(\pers{x}=t_0\), is at most \(1/16\) if \(\pers{x}\le t_0/2\), and is at least
		\(\pi/2-1/16\) if \(\pers{x}\ge2t_0\). Since the smallest value of the grid is
		selected in every run, we extended the grid down to \(10^{-7}P_0^{-4}\), and we
		denote by \(k_{\mathrm{PWG}}^{*}\) the resulting kernel. Its accuracy is
		\(86.1\%\) with the Vietoris--Rips diagrams \(D_1(P)\), \(86.3\%\) with the
		pairs and \(86.4\%\) with the alpha diagrams (Table~\ref{tab:orbitas}). With
		\(c=c_0\), every point with persistence above about \(2P_0\) receives almost
		the full weight, so the many small loops weigh as much as the few points that
		record the empty regions. The alpha diagrams, whose median persistence is about
		\(6\) times smaller, amplify this effect, which explains the loss of
		\(k_{\mathrm{PWG}}\) with them. This effect, together with the grid of \(s\) for
		\(k_{\mathrm{PSS}}\), which in \cite{Carriere2017} is fixed and not adapted to
		the scale of the diagrams, probably explains part of the low values reported
		there for these two kernels.
	\end{itemize}
	\begin{table}[!htb]
		\centering
		\footnotesize
		\setlength{\tabcolsep}{4pt}
		\begin{tabular}{lccccc}
			\toprule
			& \multicolumn{2}{c}{\(\tau\)} & & & \\
			\cmidrule(lr){2-3}
			kernel & range & most frequent & \(s\) & \(c\) & SVM \(C\) \\
			\midrule
			\(k_{\SW_1}\) & between \(2m\) and \(16m\) & \(4m\) (\(5\) of \(10\)) & -- & -- & between \(10\) and \(10^3\) \\
			\(k_{\RW_1}\) & between \(2m\) and \(8m\) & \(4m\) (\(5\) of \(10\)) & -- & -- & between \(10\) and \(10^3\) \\
			\(k_{\SW_2}\) & between \(2m\) and \(16m\) & \(2m\) (\(5\) of \(10\)) & -- & -- & between \(10\) and \(10^3\) \\
			\(k_{\RW_2}\) & between \(m\) and \(16m\) & \(2m\) (\(5\) of \(10\)) & -- & -- & between \(10\) and \(10^3\) \\
			\(k_{\mathrm{PSS}}\) & -- & -- & \(2s_0\) (\(10\) of \(10\)) & -- & \(10^3\) (\(10\) of \(10\)) \\
			\(k_{\mathrm{PWG}}\) & between \(8m\) and \(16m\) & \(16m\) (\(9\) of \(10\)) & \(s_0\) (\(10\) of \(10\)) & \(10^{-2}P_0^{-4}\) (\(10\) of \(10\)) & \(10^3\) (\(10\) of \(10\)) \\
			\bottomrule
		\end{tabular}
		\caption{\footnotesize Values selected by the protocol of Step~4 in the
			\(10\) runs, with the diagrams \(D_1(P)\) of the Vietoris--Rips
			filtration and the grids of Table~\ref{tab:orbitas_detalles}, where \(m\) is
			the median of the training distances in each run. An entry
			``between \(a\) and \(b\)'' means that the \(10\) selected values lie
			between \(a\) and \(b\), and an entry ``\(v\) (\(k\) of \(10\))'' means that
			the value \(v\) is selected in \(k\) of the \(10\) runs. For example, for
			\(k_{\mathrm{PWG}}\), the value \(\tau=16m\) is selected in \(9\) runs and
			\(\tau=8m\) in the remaining one.}
		\label{tab:seleccion}
	\end{table}
	
	\paragraph{SVMs with kernels that are not positive definite.}
	The kernels \(k_{\SW_2}\) and \(k_{\RW_2}\), which are not positive definite,
	are slightly more accurate than \(k_{\SW_1}\) and \(k_{\RW_1}\), by \(1.2\) and
	\(1.0\) points, respectively, even though their selected training Gram matrices
	have negative eigenvalues in \(10\) and \(9\) of the \(10\) runs, respectively
	(Table~\ref{tab:no_dp}). This does not contradict the theory, for three reasons.
	\begin{itemize}
		\item \underline{Positive definiteness enters the training only:} The
		SVM classifies by the sign of the function \(f\) of
		Remark~\ref{rem:kernels_aprendizaje}, that is, by a weighted vote of the
		similarities of \(A\) to the training diagrams, and neither this rule nor
		its accuracy on the test set requires \(k\) to be positive definite.
		Positive definiteness only ensures that the function \(L\) of
		Remark~\ref{rem:kernels_aprendizaje} is concave. Without it, every limit point of
		the iterates of the LIBSVM solver \cite{Fan2005}, which scikit-learn uses, is
		still a stationary point of \(L\), and such kernels often work well
		\cite{Haasdonk2005}, since what matters for the vote is that each diagram is
		more similar to the diagrams of its own class than to those of the other classes
		\cite{Balcan2008}.
		\item \underline{The problems actually solved are concave:} The function
		\(L\) of Remark~\ref{rem:kernels_aprendizaje} is concave exactly when the
		Gram matrix it uses is positive semidefinite. The \(350\times350\)
		training Gram matrix of \(k_{\SW_2}\) and \(k_{\RW_2}\) is not in \(10\) and \(9\) of the
		\(10\) runs, respectively, although its negative eigenvalues are tiny
		(Tables~\ref{tab:no_dp} and~\ref{tab:prop_datos}). However, this matrix
		is never used as a whole. Indeed, as explained in Step~4, the SVM solves one
		binary problem for each of the \(10\) pairs of classes, and each problem uses
		only the \(140\times140\) submatrix of \(K\) corresponding to the training
		diagrams of its two classes. In every run, these \(10\) submatrices are positive
		semidefinite, so the problems actually solved are concave.
		\item \underline{The advantage comes from the distances:} The \(k\) nearest
		neighbors classifier, which uses only the distances, is also more accurate with
		\(\SW_2\) and \(\RW_2\) than with \(\SW_1\) and \(\RW_1\).
	\end{itemize}
	
	\begin{table}[!htb]
		\centering
		\footnotesize
		\begin{tabular}{lcccccc}
			\toprule
			& & \multicolumn{2}{c}{accuracy (\(\%\))} & \multicolumn{2}{c}{runs with negative eigenvalues} & \\
			\cmidrule(lr){3-4}\cmidrule(lr){5-6}
			kernel & positive definite & SVM & \(k\)-NN & full matrix & submatrices & \(\lambda_{\max}(H_{500}DH_{500})\) \\
			\midrule
			\(k_{\SW_1}\) & yes & \(85.1\pm2.9\) & \(75.3\pm6.1\) & \(0\) & \(0\) & \(<10^{-15}\) \\
			\(k_{\RW_1}\) & yes & \(85.6\pm2.7\) & \(79.1\pm4.0\) & \(0\) & \(0\) & \(<10^{-15}\) \\
			\(k_{\SW_2}\) & no  & \(86.3\pm2.7\) & \(83.9\pm3.0\) & \(10\) & \(0\) & \(0.013\) \\
			\(k_{\RW_2}\) & no  & \(86.6\pm3.5\) & \(83.0\pm3.7\) & \(9\) & \(0\) & \(0.020\) \\
			\bottomrule
		\end{tabular}
		\caption{\footnotesize The four sliced kernels with the diagrams \(D_1(P)\):
			accuracy of the SVM and of \(k\) nearest neighbors (\(k\)-NN) with the
			same distances (mean \(\pm\) standard deviation over \(10\) runs); number of
			runs in which the selected \(350\times350\) training Gram matrix, or one of
			the \(10\) submatrices of size \(140\times140\) used by the binary SVMs, has
			negative eigenvalues; and largest eigenvalue of \(H_{500}DH_{500}\), where
			\(D\) is the matrix of the distance on the \(500\) diagrams
			(Remark~\ref{rem:cnd_centrado}).}
		\label{tab:no_dp}
	\end{table}
	
	\paragraph{When the Gram matrix has negative eigenvalues.}
	Table~\ref{tab:prop_datos} compares Proposition~\ref{prop:ancho_banda}, with
	\(\tau=2\sigma^2\), with the training Gram matrices of \(k_{\SW_2}\) and
	\(k_{\RW_2}\), for which \(\ell>0\). As predicted, when \(\tau\) grows, negative
	eigenvalues appear, the smallest eigenvalue \(\lambda_{\min}\) approaches
	\(-\ell/\tau\), and the largest eigenvalue \(\lambda_{\max}\) approaches \(n\),
	although the bounds of the proposition only cover values of \(\tau\) much larger
	than the selected ones. At the selected values of \(\tau\), the negative
	eigenvalues are more than \(6000\) times smaller in absolute value than the
	largest one, and for \(\tau=m\) there are none. In contrast, for \(\SW_1\) and
	\(\RW_1\) we have \(\ell=0\) up to rounding (Lemma~\ref{lem:sw_nd}, which also
	applies to the midpoint rule, as the latter is a sum of CND terms), so, by
	item~(1), their Gram matrices are positive semidefinite for every \(\tau\).
	
	\begin{table}[!htb]
		\centering
		\footnotesize
		\begin{tabular}{rcccccccc}
			\toprule
			& \multicolumn{4}{c}{\(k_{\SW_2}\)} & \multicolumn{4}{c}{\(k_{\RW_2}\)} \\
			\cmidrule(lr){2-5}\cmidrule(lr){6-9}
			\(\tau\) & sets & \(-\ell/\tau\) & \(\lambda_{\min}\) & \(\lambda_{\max}\)
			& sets & \(-\ell/\tau\) & \(\lambda_{\min}\) & \(\lambda_{\max}\) \\
			\midrule
			\(m\)  & \(0\)  & \(-0.238\) & \(0.011\)   & \(145.6\) & \(0\)  & \(-0.269\) & \(0.0085\)  & \(145.5\) \\
			\(2m\)  & \(9\)  & \(-0.119\) & \(-0.0080\) & \(217.8\) & \(10\) & \(-0.134\) & \(-0.0188\) & \(216.3\) \\
			\(4m\)  & \(10\) & \(-0.059\) & \(-0.0258\) & \(273.5\) & \(10\) & \(-0.067\) & \(-0.0309\) & \(271.8\) \\
			\(8m\)  & \(10\) & \(-0.030\) & \(-0.0204\) & \(308.6\) & \(10\) & \(-0.034\) & \(-0.0233\) & \(307.4\) \\
			\(16m\) & \(10\) & \(-0.0149\) & \(-0.0124\) & \(328.5\) & \(10\) & \(-0.0168\) & \(-0.0140\) & \(327.7\) \\
			\midrule
			\(2^{8}m\)  & \(10\) & \(-9.3\cdot10^{-4}\) & \(-9.2\cdot10^{-4}\) & \(348.6\) & \(10\) & \(-1.05\cdot10^{-3}\) & \(-1.04\cdot10^{-3}\) & \(348.5\) \\
			\(2^{12}m\) & \(10\) & \(-5.81\cdot10^{-5}\) & \(-5.81\cdot10^{-5}\) & \(349.9\) & \(10\) & \(-6.56\cdot10^{-5}\) & \(-6.55\cdot10^{-5}\) & \(349.9\) \\
			\midrule
			\multicolumn{1}{l}{item~(2)} & \multicolumn{4}{c}{\(\tau>3750\,m\) to \(6154\,m\)} & \multicolumn{4}{c}{\(\tau>5136\,m\) to \(7555\,m\)} \\
			\multicolumn{1}{l}{item~(3)} & \multicolumn{4}{c}{\(\tau\ge4.9\,m\) to \(5.5\,m\)} & \multicolumn{4}{c}{\(\tau\ge6.0\,m\) to \(6.7\,m\)} \\
			\bottomrule
		\end{tabular}
		\caption{\footnotesize Proposition~\ref{prop:ancho_banda} compared with the
			training Gram matrices of \(k_{\SW_2}\) and \(k_{\RW_2}\), with the diagrams
			\(D_1(P)\) and \(\tau\) given in multiples of the median \(m\) of each
			training set. For each \(\tau\), the column \emph{sets} gives the number of
			the \(10\) training sets, each of \(n=350\) diagrams, whose Gram matrix has a
			negative eigenvalue. The column \(-\ell/\tau\) is the value of
			\(\lambda_{\min}\) predicted by item~(3), with
			\(\ell=\lambda_{\max}(H_nDH_n)\) computed on the training set, and
			\(\lambda_{\min}\) and \(\lambda_{\max}\) are the smallest and largest
			eigenvalues of the Gram matrix. These three columns are means over the
			\(10\) training sets. The selected values of \(\tau\) lie between \(m\) and
			\(16m\) (Table~\ref{tab:seleccion}). The last two rows give, over the
			\(10\) training sets, the values of \(\tau\) beyond which item~(2) guarantees
			a negative eigenvalue and beyond which the hypothesis \(\sigma^2\ge M_D\) of
			item~(3) holds.}
		\label{tab:prop_datos}
	\end{table}
	
	\paragraph{What positive definiteness provides.}
	Positive definiteness is what provides guarantees listed in Table~\ref{tab:garantias}, alongside their accuracies. Precisely, we have the following properties of a kernel \(k\) with decision function \(f\) as in
	Remark~\ref{rem:kernels_aprendizaje}.
	\begin{itemize}
		\item \underline{Concave training:} The function \(L\) is concave for all
		parameters, so the SVM finds a global maximum.
		\item \underline{Lipschitz classifier:} There exists \(L_f>0\) such that
		\(|f(A)-f(B)|\le L_f\,d(A,B)\) for a distance \(d\) between diagrams, so a
		small perturbation of a diagram, for instance due to noise in the data, changes
		the value of the decision function only slightly.
		\item \underline{Lower bound:} On diagrams with at most \(N\) points, the distance
		that defines the kernel is bounded below by a positive multiple of \(d_p\) or
		\(d_p^r\), as in Theorems~\ref{thm:equiv_p} and~\ref{thm:equiv_p_rw}. Then two
		diagrams that are far apart in \(d_p\) remain far apart for the kernel, which
		therefore never confuses two very different diagrams.
	\end{itemize}
	For \(k_{\SW_1}\) and \(k_{\RW_1}\), Proposition~\ref{prop:kernels_aprendizaje}
	gives concave training and a Lipschitz classifier with respect to \(d_1\),
	and Theorems~\ref{thm:equiv_p} and~\ref{thm:equiv_p_rw} give the lower bound
	with respect to \(d_1\), as shown in \cite{Carriere2017} for \(\SW_1\). The
	kernels \(k_{\mathrm{PSS}}\) and \(k_{\mathrm{PWG}}\) are also positive
	definite and give Lipschitz classifiers \cite{Reininghaus2015,Kusano2018},
	but, to our knowledge, no lower bound is known for them. By contrast, for
	\(k_{\SW_2}\) and \(k_{\RW_2}\) the training is not guaranteed to be concave, no Lipschitz property
	of the classifier is known, and their lower bound, with respect to \(d_2\) and
	\(d_2^r\), gives no guarantee on the classifier without positive definiteness.
	Therefore, their good results here are a property of these data and are not
	guaranteed for other data. Moreover, some procedures do require a positive
	semidefinite Gram matrix. For instance, Gaussian process regression
	\cite{RasmussenWilliams2006} uses the Gram matrix \(K_{tt}\) of the training
	diagrams, plus \(\eta^2I\) for a noise variance \(\eta^2>0\), as a covariance
	matrix, and computes the Cholesky factorization of \(K_{tt}+\eta^2I\), which
	exists if and only if this matrix is positive definite. Hence, if \(K_{tt}\) has
	a negative eigenvalue \(\lambda_{\min}\), the factorization does not exist for
	any \(\eta^2\le-\lambda_{\min}\). By Proposition~\ref{prop:ancho_banda}(1), this
	never happens with \(k_{\SW_1}\) and \(k_{\RW_1}\), since every Gram matrix of
	these kernels, and hence every principal submatrix of it, is positive
	semidefinite. Thus, a kernel that is not positive definite may work well on some
	data and in some procedures and fail in others, and in general one does not know
	in advance which case occurs. In summary, positive definiteness does not determine the accuracy of an SVM, but it is what turns \(k_{\SW_1}\) and
	\(k_{\RW_1}\) into kernels with theoretical guarantees. For this reason, when it
	is not known in advance how the data will behave, it is advisable to use kernels
	built from CND distances, such as \(k_{\SW_1}\) and \(k_{\RW_1}\).
	
	\begin{table}[!htb]
		\centering
		\footnotesize
		\begin{tabular}{lccccc}
			\toprule
			kernel & accuracy, \(D_1\) & positive definite & concave training & Lipschitz classifier & lower bound \\
			\midrule
			\(k_{\SW_1}\) & \(85.1\) & \textcolor{green!45!black}{yes} & \textcolor{green!45!black}{yes} & \textcolor{green!45!black}{yes}, \(d_1\) & \textcolor{green!45!black}{yes}, \(d_1\) \\
			\(k_{\RW_1}\) & \(85.6\) & \textcolor{green!45!black}{yes} & \textcolor{green!45!black}{yes} & \textcolor{green!45!black}{yes}, \(d_1\) & \textcolor{green!45!black}{yes}, \(d_1\) \\
			\(k_{\SW_2}\) & \(86.3\) & \textcolor{red!70!black}{no} & \textcolor{red!70!black}{no} & not known & \textcolor{gray}{yes, \(d_2\)} \\
			\(k_{\RW_2}\) & \(86.6\) & \textcolor{red!70!black}{no} & \textcolor{red!70!black}{no} & not known & \textcolor{gray}{yes, \(d_2^r\)} \\
			\(k_{\mathrm{PSS}}\) & \(82.6\) & \textcolor{green!45!black}{yes} & \textcolor{green!45!black}{yes} & \textcolor{green!45!black}{yes} \cite{Reininghaus2015} & not known \\
			\(k_{\mathrm{PWG}}\) & \(84.9\) & \textcolor{green!45!black}{yes} & \textcolor{green!45!black}{yes} & \textcolor{green!45!black}{yes} \cite{Kusano2018} & not known \\
			\bottomrule
		\end{tabular}
		\caption{\footnotesize Accuracy with the Vietoris--Rips diagrams
			\(D_1(P)\) (Table~\ref{tab:orbitas}) and the guarantees defined in the
			text. In the last two columns, each entry is the distance with respect to
			which the property holds. For \(k_{\SW_2}\) and \(k_{\RW_2}\), the lower bound
			holds, but without positive definiteness it gives no guarantee on the
			classifier.}
		\label{tab:garantias}
	\end{table}
	
	\section{Concluding remarks}
	\label{sec:conclusiones}
	
	This work starts from the Sliced Wasserstein kernel of Carri\`ere, Cuturi and
	Oudot \cite{Carriere2017} and extends it in two directions: to every order
	\(p\ge1\) and to a reflected variant. The picture that emerges is that the
	order \(p=1\) is special. Indeed, for \(p=1\) the distances \(\SW_1\) and
	\(\RW_1\) are metrics and CND, and the kernels \(k_{\SW_1}\) and \(k_{\RW_1}\) are
	positive definite, universal on compact sets and stable with respect to
	\(d_1\), so the distance \(\RW_1\) provides a new kernel with the same guarantees
	as \(\SW_1\). In contrast, for \(p>1\) the triangle inequality fails, and the
	distances \(\SW_2\) and \(\RW_2\) are not CND. Along the way, we prove that
	\(d_p\) admits the two equivalent formulations of
	Theorem~\ref{thm:dp_equivalencias}, the second of which is stated without proof
	in \cite{BubenikElchesen2022}. We also show that the reflected distance satisfies
	\(2^{1/p}d_p\le d_p^r\le2\,d_p\) with sharp constants, and we obtain constants of
	metric equivalence between the sliced distances and \(d_p\) and \(d_p^r\) that
	are sharp for the upper bound and improve those of \cite{Carriere2017} for
	\(p=1\).
	
	On the orbit recognition benchmark, the kernels \(k_{\SW_1}\) and
	\(k_{\RW_1}\) reach accuracies of \(85.1\%\) and \(85.6\%\), respectively, in a
	five-class task where chance gives \(20\%\). With the protocol of
	\cite{Carriere2017}, they outperform the persistence scale space kernel and are
	on par with the persistence weighted Gaussian kernel, which reaches \(86.1\%\) if
	the grid of its weight parameter is extended downward. Furthermore, they are computed
	in milliseconds, and their accuracies change by about one point when the
	Vietoris--Rips diagrams are replaced by the alpha diagrams used in
	\cite{Carriere2017}. The kernels built from \(\SW_2\) and \(\RW_2\), which are
	not positive definite, are slightly more accurate on these data, but they lack
	the guarantees of Proposition~\ref{prop:kernels_aprendizaje}. Moreover, among
	the positive definite kernels in the comparison, the kernels \(k_{\SW_1}\) and
	\(k_{\RW_1}\) are the only ones for which a lower bound in terms of \(d_1\) is
	known.
	
	Three questions remain open. The first is whether \(\SW_p\) and \(\RW_p\) are
	CND for \(p\in(1,2)\) and \(p>2\); the experiments suggest that they are not
	above a threshold \(p^*\in(1,1.2]\). The second is whether the order
	\(1/N^2\) of the lower constant in Theorems~\ref{thm:equiv_p}
	and~\ref{thm:equiv_p_rw} can be improved. The third is on which data and in
	which procedures the failure of the CND property has consequences; answering it
	requires statistical tools, and we leave it for future work.
	
	\paragraph{Acknowledgements.}
	M.P. acknowledges support from the Chilean ANID basal center CenIA, code FB210017.
	
	\paragraph{Use of AI tools.}
	The authors used Claude (Anthropic), an AI assistant, during the preparation
	of this work. It was used to help write and revise the text of the
	manuscript, to write and check the Python code of the numerical experiments
	and of the interval-arithmetic computations in the supplementary material,
	and to check proofs and references. All results, proofs, code and text were
	reviewed by the authors, who take full responsibility for the content of
	the article.
	
	\newpage
	\begin{appendices}
		
		\section{Proof of Theorem~\ref{thm:dp_equivalencias}}
		\label{app:D}
		
		Theorem~\ref{thm:dp_equivalencias} follows from two lemmas.
		Lemma~\ref{lem:formulacion_1} proves formulation~(1), and
		Lemma~\ref{lem:formulacion_2} derives formulation~(2) from the first.
		
		\paragraph{Notation.} Fix $A,B\in\mathcal{D}_{\mathrm{fin}}$. As in
		Definition~\ref{def:diagrama_finito}, copies of the same point in a diagram
		are treated as distinct elements.
		\begin{itemize}
			\item An \emph{admissible triple} is a triple $(A_0,B_0,s)$ with
			$A_0,B_0\subseteq\Delta\times\N$ finite, $|A_0|=|A|$, $|B_0|=|B|$, and
			$s:A\cup B_0\to A_0\cup B$ a bijection.
			\item An \emph{admissible decomposition} is a triple $(A_1,B_1,\gamma)$ with
			$A_1\subseteq A$, $B_1\subseteq B$, and $\gamma:A_1\to B_1$ a bijection.
			We write $A_2\coloneqq A\setminus A_1$ and $B_2\coloneqq B\setminus B_1$.
			\item For a bijection $\beta$ between finite sets, we define
			$c_p(\beta)\coloneqq\bigl(\sum_{x}\|x-\beta(x)\|_\infty^p\bigr)^{1/p}$,
			where the sum runs over the domain of $\beta$. For an admissible
			decomposition, we define
			\[
			S(A_1,B_1,\gamma)\coloneqq
			\sum\limits_{a\in A_1}\|a-\gamma(a)\|_\infty^p
			+\sum\limits_{a\in A_2}\delta(a)^p
			+\sum\limits_{b\in B_2}\delta(b)^p .
			\]
		\end{itemize}
		\paragraph{Two elementary facts.} By Remarks~\ref{rem:norma_diagonal}
		and~\ref{rem:prop_diagonal}(1), the following hold.
		\begin{enumerate}
			\item[(H1)] If $x\in\Omega$ and $y\in\Delta\times\N$, then
			$\|x-y\|_\infty\ge\delta(x)$, with equality when the first two
			coordinates of $y$ are those of $\pi_\Delta(x)$.
			\item[(H2)] Two points of $\Delta\times\N$ with the same first two
			coordinates are at distance zero.
		\end{enumerate}
		
		\paragraph{The case $p=\infty$.} We write the proofs for $p\in[1,\infty)$.
		For $p=\infty$, they are identical once each sum of $p$-th powers is
		replaced by a maximum, since each individual cost $\|x-s(x)\|_\infty$
		appearing in them is either preserved or bounded term by term.
		
		\begin{lemma}[\textbf{Formulation~(1)}]
			\label{lem:formulacion_1}
			\[
			d_p(A,B)=\min_{(A_0,B_0,s)\ \text{admissible}} c_p(s).
			\]
		\end{lemma}
		
		\begin{proof}
			Let $I\coloneqq\inf c_p(s)$, where the infimum is taken over all
			admissible triples. We show that $I=d_p(A,B)$ and that the infimum is
			attained.
			
			\textbf{$\blacktriangleright$ Step 1.} $d_p(A,B)\le I$ (extension of
			admissible triples). Fix an admissible triple $(A_0,B_0,s)$. Since
			$\overline{A}$ is the disjoint union of $A\cup B_0$ and
			$(\Delta\times\N)\setminus B_0$, and $\overline{B}$ is the disjoint union
			of $A_0\cup B$ and $(\Delta\times\N)\setminus A_0$, it suffices to
			construct a zero-cost bijection
			$\phi:(\Delta\times\N)\setminus B_0\to(\Delta\times\N)\setminus A_0$.
			For each $z\in\R$, let $F_z\coloneqq\{(z,z,n):n\in\N\}$. Since $A_0$ and
			$B_0$ are finite, the sets $F_z\setminus B_0$ and $F_z\setminus A_0$ are
			countably infinite, so there is a bijection $\phi_z$ between them, which
			has zero cost by~(H2). Together, the maps $\phi_z$ define $\phi$, and the
			map $\gamma_s$ that equals $s$ on $A\cup B_0$ and $\phi$ elsewhere is a
			bijection that moves only points of $A\cup B_0$ with positive cost.
			Hence $\gamma_s$ is a partial bijection with $c_p(\gamma_s)=c_p(s)$, and
			so $d_p(A,B)\le c_p(s)$. Taking the infimum over admissible triples
			proves Step~1.
			
			\textbf{$\blacktriangleright$ Step 2.} $I\le d_p(A,B)$ (reduction of
			partial bijections). Let $\gamma:\overline{A}\to\overline{B}$ be a partial
			bijection. The set
			$X\coloneqq\{x\in\Delta\times\N:\|x-\gamma(x)\|_\infty>0\text{ or }\gamma(x)\in B\}$
			is finite, since $\gamma$ moves only finitely many points with positive
			cost and at most $|B|$ points are sent to $B$. Moreover, the preimage of
			every point of $B$ lies in $A\cup X$. Hence, with
			$Y\coloneqq\gamma(A\cup X)\cap(\Delta\times\N)$, the restriction
			$\gamma|_{A\cup X}:A\cup X\to Y\cup B$ is a bijection with cost at most
			$c_p(\gamma)$, whence $|A|+|X|=|B|+|Y|$.
			\begin{itemize}
				\item \emph{Reducing $|X|$.} If $|X|>|B|$, then $|Y|>|A|$, and by the
				pigeonhole principle some $x\in X$ satisfies $\gamma(x)\in Y$.
				Removing the pair $(x,\gamma(x))$ yields a bijection between
				$A\cup(X\setminus\{x\})$ and $(Y\setminus\{\gamma(x)\})\cup B$ whose
				cost is no larger. Iterating this step, we obtain $X_0\subseteq X$,
				$Y_0\subseteq Y$, and a bijection $\beta_0:A\cup X_0\to Y_0\cup B$ with
				$c_p(\beta_0)\le c_p(\gamma)$ and $|X_0|\le|B|$, and hence
				$|Y_0|=|A|+|X_0|-|B|\le|A|$.
				\item \emph{Equalizing cardinalities.} Let $r\coloneqq|B|-|X_0|$.
				Choose points $w_1,\dots,w_r$ and $w_1',\dots,w_r'$ in
				$\Delta\times\N$ with new indices such that $w_i$ and $w_i'$ have
				the same first two coordinates. Define
				$B'\coloneqq X_0\cup\{w_1,\dots,w_r\}$ and
				$A'\coloneqq Y_0\cup\{w_1',\dots,w_r'\}$, and let $\beta$ be equal to
				$\beta_0$ on $A\cup X_0$ and to $w_i\mapsto w_i'$ elsewhere. Then the
				triple $(A',B',\beta)$ is admissible, since $|B'|=|B|$ and
				$|A'|=|Y_0|+r=|A|$, and by~(H2) we have
				$c_p(\beta)=c_p(\beta_0)\le c_p(\gamma)$.
			\end{itemize}
			Hence $I\le c_p(\gamma)$, and taking the infimum over $\gamma$ proves
			Step~2.
			
			\textbf{$\blacktriangleright$ Step 3.} The infimum is attained. Given an
			admissible triple $(A_0,B_0,s)$, we modify it without increasing its cost.
			If $a\in A$ and $s(a)\in A_0$, then we replace $s(a)$ by $\pi_\Delta(a)$
			with a new index; if $b\in B$ and $s^{-1}(b)\in B_0$, then we replace
			$s^{-1}(b)$ by $\pi_\Delta(b)$ with a new index. By~(H1), these
			replacements do not increase the cost. Moreover, each pair
			$x\mapsto s(x)$ with $x\in B_0$ and $s(x)\in A_0$ is replaced by a pair of
			points with new indices and the same first two coordinates, which has
			zero cost by~(H2). The resulting triple is still admissible and, with
			$A_1\coloneqq\{a\in A:s(a)\in B\}$, its cost raised to the power $p$ is
			$S(A_1,s(A_1),s|_{A_1})$, which depends only on $A_1$ and $s|_{A_1}$.
			Since there are only finitely many choices for these data, it follows
			that $I$ is a minimum over a finite set of values, and so it is attained.
		\end{proof}
		
		\begin{lemma}[\textbf{Formulation~(2)}]
			\label{lem:formulacion_2}
			\[
			d_p(A,B)^p=\min_{(A_1,B_1,\gamma)\ \text{admissible}} S(A_1,B_1,\gamma).
			\]
		\end{lemma}
		
		\begin{proof}
			Let $M$ denote the right-hand side, which is a minimum over a finite set.
			
			\textbf{$\blacktriangleright$ Step 1.} $d_p(A,B)^p\le M$. Fix an
			admissible decomposition $(A_1,B_1,\gamma)$. Choose pairwise distinct
			indices $\alpha_a,\beta_b,\beta'_b,\beta''_b$ and set
			\[
			\begin{aligned}
				A_0&\coloneqq\{(\pi_\Delta(a),\alpha_a):a\in A_2\}
				\cup\{(\pi_\Delta(b),\beta'_b):b\in B_1\},\\
				B_0&\coloneqq\{(\pi_\Delta(b),\beta_b):b\in B_2\}
				\cup\{(\pi_\Delta(b),\beta''_b):b\in B_1\}.
			\end{aligned}
			\]
			Since $|A_1|=|B_1|$, we have $|A_0|=|A_2|+|B_1|=|A|$ and
			$|B_0|=|B_2|+|B_1|=|B|$. Define $s:A\cup B_0\to A_0\cup B$ by
			\[
			\begin{aligned}
				s(a)&=\gamma(a)\quad(a\in A_1), & s(a)&=(\pi_\Delta(a),\alpha_a)\quad(a\in A_2),\\
				s((\pi_\Delta(b),\beta_b))&=b\quad(b\in B_2), &
				s((\pi_\Delta(b),\beta''_b))&=(\pi_\Delta(b),\beta'_b)\quad(b\in B_1).
			\end{aligned}
			\]
			The four blocks of the domain are mapped bijectively onto $B_1$,
			$\{(\pi_\Delta(a),\alpha_a):a\in A_2\}$, $B_2$, and
			$\{(\pi_\Delta(b),\beta'_b):b\in B_1\}$, which are pairwise disjoint and
			cover $A_0\cup B$, so $(A_0,B_0,s)$ is an admissible triple. Moreover,
			by~(H1) and~(H2), the elements of the second and third blocks cost
			$\delta(a)$ and $\delta(b)$, respectively, and those of the fourth cost
			zero, so $c_p(s)^p=S(A_1,B_1,\gamma)$. Hence, by
			Lemma~\ref{lem:formulacion_1}, we have
			$d_p(A,B)^p\le c_p(s)^p=S(A_1,B_1,\gamma)$, and taking the minimum over
			admissible decompositions proves Step~1.
			
			\textbf{$\blacktriangleright$ Step 2.} $M\le d_p(A,B)^p$. By
			Lemma~\ref{lem:formulacion_1}, there is an admissible triple
			$(A_0,B_0,s)$ with $d_p(A,B)^p=c_p(s)^p$. Set
			$A_1\coloneqq\{a\in A:s(a)\in B\}$ and $B_1\coloneqq s(A_1)$, let $A_2$
			and $B_2$ be as in the notation, and set
			$B_0'\coloneqq\{b_0\in B_0:s(b_0)\in B_2\}$. Then
			$\gamma\coloneqq s|_{A_1}:A_1\to B_1$ and $s|_{B_0'}:B_0'\to B_2$ are
			bijections. If $a\in A_2$, then $s(a)\in A_0$ and, by~(H1),
			we have $\|a-s(a)\|_\infty\ge\delta(a)$. If $b_0\in B_0'$, then (H1) gives
			$\|b_0-s(b_0)\|_\infty\ge\delta(s(b_0))$; reindexing by $b=s(b_0)$, we see
			that the sum over $B_0'$ is at least $\sum_{b\in B_2}\delta(b)^p$.
			Finally, the terms with $b_0\in B_0\setminus B_0'$ are nonnegative.
			Therefore $d_p(A,B)^p=c_p(s)^p\ge S(A_1,B_1,\gamma)\ge M$.
		\end{proof}
		
		\begin{proof}[Proof of Theorem~\ref{thm:dp_equivalencias}]
			Formulation~(1) is Lemma~\ref{lem:formulacion_1}, and formulation~(2) is
			Lemma~\ref{lem:formulacion_2}. By the convention adopted for the case
			$p=\infty$, both lemmas also hold for $p=\infty$.
		\end{proof}

		\section{Deferred proofs from Section~\ref{sec:kernel}}
		\label{app:A}

		\begin{corollary}[\textbf{Wasserstein distance for empirical measures ($p \ge 1$)}]
			\label{cor:empirico_p}
			Let $p\in[1,\infty)$, let $x_1\le\cdots\le x_n$ and $y_1\le\cdots\le y_n$
			be real numbers, and let $\mu=\sum\limits_{i=1}^{n}\delta_{x_i}$ and
			$\nu=\sum\limits_{i=1}^{n}\delta_{y_i}$. Then the following hold.
			\begin{enumerate}
				\item $\displaystyle\mathcal{W}_p(\mu,\nu)=\Bigl(\sum_{i=1}^{n}|x_i-y_i|^p\Bigr)^{1/p}$.
				\item For every permutation $\sigma$ of $\{1,\dots,n\}$, we have
				$\displaystyle\sum_{i=1}^{n}|x_i-y_i|^p\le\sum_{i=1}^{n}|x_i-y_{\sigma(i)}|^p$.
			\end{enumerate}
		\end{corollary}
		
		\begin{proof}
			The generalized quantile function of the measure $\mu/n$ is
			$F_{\mu/n}^{-1}(t)=x_i$ for $t\in\bigl(\frac{i-1}{n},\frac{i}{n}\bigr]$, and
			similarly for $\nu/n$. Then, by Proposition~\ref{prop:equiv_p}(1), we have
			$\mathcal{W}_p(\mu,\nu)^p=n\int_0^1|F_{\mu/n}^{-1}(t)-F_{\nu/n}^{-1}(t)|^p\,\dd t
			=\sum\limits_{i=1}^n|x_i-y_i|^p$, which is~(1). For~(2), note that the
			coupling $\sum\limits_{i=1}^n\delta_{(x_i,y_{\sigma(i)})}\in\Pi(\mu,\nu)$
			has cost $\sum\limits_{i=1}^n|x_i-y_{\sigma(i)}|^p\ge\mathcal{W}_p(\mu,\nu)^p$.
		\end{proof}

		\paragraph{Proof of Lemma~\ref{lem:estructura_medidas_p}.}
		\mbox{}
		\begin{proof}
			We identify each triple $(b,d,j)$ with the point $(b,d)$. Item~(1) is a
			restatement of Definition~\ref{def:medidas_proyectadas}, since $P$ is the
			disjoint union of $A$ and $r_\Delta(B)$, and $Q$ is the disjoint union of
			$B$ and $r_\Delta(A)$. In particular, both measures have total mass
			$r=|A|+|B|$, so item~(2) is Corollary~\ref{cor:empirico_p}(1) applied to
			their ordered supports. For~(3), we write $P=\{x_1,\dots,x_r\}$ and
			$Q=\{y_1,\dots,y_r\}$ with $\pi_\theta(x_i)=\alpha_i^\theta$ and
			$\pi_\theta(y_i)=\beta_i^\theta$, and we take $s^\theta(x_i)=y_i$. Since
			$\pi_\theta(x)-\pi_\theta(y)=\langle x-y,u_\theta\rangle$, item~(3)
			follows from~(2). Finally, every bijection $\sigma:P\to Q$ has the form
			$\sigma(x_i)=y_{\tau(i)}$ for some permutation $\tau$ of $\{1,\dots,r\}$,
			so item~(4) follows from Corollary~\ref{cor:empirico_p}(2).
		\end{proof}

		\section{Proof of Proposition~\ref{prop:ancho_banda}}
		\label{app:ancho_banda}
		
		In this appendix, we prove Proposition~\ref{prop:ancho_banda} of
		Subsection~\ref{subsec:orbitas}, using the notation introduced there.
		The idea of the proof is as follows. If \(t=1/(2\sigma^2)\) is small, then
		the Gram matrix \(K_\sigma\) is close to \(\mathbf{1}\mathbf{1}^\top-tD\). In the
		direction of \(\mathbf{1}\), this matrix is large, of order \(n\), whereas on the
		subspace \(V_n=\{w\in\R^n:w\cdot\mathbf{1}=0\}\) it behaves like \(-tD\), and the
		largest value of \(w^\top Dw\) over the unit vectors of \(V_n\) is \(\ell\).
		Hence, we expect one eigenvalue close to \(n\) and one close to \(-t\ell\). The
		following lemmas control the two errors in this approximation, namely the
		remainder of the exponential (Lemma~\ref{lem:resto_exp}) and the coupling
		between the direction of \(\mathbf{1}\) and the subspace \(V_n\)
		(Lemmas~\ref{lem:cotas_D}, \ref{lem:cota_D}, and~\ref{lem:cota_E}), by means of the
		Rayleigh quotient (Remark~\ref{rem:rayleigh}). The positive
		semidefinite case, on the other hand, relies on Schoenberg's theorem for a fixed collection of points
		(Lemma~\ref{lem:schoenberg_finito}).
		
		\begin{lemma}[\textbf{Schoenberg's theorem for a finite collection}]
			\label{lem:schoenberg_finito}
			Let \(D\) be a real symmetric \(n\times n\) matrix such that \(D_{ii}=0\) for
			every \(i\) and \(a^\top Da\le0\) for every \(a\in\R^n\) with
			\(\sum_ia_i=0\). Then, for every \(t>0\), the matrix
			\(E_t=\bigl(e^{-tD_{ij}}\bigr)_{i,j}\) is positive semidefinite.
		\end{lemma}
		
		\begin{proof}
			This is Schoenberg's theorem \cite[Ch.~3, Theorem~2.2]{Berg1984}, which
			holds for kernels on an arbitrary nonempty set, applied to the set
			\(\{1,\dots,n\}\) and the kernel \((i,j)\mapsto D_{ij}\). In
			\cite{Berg1984}, negative definite kernels are defined
			with complex coefficients, but for a real symmetric matrix this condition is equivalent
			to the one in the statement. Indeed, if
			\(c=a+\mathrm{i}\,b\) with \(a,b\in\R^n\), then
			\(\sum_{i,j}c_i\bar c_jD_{ij}=a^\top Da+b^\top Db\), and \(\sum_ic_i=0\) if and only if
			\(\sum_ia_i=\sum_ib_i=0\).
		\end{proof}
		
		\begin{lemma}[\textbf{Taylor remainder of the exponential}]
			\label{lem:resto_exp}
			Let \(D\) be a real symmetric \(n\times n\) matrix with \(0\le D_{ij}\le M\),
			and let \(t>0\). Then
			\[
			E_t=\bigl(e^{-tD_{ij}}\bigr)_{i,j}=\mathbf{1}\mathbf{1}^\top-tD+R,
			\]
			where \(0\le R_{ij}\le t^2M^2/2\), and, for every unit vector \(x\in\R^n\),
			\[
			|x^\top Rx|\le\frac{nt^2M^2}{2}.
			\]
		\end{lemma}
		
		\begin{proof}
			Write \(e^{-s}=1-s+r(s)\). By Taylor's theorem with the Lagrange form of the remainder, we have
			\(r(s)=e^{-\xi}s^2/2\) for some \(\xi\in(0,s)\), and so
			\(0\le r(s)\le s^2/2\) for \(s\ge0\). Therefore, the entries \(R_{ij}=r(tD_{ij})\)
			satisfy \(0\le R_{ij}\le t^2M^2/2\). If \(x\) is a unit vector, then
			\(\bigl(\sum_i|x_i|\bigr)^2\le n\) by the Cauchy--Schwarz inequality, and
			therefore
			\[
			|x^\top Rx|\le\sum_{i,j}|x_i||x_j|R_{ij}
			\le\frac{t^2M^2}{2}\Bigl(\sum_i|x_i|\Bigr)^2
			\le\frac{nt^2M^2}{2}. \qedhere
			\]
		\end{proof}
		
		\begin{remark}[\textbf{Rayleigh quotient}]
			\label{rem:rayleigh}
			Let \(S\) be a real symmetric \(n\times n\) matrix. By the spectral
			theorem, the matrix \(S\) has an orthonormal basis of eigenvectors \(q_1,\dots,q_n\),
			with eigenvalues \(\lambda_1,\dots,\lambda_n\). If \(z=\sum_kc_kq_k\), then
			\(z^\top Sz=\sum_k\lambda_kc_k^2\) and \(\|z\|^2=\sum_kc_k^2\), and therefore
			\[
			\lambda_{\min}(S)\,\|z\|^2\le z^\top Sz\le\lambda_{\max}(S)\,\|z\|^2
			\qquad\text{for every }z\in\R^n,
			\]
			with equality when \(z\) is an eigenvector associated with \(\lambda_{\min}(S)\) or
			\(\lambda_{\max}(S)\), respectively. In particular, the minimum and the maximum of
			\(z^\top Sz\) over the unit vectors \(z\) are \(\lambda_{\min}(S)\) and
			\(\lambda_{\max}(S)\), respectively \cite[Section~4.2]{HornJohnson2013}.
		\end{remark}
		
		\begin{lemma}[\textbf{Elementary bounds for \(D\)}]
			\label{lem:cotas_D}
			Let \(n\ge2\) and let \(D\) be a real symmetric \(n\times n\) matrix with
			\(D_{ii}=0\) and \(0\le D_{ij}\le M\). Let \(u=\mathbf{1}/\sqrt n\),
			\(V_n=\{w\in\R^n:w\cdot\mathbf{1}=0\}\), \(\ell=\lambda_{\max}(H_nDH_n)\), and
			\(c=(n-1)M\). Then the following hold.
			\begin{enumerate}
				\item For every \(w\in V_n\), we have \(w^\top Dw\le\ell\,\|w\|^2\).
				\item \(u^\top Du\le c\).
				\item For every \(w\in V_n\), we have \(|u^\top Dw|\le c\,\|w\|\).
			\end{enumerate}
		\end{lemma}
		
		\begin{proof}
			(1) Since \(H_n\) is the orthogonal projection onto \(V_n\)
			(Remark~\ref{rem:cnd_centrado}) and \(w\in V_n\), we have \(H_nw=w\). Moreover, since
			\(H_nDH_n\) is symmetric, Remark~\ref{rem:rayleigh} gives
			\[
			w^\top Dw=w^\top H_nDH_nw\le\lambda_{\max}(H_nDH_n)\,\|w\|^2=\ell\,\|w\|^2.
			\]
			
			(2) Since \(D_{ii}=0\), the sum \(\sum_{i,j}D_{ij}\) contains only the
			\(n(n-1)\) off-diagonal entries, each of which is at most \(M\).
			Therefore,
			\[
			u^\top Du=\frac1n\sum_{i,j}D_{ij}\le(n-1)M=c.
			\]
			
			(3) Since \(D_{ij}\ge0\) and every column of \(D\) sums to at most \(c\),
			the triangle inequality and the Cauchy--Schwarz inequality
			\(\sum_j|w_j|\le\sqrt n\,\|w\|\) yield
			\[
			|u^\top Dw|\le\frac1{\sqrt n}\sum_j|w_j|\sum_iD_{ij}
			\le\frac{c}{\sqrt n}\sum_j|w_j|\le c\,\|w\|. \qedhere
			\]
		\end{proof}
		
		\begin{lemma}[\textbf{Bound for \(x^\top Dx\)}]
			\label{lem:cota_D}
			Under the hypotheses of Lemma~\ref{lem:cotas_D}, let \(x\in\R^n\) be a unit
			vector. Since \(w\cdot u=(w\cdot\mathbf{1})/\sqrt n\), we have \(V_n=u^\perp\), and
			therefore we can write \(x=\alpha u+w\), with \(\alpha=x\cdot u\) and \(w\in V_n\). If
			\(\rho=\|w\|\), then \(\alpha^2+\rho^2=1\) and
			\[
			x^\top Dx\le c\alpha^2+2c\,|\alpha|\,\rho+\ell\,\rho^2.
			\]
		\end{lemma}
		
		\begin{proof}
			Since \(u\perp w\), we have \(\alpha^2+\rho^2=\|x\|^2=1\). Expanding, we get
			\[
			x^\top Dx=\alpha^2\,u^\top Du+2\alpha\,u^\top Dw+w^\top Dw.
			\]
			By Lemma~\ref{lem:cotas_D}, the first term is at most \(c\alpha^2\), the
			second is at most \(2c|\alpha|\rho\), and the third is at most
			\(\ell\rho^2\). \qedhere
		\end{proof}
		
		\begin{lemma}[\textbf{Lower bound for \(x^\top E_tx\)}]
			\label{lem:cota_E}
			Under the hypotheses of Lemma~\ref{lem:cotas_D}, let \(t>0\) with
			\(tM\le1/2\), and let \(E_t=\bigl(e^{-tD_{ij}}\bigr)_{i,j}\). Then, for every
			unit vector \(x\in\R^n\),
			\[
			x^\top E_tx\ge-t\ell-\frac52\,nt^2M^2.
			\]
		\end{lemma}
		
		\begin{proof}
			Write \(x=\alpha u+w\) as in Lemma~\ref{lem:cota_D}. By
			Lemma~\ref{lem:resto_exp}, we have \(E_t=\mathbf{1}\mathbf{1}^\top-tD+R\). Since
			\(\mathbf{1}^\top u=\sqrt n\) and \(\mathbf{1}^\top w=0\), it follows that
			\((\mathbf{1}^\top x)^2=n\alpha^2\), and therefore
			\[
			x^\top E_tx=n\alpha^2-t\,x^\top Dx+x^\top Rx.
			\]
			By Lemma~\ref{lem:cota_D} and the bound \(|x^\top Rx|\le nt^2M^2/2\), we obtain
			\[
			x^\top E_tx\ge(n-tc)\alpha^2-2tc\,|\alpha|\,\rho-t\ell\rho^2-\frac{nt^2M^2}{2}.
			\]
			The cross term \(-2tc|\alpha|\rho\) is the only one that can be large, so
			we absorb it into the term \((n-tc)\alpha^2\). Since \(c=(n-1)M\) and
			\(tM\le1/2\), we have \(tc=(n-1)\,tM\le(n-1)/2\), and therefore
			\(n-tc\ge n-(n-1)/2=(n+1)/2\). Set \(\eta=(n+1)/2\), so that
			\(n-tc\ge \eta\). By the
			inequality \(2ab\le\varepsilon a^2+b^2/\varepsilon\) with \(a=|\alpha|\),
			\(b=tc\rho\), and \(\varepsilon=\eta\), we have
			\[
			2tc\,|\alpha|\,\rho\le \eta\alpha^2+\frac{t^2c^2}{\eta}\rho^2 .
			\]
			Therefore, since \(\rho^2\le1\), we obtain
			\[
			x^\top E_tx\ge-\frac{t^2c^2}{\eta}\rho^2-t\ell\rho^2-\frac{nt^2M^2}{2}
			\ge-t\ell-\frac{t^2c^2}{\eta}-\frac{nt^2M^2}{2}.
			\]
			Finally, since \((n-1)^2\le n(n+1)\), we have
			\(c^2/\eta=2(n-1)^2M^2/(n+1)\le 2nM^2\), and therefore
			\[
			x^\top E_tx\ge-t\ell-2nt^2M^2-\frac{nt^2M^2}{2}
			=-t\ell-\frac52\,nt^2M^2. \qedhere
			\]
		\end{proof}
		
		\begin{proof}[Proof of Proposition~\ref{prop:ancho_banda}]
			Since \(K_\sigma\) is symmetric, Remark~\ref{rem:rayleigh} shows that
			\(\lambda_{\min}(K_\sigma)\) and \(\lambda_{\max}(K_\sigma)\) are the minimum
			and the maximum of \(x^\top K_\sigma x\) over the unit vectors \(x\).
			
			(1) If \(\ell=0\), then every eigenvalue of \(H_nDH_n\) is at most \(0\);
			that is, \(H_nDH_n\) is negative semidefinite. Thus, if \(a\in\R^n\) satisfies
			\(\sum_ia_i=0\), then \(H_na=a\), and therefore
			\(a^\top Da=a^\top H_nDH_na\le0\). Moreover, the diagonal entries are \(D_{ii}=\delta(A_i,A_i)=0\). Then, by
			Lemma~\ref{lem:schoenberg_finito} with \(t=1/(2\sigma^2)\), the matrix
			\(K_\sigma\) is positive semidefinite.
			
			For (2) and (3), let \(t=1/(2\sigma^2)\). By Lemma~\ref{lem:resto_exp} with
			\(M=M_D\), we have \(K_\sigma=\mathbf{1}\mathbf{1}^\top-tD+R\), where
			\(|x^\top Rx|\le nt^2M_D^2/2\) for every unit vector \(x\).
			
			(2) Let \(v\) be a unit eigenvector of \(H_nDH_n\) associated with the eigenvalue
			\(\ell>0\). Since \(v=H_n(DH_nv)/\ell\), the vector \(v\) lies in the image of
			\(H_n\), which is \(V_n=\{a\in\R^n:a\cdot\mathbf{1}=0\}\). Therefore, we have
			\(\mathbf{1}^\top v=0\) and \(H_nv=v\), and so \(v^\top Dv=v^\top H_nDH_nv=\ell\).
			Hence
			\[
			\lambda_{\min}(K_\sigma)\le v^\top K_\sigma v
			=(\mathbf{1}^\top v)^2-t\,v^\top Dv+v^\top Rv
			=-t\ell+v^\top Rv\le-t\ell+\frac{nt^2M_D^2}{2}.
			\]
			Since \(t=1/(2\sigma^2)\), this is the inequality in (2). If, moreover,
			\(\sigma^2>nM_D^2/(4\ell)\), then the right-hand side is negative, and so
			\(K_\sigma\) has a negative eigenvalue.
			
			(3) Assume that \(\sigma^2\ge M_D\), that is, \(tM_D\le1/2\). Then, by
			Lemma~\ref{lem:cota_E} with \(M=M_D\), we have
			\(x^\top K_\sigma x\ge-t\ell-\tfrac52\,nt^2M_D^2\) for every unit vector \(x\),
			and therefore
			\[
			\lambda_{\min}(K_\sigma)\ge-t\ell-\frac52\,nt^2M_D^2.
			\]
			Combining this with (2), we obtain
			\[
			-t\ell-\frac52\,nt^2M_D^2\le\lambda_{\min}(K_\sigma)\le-t\ell+\frac12\,nt^2M_D^2,
			\]
			and therefore
			\[
			\bigl|\lambda_{\min}(K_\sigma)+t\ell\bigr|
			\le\frac52\,nt^2M_D^2=\frac{5nM_D^2}{8\sigma^4}.
			\]
			Since \(t\ell=\ell/(2\sigma^2)\), this is the first inequality in (3).
			
			For the second inequality, since \(0\le(K_\sigma)_{ij}\le1\), we have, for every
			unit vector \(x\),
			\[
			x^\top K_\sigma x\le\sum_{i,j}|x_i||x_j|
			=\Bigl(\sum_i|x_i|\Bigr)^2\le n,
			\]
			and therefore \(\lambda_{\max}(K_\sigma)\le n\). On the other hand, let
			\(u=\mathbf{1}/\sqrt n\). Since \(u\) is a unit vector and
			\((K_\sigma)_{ij}\ge e^{-tM_D}\) with \(tM_D\le1/2\), we have
			\[
			\lambda_{\max}(K_\sigma)\ge u^\top K_\sigma u
			=\frac1n\sum_{i,j}(K_\sigma)_{ij}
			\ge n\,e^{-tM_D}\ge e^{-1/2}\,n. \qedhere
			\]
		\end{proof}
		
	\end{appendices}

	\newpage
	\phantomsection

\end{document}